\documentclass[preprint,11pt]{aastex}

\usepackage{wrapfig}
\usepackage{multicol}

\usepackage{graphicx}
\usepackage{natbib}
\usepackage{pdflscape} 
\usepackage{amssymb}
\usepackage{ulem} 
\usepackage{color} 
\usepackage{amsmath} 

\newcommand{\be}{\begin{equation}}
\newcommand{\ee}{\end{equation}}
\newcommand{\nn}{\mbox{} \nonumber \\ \mbox{} }
\newcommand{\ba}{\begin{eqnarray}}
\newcommand{\ea}{\end{eqnarray}}

\newcommand{\Alfven}{ Alfv\'{e}n }

\newcommand{\E}{{\bf E}}
\newcommand{\B}{{\bf B}}

\renewcommand{\div}{{\rm \,div\,}}

\newcommand\eg{{\it{{e.g., }}}}

\newcommand{\LC}{light cylinder}
\newcommand{\Lf}{{Lorentz factor}}

\newcommand{\Bf}{{magnetic field}}

\newcommand{\NS}{neutron star}

\newcommand{\EM}{electromagnetic}

\newcommand{\ms}{magnetosphere}

\def\1023{PSR J1023+0038}

\def\m28{IGR J18245--2452}

\begin{document}

\title{Relativistic magnetic explosions}

\author{Maxim  V.~Barkov $^{1}$, Praveen Sharma$^{2}$, Konstantinos N. Gourgouliatos $^{3}$, Maxim Lyutikov$^{2}$ \\
$^{1}$ Institute of Astronomy, Russian Academy of Sciences, Moscow, 119017, Russian Federation \\
$^2$ Department of Physics, Purdue University,  525 Northwestern Avenue, West Lafayette, IN, USA \\
$^3$ Department of Physics, University of Patras,  26504, Patras, Greece}

\begin{abstract}
Many  explosive astrophysical  events,  like  magnetars' bursts and flares, are magnetically driven.  We consider dynamics of such  magnetic explosions -  relativistic expansion   of highly magnetized and highly magnetically over-pressurized  clouds. The corresponding dynamics  is   qualitatively   different   from  fluid explosions due to the topological constraint of the  conservation of the magnetic flux.  Using analytical, relativistic MHD as well as  force-free calculations, we find that the creation of a relativistically  expanding, causally disconnected flow obeys a threshold condition: it requires sufficiently high initial over-pressure and sufficiently quick decrease of the pressure in the  external medium (the  pre-explosion wind). In the  subcritical case the  magnetic cloud just ``puffs-up" and quietly expands with the pre-flare wind. We also find a compact  analytical solution to the Prendergast's problem -  expansion of  force-free plasma into vacuum.
\end{abstract}

\section{Introduction}

Magnetically powered outflows form  the basis of  our understanding of pulsars \citep{GJ,1973ApJ...180L.133M,2010PASJ...62.1093M} and  Active Galactic Nuclei   \citep[AGNe,][]{1977MNRAS.179..433B,2007MNRAS.380...51K}. In  these cases the flows are typically approximated as (quasi)-stationary. Conservation of the magnetic flux then requires 
that the  toroidal \Bf\ {scales as} $B_\phi \propto 1/r$ while {the} poloidal $B_p  \propto 1/r^2$. 


Dynamics of magnetically-driven   {\it explosions} is qualitatively  different from fluid explosions,     quasi-stationary MHD flows,  and from the  1D magnetized models. The  two key ingredients are: (i) non-stationarity;  (ii)    the requirement of conservation of the  magnetic flux. They    change completely   the overall dynamics of magnetic  explosions,  if compared with magnetized steady flows,  fluid explosions and/or 1D models. Describing dynamics of multi-dimensional magnetic explosions is the main goal of the present paper.

First, the  acceleration properties   are  different between steady-state and explosive events. For example, in steady-state pulsar winds
the acceleration stops when the wind  four-velocity equals approximately the  
fast magneto-sonic 
four-velocity \citep{1969ApJ...158..727M,1970ApJ...160..971G,HeyvaertsNorman,2009ApJ...698.1570L}. At this point the flow remains magnetically dominated, hence most of the  initial magnetic power  is not spent on acceleration. 
If the initial magnetization was $\sigma_0 \equiv B^2/(4\pi \rho c^2) \gg 1$, the flow reaches terminal \Lf\  $\Gamma_w \sim \sigma_0^{1/3}$, while  the final magnetization of the flow has $\sigma_w \sim  \sigma_0^{2/3}$ - magnetization decreases during acceleration, but remains high in the coasting stage. 

Non-steady flows behave differently.
\cite{2010PhRvE..82e6305L} found  a fully analytic {\it one-dimensional}  solution, a simple wave,  for  expansion of {a} highly magnetized plasma into vacuum and/or low density medium. 
The result is somewhat surprising:   initially the plasma accelerates as $\Gamma \propto t^{1/3}$ and can reach (locally, near the head  part of the flow) terminal \Lf\ $\Gamma_f=1+2 \sigma_0$. 
  \citep[see also][]{2010ApJ...720.1490L}.  
 Thus, it was shown, that time-dependent explosions can achieve much larger {\Lf}s that the steady state flows: $1+2 \sigma_0$  for non-steady versus $\sigma_0^{1/3} $ for steady-state.
   \cite[The cold MHD solution were generalized to include pressure effects using a mathematically tricky  hodograph transformation by][]{2012PhRvE..85b6401L}.

\begin{figure}[th!]
\includegraphics[width=.99\linewidth]{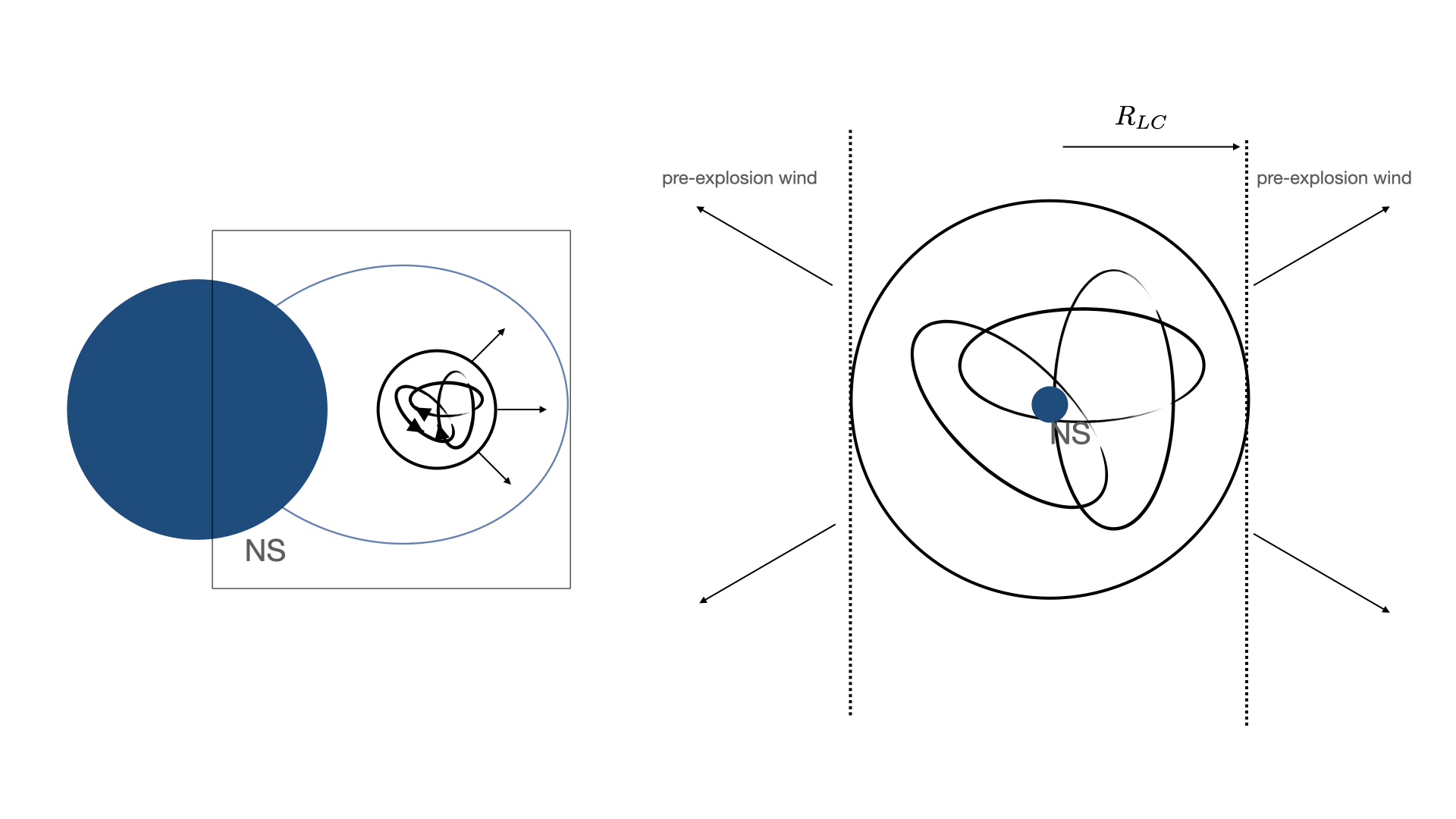}
  \caption{Cartoon of magnetic explosion  generated by a magnetar \protect\citep{2022MNRAS.509.2689L}.  A magnetic bubble with tangled \Bf\ is created near the \NS; it expands within the \ms\ and enters  the pre-explosion wind when its size reaches the size of the \LC. At this point the blob is  over-pressurized with respect to the preceding wind.  We neglect the asymmetry due to the linear motion of the blob, and the polar-angle dependence of the preceding wind.}
\label{magnetictube1}
\end{figure}

The second  important point, missed by 1D  magnetized models, is the requirement  of conservation of  the magnetic flux.   Consider a  magnetized  ball ejected during a  burst, Fig.~\ref{magnetictube1}. Fields are tangled, and there is no clear separation between poloidal and toroidal components: they are linked and must evolve similarly. The flux conservations then  requires that field scales as $B \propto 1/R(t)^2$, where $R(t)$ is the overall size of an  expanding cloud. 

 The scaling  $B \propto 1/R(t)^2$  works for more structured  solutions as well, see below, when there is a clear separation between toroidal and poloidal fields.
There is a number of analytical solution that confirm the  $B \propto 1/R(t)^2$ scaling: the Prendergast solution \citep[][see also \S \ref{Prend1}]{2005MNRAS.359..725P}, and the expanding spheromak solution  \citep{2011SoPh..270..537L}.

Thus, multi-dimensional magnetic explosions  are qualitatively different from both hydrodynamics explosions and 1D MHD flows:  the requirement of conservation of the  magnetic flux dominates  the overall dynamics of magnetic  explosions.   
This is missed in conventional hydrodynamic models of the ejections as expanding shell with parametrically  added \Bf,  as well as   one-dimensional models of magnetic disturbances.  Qualitatively, this is the same issue as the ``sigma problem'' in Pulsar Wind Nebulae: as \cite{1984ApJ...283..694K}  argued,  the presence of the global  \Bf\ changes  the dynamics completely if compared with the fluid case. \cite{2002luml.conf..381B} reformulated the problem in terms of the magnetic flux: the  $\sigma$-problem is the problem of the (properly defined) toroidal magnetic flux conservation.   Below we apply those ideas to the launching region of magnetic explosions.

To illustrate the effect of requirement of magnetic flux conservation  consider  a Solar flare-type event near the surface of the \NS\ that  produces a magnetically  disconnected magnetized cloud  with complicated  (linked)  internal magnetic structure  and total internal  pressure (\Bf\ and pairs),  slightly  exceeding the local dipole field \citep[][]{2022MNRAS.509.2689L}. Let the initial energy associated with the blob be  $E_0 \sim B_{NS}^2 R_0^3$  (where $R_0 \leq R_{NS}$ is the initial size of the blob; we assume that that  the blob originates near the surface). As the magnetic blob expands its pressure quickly becomes larger than of the surrounding dipolar one. Expansion  quickly becomes relativistic (within the \ms). Still, the energy within the blob decreases with increasing radius.

By the time the blob expands to the size of the \LC\ it's energy is $\sim E_0 (R_0/R_{LC})$. This energy should be compared with the  
energy of the  dipolar field, as measured at  the \LC, $E_0 \gg E_{LC}$, where $E_{LC} \sim  B_{LC}^2 R_{LC}^3$. 
For 
\be
\frac{ R_0 }{ R_{NS}} \geq \sqrt{  \frac{R_{NS}} {R_{LC}}}
\label{condition}
\ee
by the time the blob expanded to the size of the \LC\ it's energy is still larger than the dipolar  magnetic  energy  at the \LC.

What is the ensuing fate of the explosion? As we discuss in the present paper two outcomes are possible. First,  as the blob  expands into the wind,  magnetic field in the blob scales as $B_b \propto  1/R^2$, decreasing much faster than the \Bf\ in the wind,  $B_{\rm wind} \propto  1/R$. As a result, the expanding bubble comes into force-balance with the preceding wind near the light cylinder,
at 
\be
\frac{R_{eq}}{ R_{LC}} =  \left( \frac{ R_0}{R_{NS}} \right) ^{3/2} \frac{ R_{LC}}{R_{NS}} \sim \, \mbox{few} 
\label{Req}
\ee
(since $R_0 \leq R_{NS}$).
 Thus,  the  expelled blob quickly reaches pressure  equipartition with the wind flow, at $\sim $ few \LC\ radii.  It is  then  advected  (puffs-up) ``quietly''  with the wind. 
  
 Alternatively, as we demonstrate in this paper,   if  the explosion is powerful enough the expelled blob detonates, and is totally disrupted. The condition for the explosion is 
 \be
 E_0 \frac{R_0}{R_{LC} }\gg E_{LC}
 \label{condition2}
 \ee

The plan of the paper as follows. In \S \ref{MHD} we perform a number of 3D relativistic MHD simulations of highly-over-pressurized spheromak configurations into expanding pre-explosion wind. In \S \ref{PHAEDRA} we describe corresponding 2D force-free simulations, with 
 external \Bf\ decreasing with radius, (either as $1/r$ or $1/r^2$ - this mimics the dependance of the decreasing \Bf\ in the accelerating pulsar wind). 
These two types of simulations probe somewhat different aspects of the problem: MHD is better suited to take account of dissipative plasma effects, while force-free limit studies the extreme magnetic dynamics. In \S \ref{Prend1} we discuss a compact analytical solution to the Prendergast  problem: relativistic self-similar expansion of force-free plasma.

\section{Relativistic MHD simulations  of magnetic explosions}
\label{MHD}

\subsection{Model set-up and code}

We model the magnetized cloud as a spheromak.   Spheromaks \citep{1957ApJ...126..457C,1979NucFu..19..489R,BellanSpheromak} are axisymmetric force-free configurations. 
Two types of magnetic structures are considered: basic spheromak, \S \ref{Basicspheromak}, and higher order (3D) spheromak, \S
\ref{Higherorder}. The higher order spheromak mimics a fully entangled quasi-spherical magnetic cloud.

 The main part of the simulations' set up is the properties of the external medium. We consistently  explore several possibilities.  First,  for magnetically confined spheromak the   {\it static} external field is reduced with respect to the internal \Bf\ - this mimics over-pressurized magnetic cloud ready to explode, \S \ref{static}. Second, in addition to magnetic pressure drop at the edge of a spheromak we introduce a Hubble-like flow outside, \S \ref{Spheromakinthewind} (this mimics the  pre-explosion wind, see Fig. \ref{magnetictube1}).  We investigate various combinations of these set-ups, the dependence of the dynamics on the magnetization and the properties of the field jump at the boundary.


The simulations are  performed  with {\it PLUTO} code \citep{mbm07} in 3D relativistic MHD (RMHD) approximation, which used the WENO3 based method \citep{2009JCoPh.228.4248Y}, $3^{rd}$-order TVD Runge Kutta time stepping procedure, and an HLLD Riemann Solver \citep{mub09}. {\it PLUTO} is a public modular Godunov-type code entirely written in C intended mainly for astrophysical applications, and high Mach number flows in multiple spatial dimensions. 

 Energy, and not entropy, is the independent variable, and an Eulerian scheme is adopted when solving the equations of RMHD. Special relativity is sufficient to describe the problem at hand. The duration of the time steps is determined by the adopted CFL number of 0.2, to keep the stability of the simulations. A static uniform smooth grid  allows an adequate control of the resolution in the whole computational grid during a whole run.  
The simulated flow was approximated by ideal gas with polytropic index $\Gamma_g = 4/3$.  In this section we used non-dimensional relativistic units $p_{g,0} = \rho_0 c^2$.

\subsection{Basic spheromak}
\label{Basicspheromak}
The internal magnetic field of a basic spheromak in spherical coordinates $r-\theta-\phi$   can be expressed in terms of the flux function  $\Psi_{in}$.
\ba && 
\Psi_{in}= \left(\cos \left(\frac{\alpha  r}{a}\right)-\frac{a \sin
   \left(\frac{\alpha  r}{a}\right)}{\alpha  r}\right)  \frac{a^2 \sin ^2(\theta )}{\alpha ^2}B_0^{(in)}
   \nn &&
   \mathbf{B}_{in}= \nabla \Psi_{in} \times  \nabla \phi+  \frac{\alpha}{a}  \Psi_{in}  \nabla \phi
\label{eq:Bsp}
\ea
where $\alpha = {4.49341}$, $a$ is the radius of a spheromak, Fig.~\ref{basicSpher}. Explicit  expressions for \Bf\ are given in Appendix \ref{explicit}. 
  The spheromak is set as a sphere with radius $a = 1$. Inside  the spheromak we set initial density and pressure $\rho = p_g =  1$.

\begin{figure}[!ht]
\includegraphics[width=.99\linewidth]{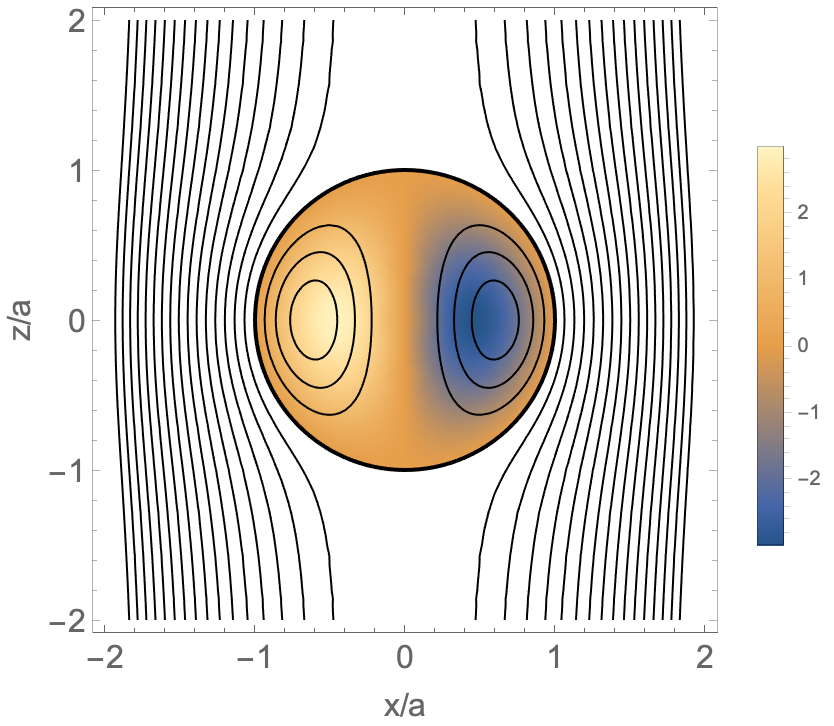}
  \caption{Magnetically confined basic spheromak ($x-z$ cut). Solid lines show  magnetic flux surfaces, color indicates the toroidal (out of the board) component of the \Bf. }
\label{basicSpher}
\end{figure}

We start with a model of magnetically confined spheromak with external \Bf\
\ba && 
\label{eq:Bsp1}
\Psi_{out} =  -\left(1-\frac{r^3}{a^3}\right) \frac{a  \sin ^2(\theta )}{2 r} B_0^{(out)}
\nn && 
\mathbf{B}^{(out)} = \nabla \Psi_{out} \times  \nabla \phi
\label{basic1}
\ea
the sketch of the configuration is shown on Fig.~\ref{basicSpher}. For magnetically confined spheromak $B^{(in)} (r=a) = B^{(out)} (r=a)$.
At large cylindrical radius  the field  corresponds  to uniform constant $B_z= B_0^{(out)}$.

\subsection{Magnetically over-pressured  basic spheromak in static  medium}
\label{static}
To initialize an explosion 
we  decrease the normalization of  external   \Bf\  
\be
B^{(out)}_0 =  \lambda B_0,
\label{lambda}
\ee  This introduces a  jump at the surface so that the spheromak is  magnetically over-pressurized. Outside of the spheromak we set  the magnetic field with  $  \lambda= 10^{-3}$.
The initial  velocities were set equal to 0. 
 

\begin{figure}[!ht]
\includegraphics[width=.83\linewidth]{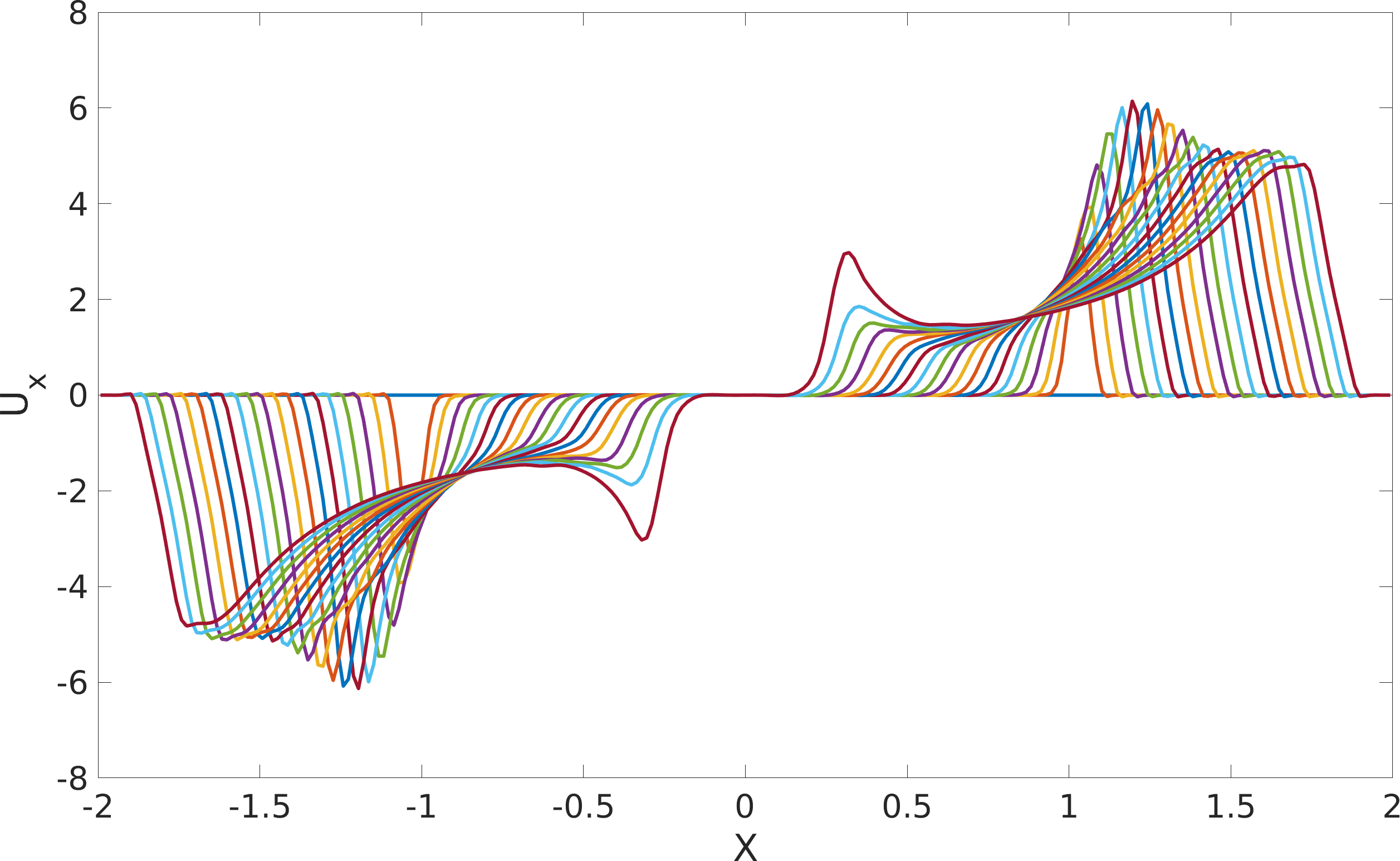}
\includegraphics[width=.83\linewidth]{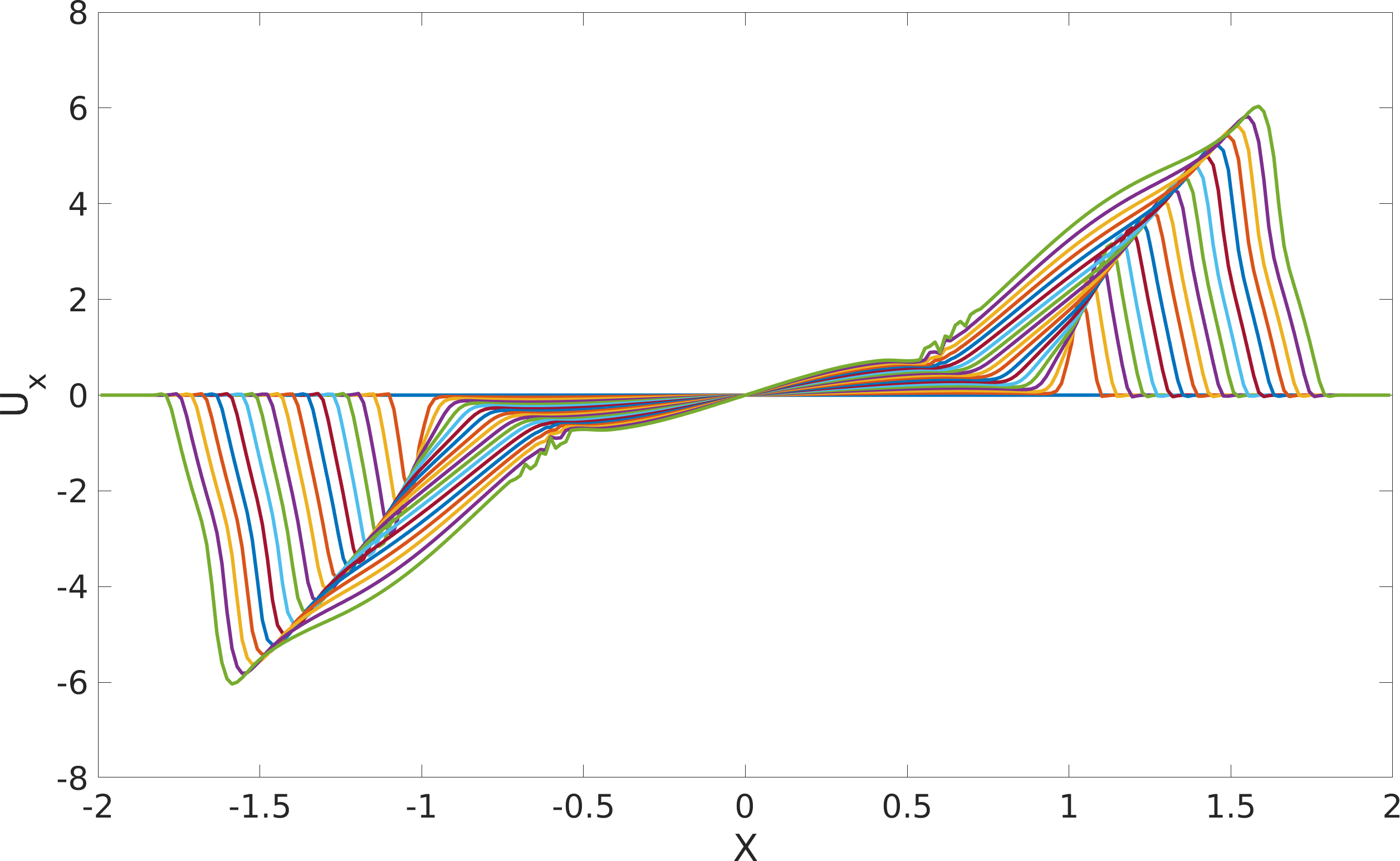}
\caption{ Comparison of the hydrodynamic flow four-velocity $U_x=\Gamma v_x$ for early stages of expansion of a relativistic  over-pressurized  bomb (profiles of four-momenta as functions of time) for different magnetizations: upper panel - high magnetization $\sigma_0 = 60$, lower panel - low magnetization  $\sigma_0 = 10^{-3}$; MHD simulations, same total over-pressure. The dynamics is drastically different: the fluid case expands much faster, and keeps accelerating; the magnetized case expands slower and is already decelerating (due to the requirement of conservation of magnetic flux.)}
\label{fig:pulseff}
\end {figure}

\begin{figure}[!ht]
\includegraphics[width=.49\linewidth]{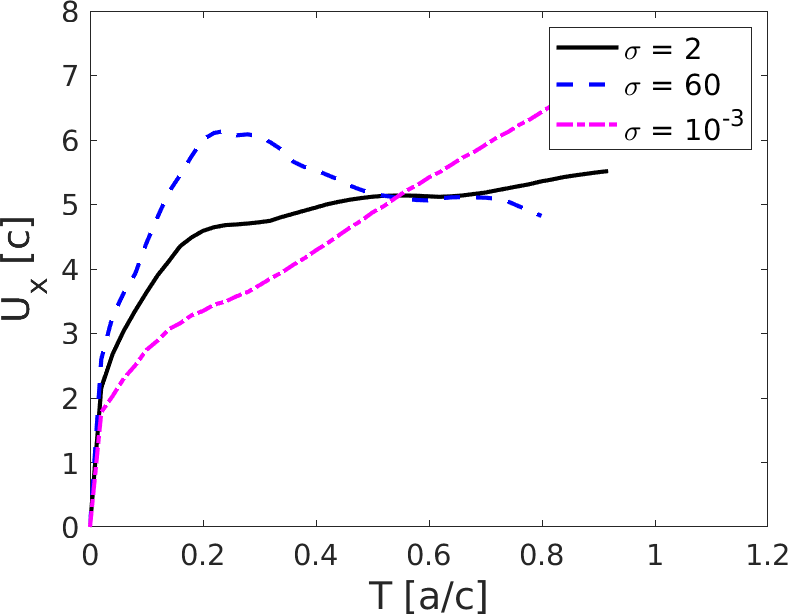}
\includegraphics[width=.49\linewidth]{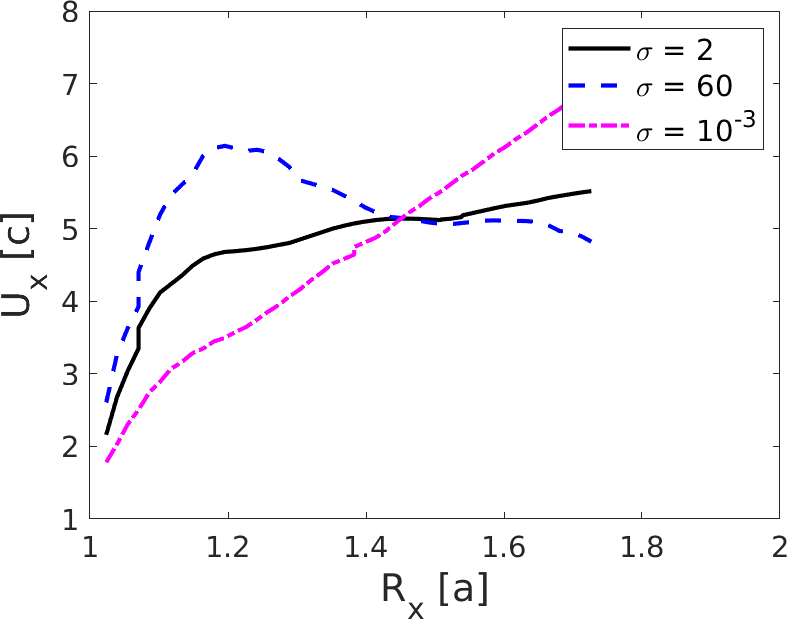}
  \caption{The maximal the hydrodynamic flow four-velocity as function of   time (left panel) and position of maximal velocity (right panel) for different initial magnetization for the case of low external pressure.}
\label{fig:sphtv}
\end{figure}


 We performed small scale simulations of spheromak evolution in rarefied surrounding media $\rho = p_g =  3\times10^{-3}$.   We choose magnetic field which corresponds to high magnetization $\sigma = B^2/4\pi(\epsilon_g +p_g) = 60$, medium magnetization $\sigma = 2$ and low magnetization $\sigma = 1/10^3$. The initial total pressure was kept the same in all simulations $\epsilon_m+p_m+\epsilon_g+p_g ={\rm  const.}$ and for $\sigma = 60$ we choose $p_g = 1$. So, reducing the magnetic pressure we increase the plasma pressure respectively keeping total energy density the same. Here we performed so-called  small scale simulations with the grid size was taken to be  $x, y \mbox{ and } z \in [-2, 2]$, in initial spheromak radius units ``a".  The total number of cells in each direction: $ N_x = N_y  = N_z =256$. 

The examples of velocities profile in X-direction (equatorial plane of spheromak) are shown in Fig.~\ref{fig:pulseff}. The strongly magnetized solution significantly differ from low magnetized one. Low magnetized solution shows  graduate acceleration of flow on its boundary  and due to absence of hoop stress the  relatively slow acceleration in the central zone. The highly magnetized case shows no expansion inside the spheromak so far rarefaction wave did not reach the given radius. The shock formed on the spheromak's surface accelerated quickly at first, but at some point outflow start to decelerate.  

The maximal velocity is achieved at the outer layers of the spheromak Fig.~\ref{fig:pulseff}. So the time evolution of maximal velocity and radial evolution of maximal velocity position depends proportionally Fig.~\ref{fig:sphtv}. In the low magnetized case the maximal velocity of the flow grows with the time. The middle magnetized case shows very slow growth with time. The strongly magnetized case shows fast initial rise and slow down later on. In late phases the  speed of outflow is {\it  inversely}  proportional to the magnetization of spheromak.

\subsection{Over-pressured spheromak in external Hubble-like  flow (the wind)}
\label{Spheromakinthewind}

Next we model interaction of the ejected magnetic blob with the preexisting wind. We assume that a magnetic cloud reached a size of the \LC, and starts  to interact with the preexisting wind. The wind is accelerating linearly, with $v \propto r$. Instead of varying the  properties of the wind (\eg mass loss rate) it is more convenient to vary {the} velocity of the wind. We use different scalings $v = \eta_v c r /a$,  with $\eta_v $ in the range $ 0.01 \leq \eta_v \leq   0.1$. Small values of  $\eta_v $ correspond to nearly static external configurations (this allows comparison with results in \S  \ref{Basicspheromak}) , while large $\eta_v \to 1$ mimic realistic pulsar winds. 

In this section we used  large scale simulations the grid size was taken to be  $x, y \mbox{ and } z \in [-4.5, 4.5]$, in units of initial spheromak radius.  To probe different velocity expansion, we performed simulations with total number of cells in each direction: $ N_x = N_y  = N_z =384$. 

\begin{figure}[th!]
	\includegraphics[width=.49\linewidth]{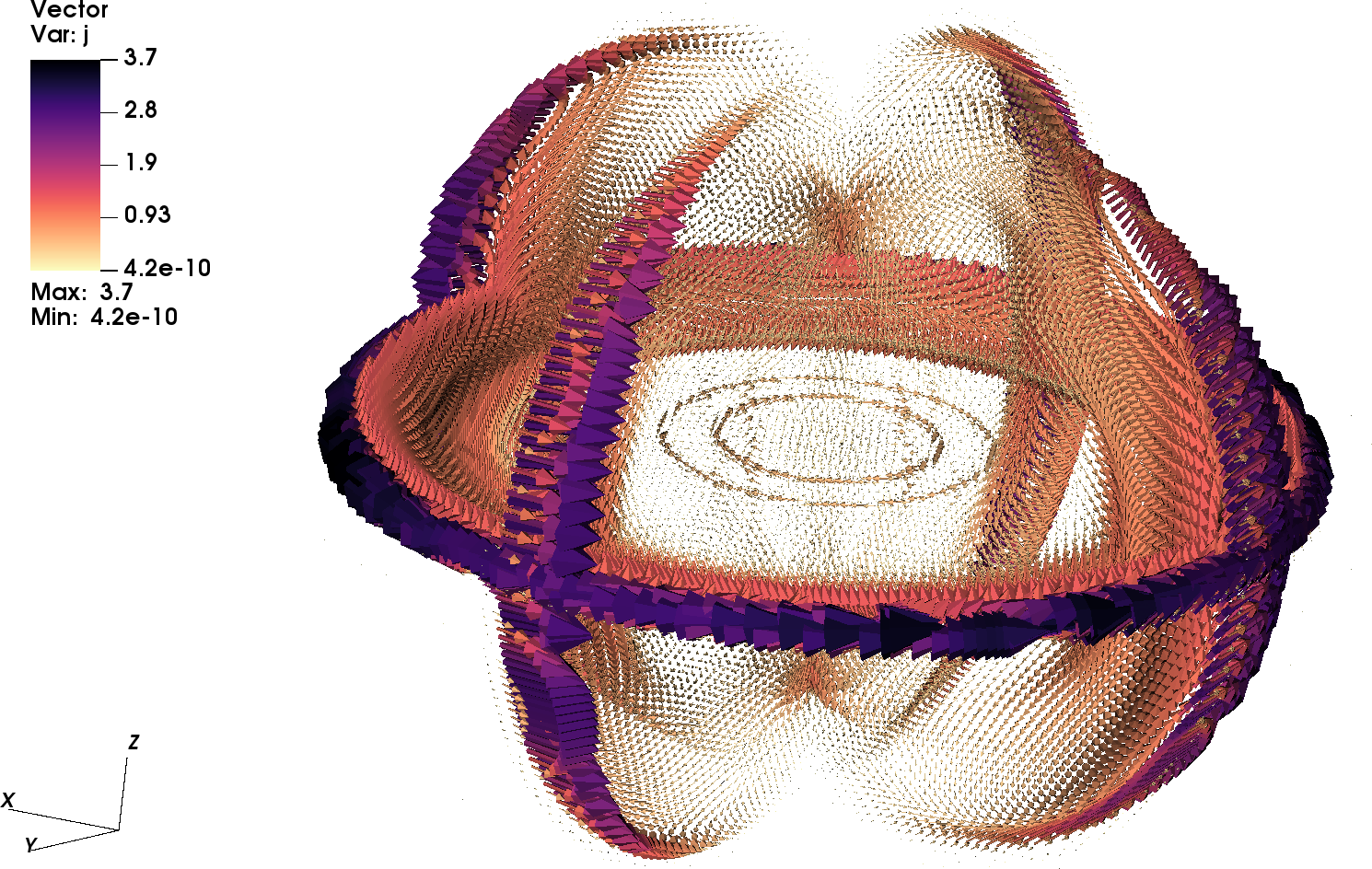}
	\includegraphics[width=.49\linewidth]{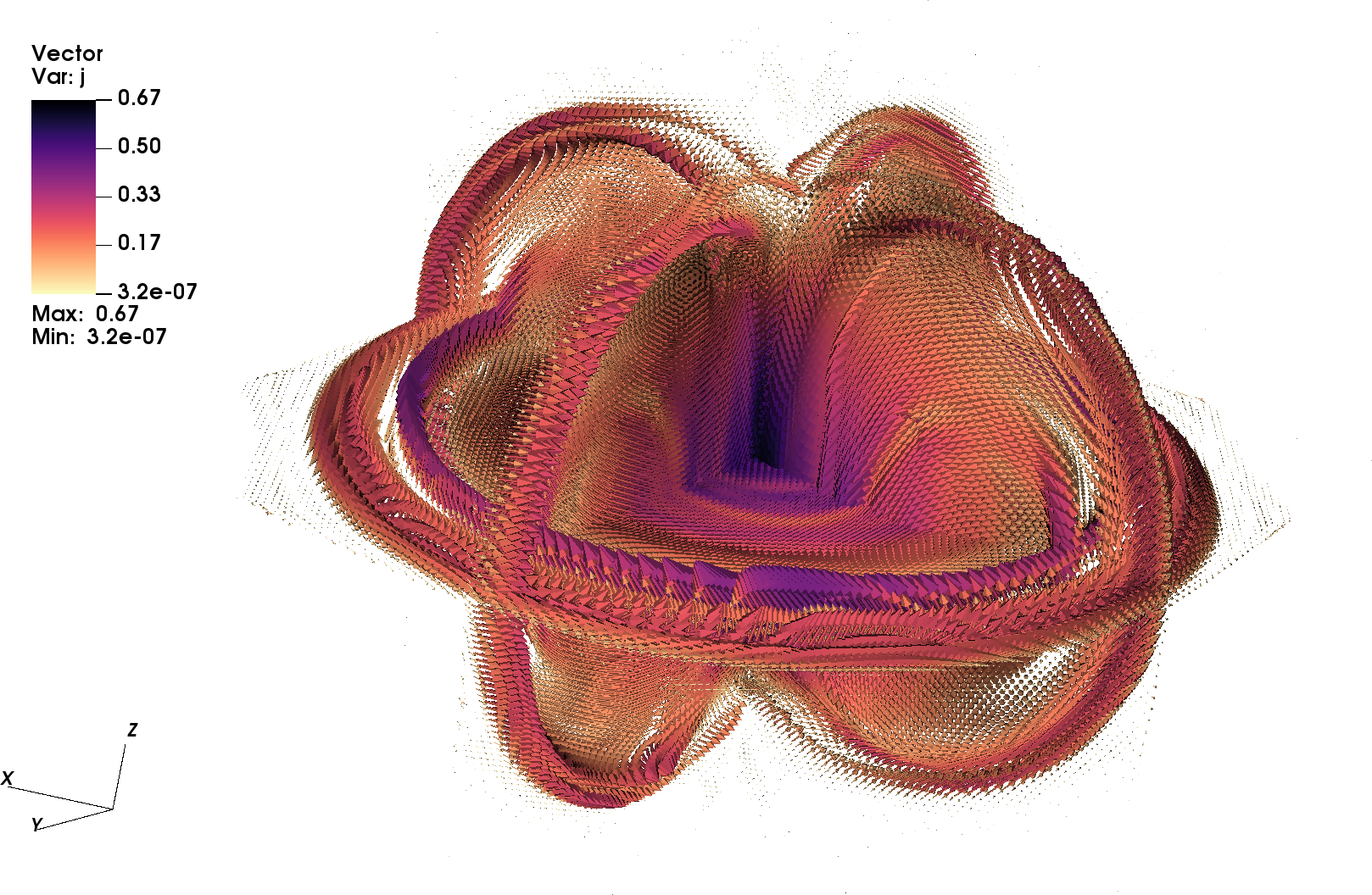}
\\
\includegraphics[width=.49\linewidth]{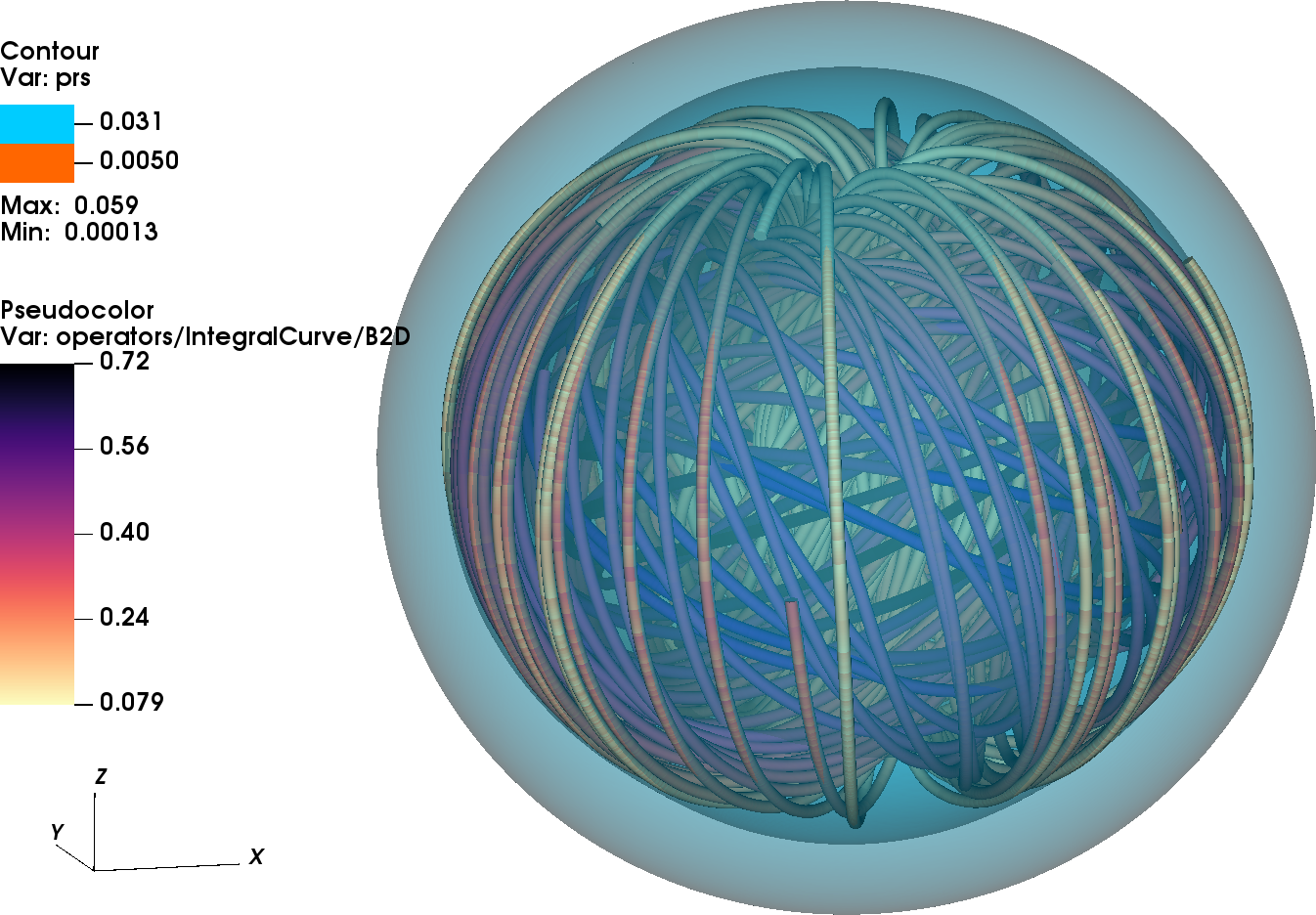}
	\includegraphics[width=.49\linewidth]{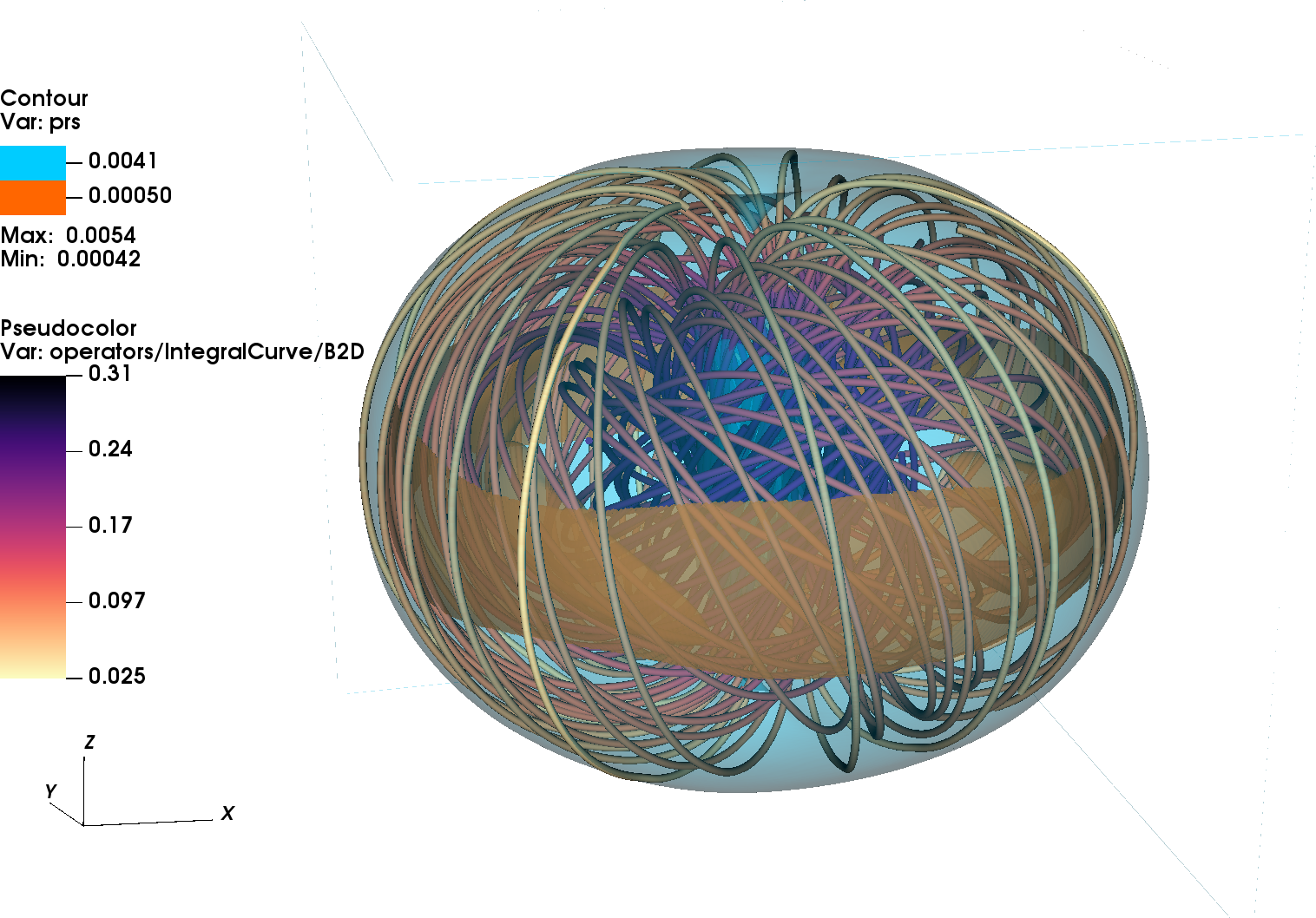}
	\caption{Current density  (top row) and  pressure distribution (by contours; \Bf\ lines are also shown - bottom row)   for basic spheromak  in external  for Hubble expanding environment with $\eta_v =  0.1$  (left, exploding) at $t=7 [a/c]$,  and $\eta_v  = 0.01 $  (right, puffing-up) at $t=70 [a/c]$. In the exploding case most of the current is concentrated near the surface - see  \S \protect\ref{Prend1} for analytical explanation.}
	\label{fig:sbjvec}
\end{figure}

In the wind zone spheromak can be described in the comoving rest frame. In the wind zone the toroidal component of magnetic field dominates with radius, so in this section we can choose Z axis direction along large scale toroidal magnetic field \citep{2007MNRAS.380...51K,2009MNRAS.394.1182K}.  The magnetic strength of the spheromak was chosen to set $\sigma = 3$ at the spheromak center and uniform density and pressure $\rho = p_g =  1$. The current density  and  pressure distribution  are  plotted in  Fig.~\ref{fig:sbjvec}. 

\begin{figure}[!ht]
	\includegraphics[width=.49\linewidth]{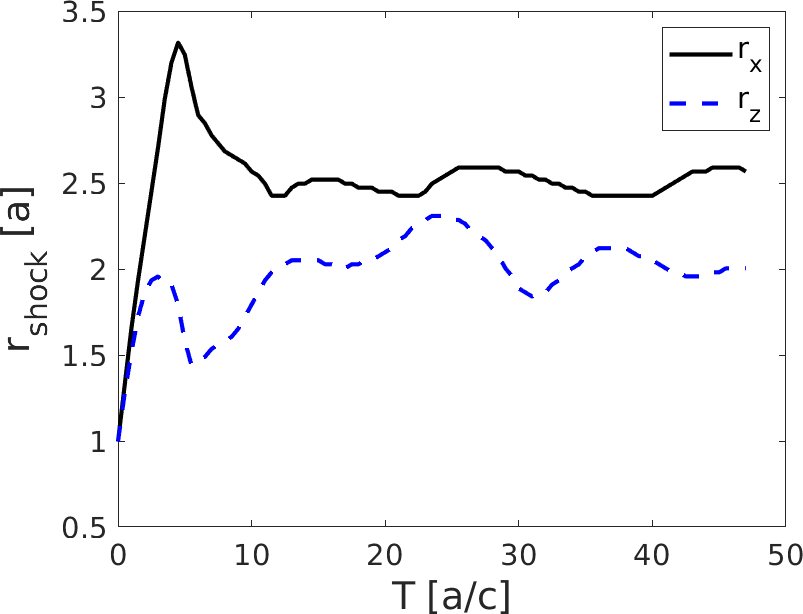}
	\caption{Evolution of  over-pressured  basic spheromak in static  medium.  Solid line: extension  in $x$-direction and dashed line in $z$-direction for  initial magnetization $\sigma = 3$, initial external density and pressure $\rho = p_g =  1/16$. The spheromak just puff-up, no explosion is generated.}
	\label{fig:sbb}
\end{figure}

 In the absence of the expansion (see \S \ref{Spheromakinthewind})  the  highly magnetized over-pressurized spheromak expands up to new equilibrium radius which can be estimated as $r_{eq}\approx a(p_{tot,in}/P_{tot,out})^{1/4}$  \citep{2011SoPh..270..537L} and for chosen parameters  $r_{eq}\approx 2a$. The radial evolution of spheromak in stationary environment with a drop of the external pressure in 16 times are shown on Fig.~\ref{fig:sbb}.  The initial expansion in equatorial plane (X-direction) after passing the equilibrium radius slows down, and later switched to shrinking and saturation near $2.5a$. In polar direction (Z-axis) expansion was not so strong and saturates near  $2a$.

\begin{figure}[th!]
\includegraphics[width=.49\linewidth]{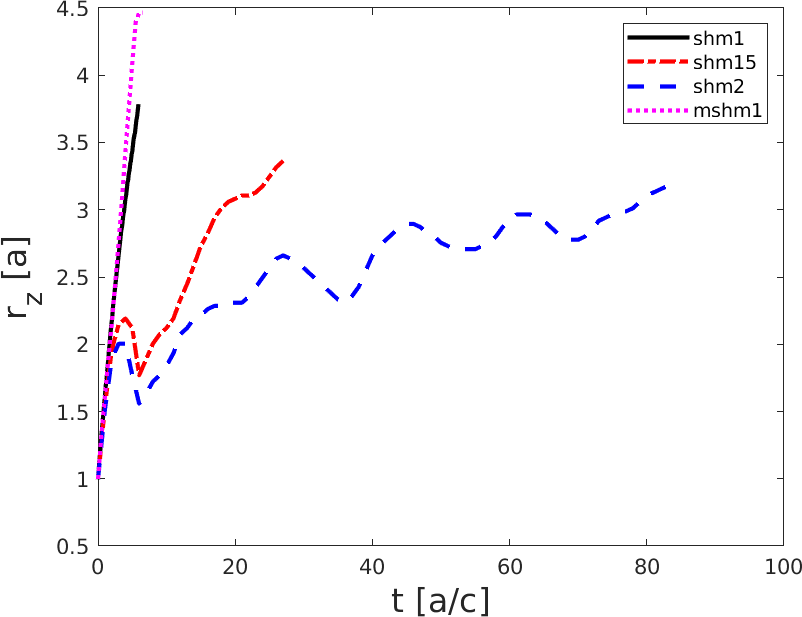}
\includegraphics[width=.49\linewidth]{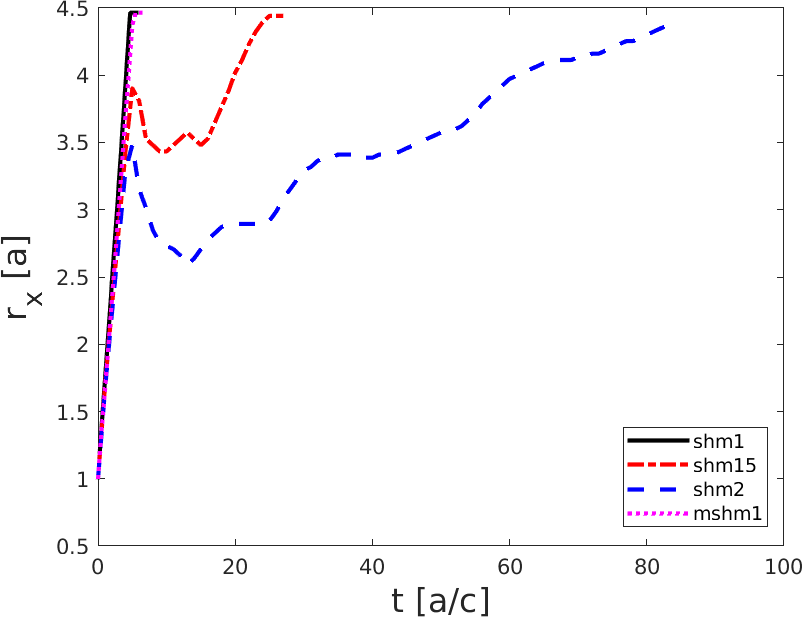}
  \caption{Radius of spheromak  (along $z$ direction left panel,  and $x$ direction right panel) for different Hubble expanding environment,  $\eta_v = 0.1 $ -- the solid line,  $\eta_v = 0.032$ -- dot-dashed line, 
 $\eta_v = 0.01$ -- dashed line and the ``high order'' spheromak with  $\eta_v = 0.1$ -- the doted line. Here we clearly see two different kind of solutions: 1) explosion $\eta_v = 0.1 $ like and 2) hydro-static puffing up  $\eta_v \le 0.032 $.
}
\label{fig:shubblert}
\end{figure}

\begin{figure}[th!]
	\includegraphics[width=.49\linewidth]{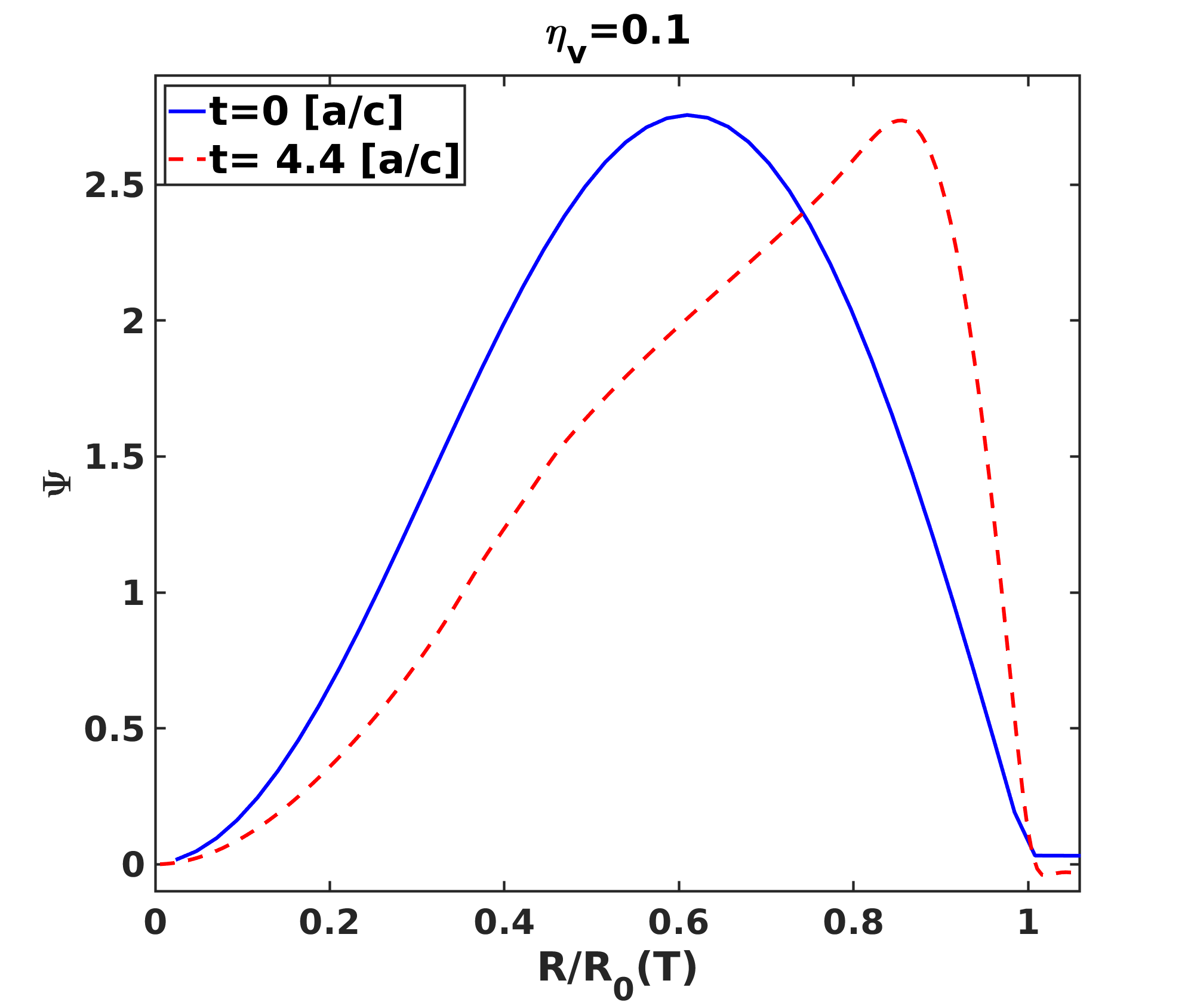}
	\includegraphics[width=.49\linewidth]{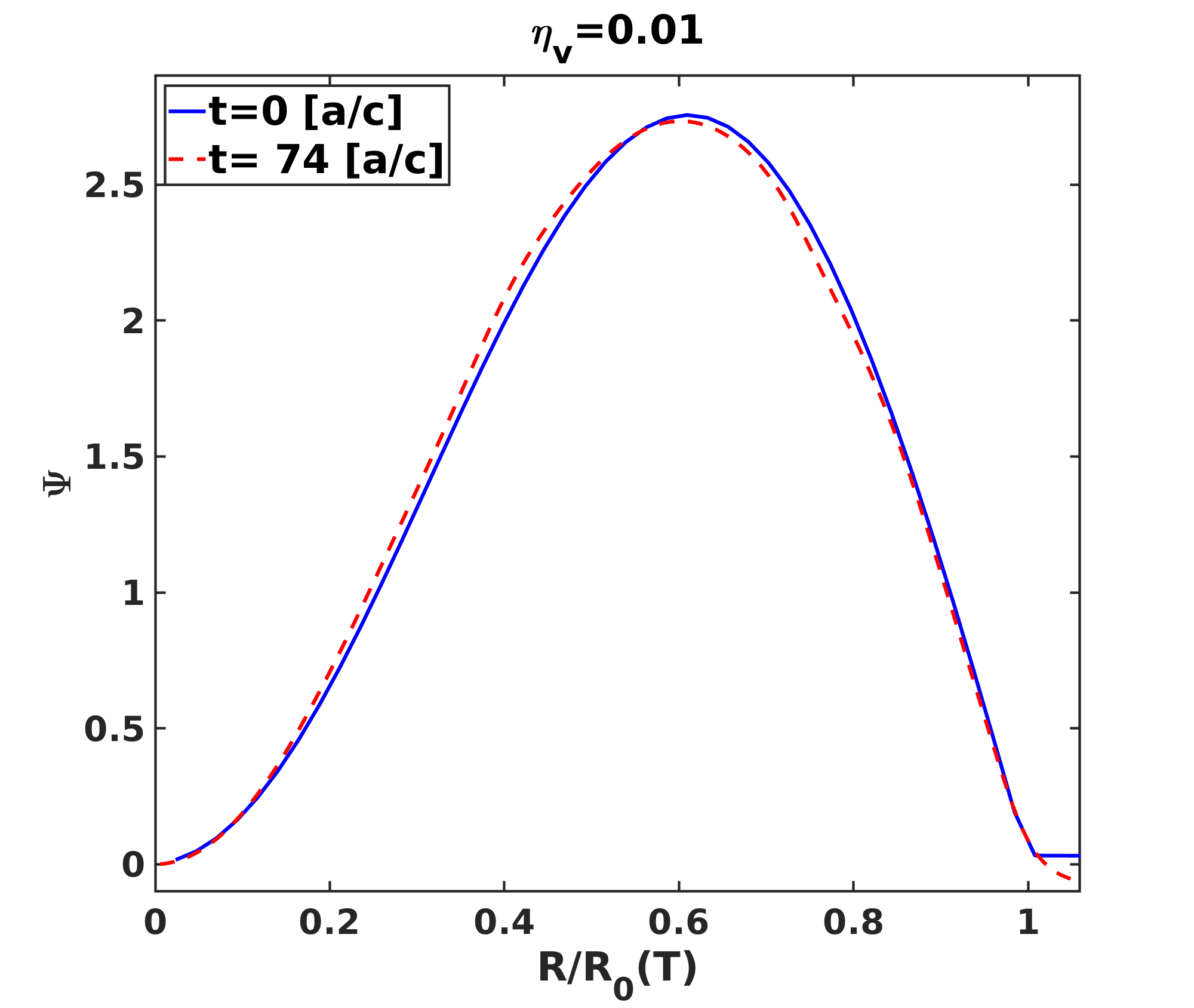}
	\caption{Evolution of the distribution of the magnetic flux $\Psi = -2\pi\int_{0}^{R} B_z(r) r dr$  in the spheromak  for different Hubble expanding environment,  $\eta_v = 0.1 $ (left panel - exploding case, flux is concentrated near the surface),  $\eta_v = 0.01$ -- (right panel - non-exploding case, flux distribution remains mostly unchanged).}
	\label{fig:sbhMF}
\end{figure}

\begin{figure}[th!]
\includegraphics[width=.33\linewidth]{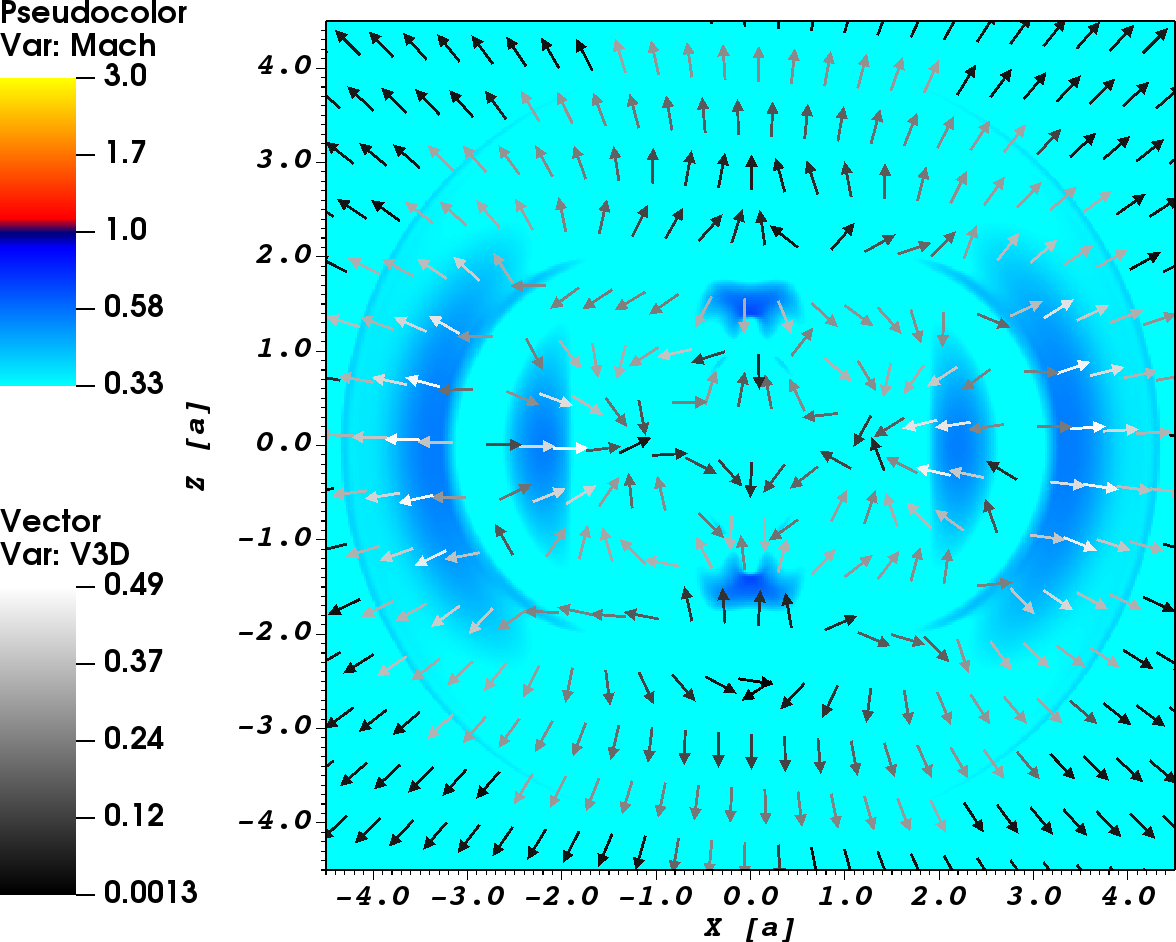}
\includegraphics[width=.33\linewidth]{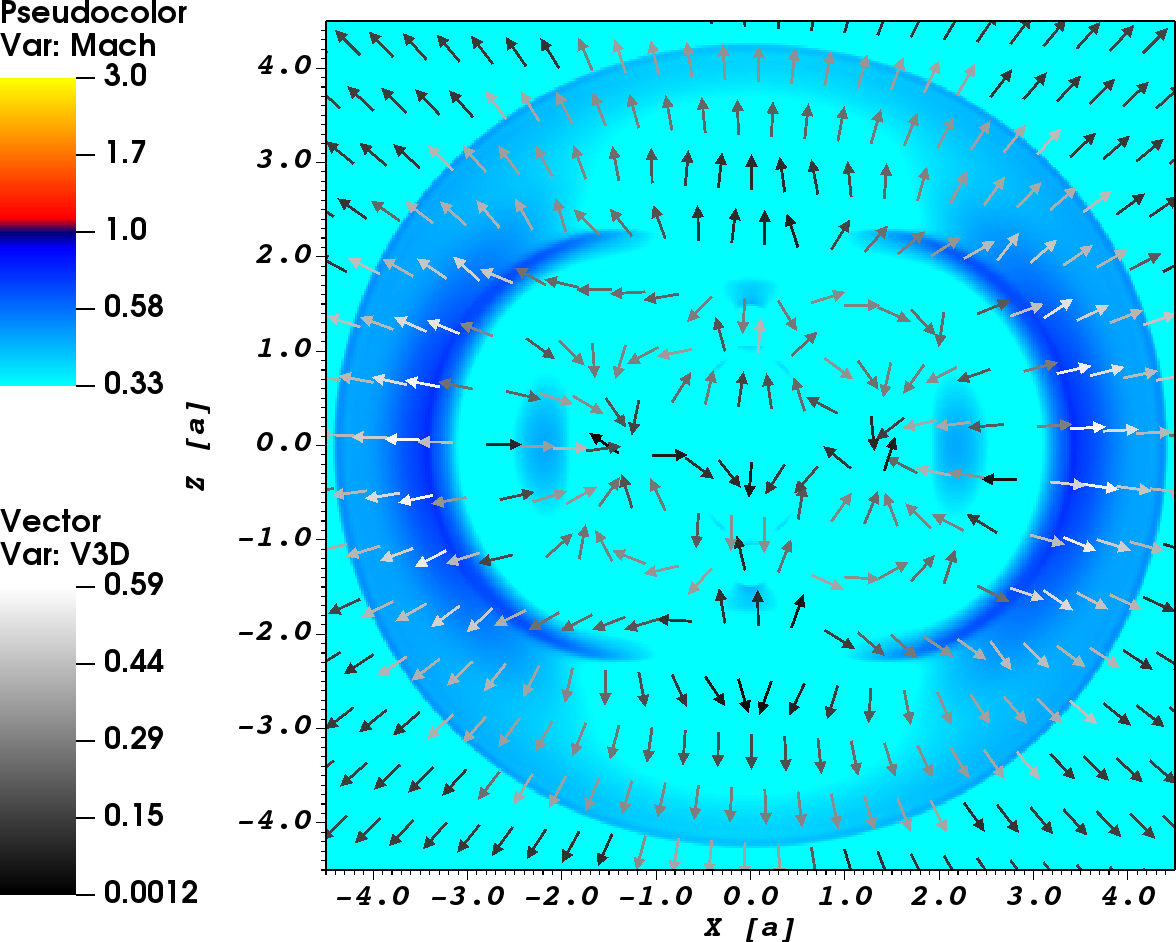}
\includegraphics[width=.33\linewidth]{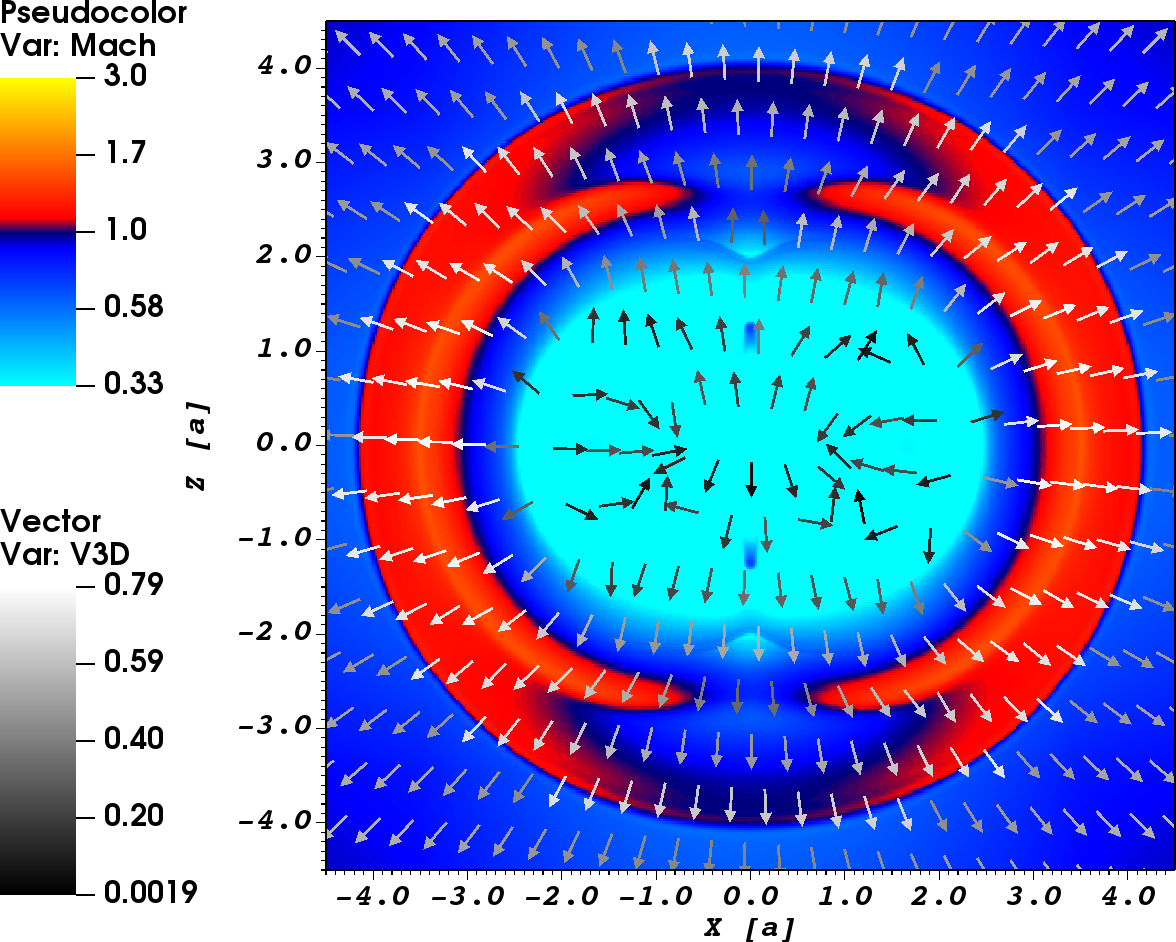}
  \caption{Mach number for fast magnetosonic speed by color and velocity field for spheromaks expanding into different Hubble-like   environment,  $\eta_v = 0.01 $  (left),  $\eta_v = 0$ (center) , 
 $ \eta_v = 0.1$  (right). (Qualitatively, red colors indicate formation of causally disconnected expanding regions.)}
\label{fig:shmach}
\end{figure}   


The evolution of $B_z$ magnetic field component distribution and velocity field are shown in  Appendix~\ref{sec:mhdBz}. 
An interesting fact is the small dependence of explosion process on the order of the spheromak. The spheromak transforms to shell like structure with bounced back central part. After bouncing off the external layers, the spheromak shrinks and oscillates continuing its expansion. This process clearly can be see on the  evolution of X and Z-direction spheromak radii on  Fig.~\ref{fig:shubblert}. 
The evolution of magnetic flux ( $\Psi = -2\pi\int_{0}^{R} B_z(r) r dr$) is presented on Fig.~\ref{fig:sbhMF}. Here we clearly see difference  in magnetic flux distribution for the explosion like solution ($\eta_v = 0.1$) between initial and evolved ones. In the case of slow expansion ($\eta_v=0.01$) magnetic flux of the spheromak preserve its shape for many tens of dynamical times. The explosion like solutions leads to strong redistribution of magnetic field in spheromak. Fig.~\ref{fig:shmach} 
illustrates the key difference in the expansion process:   the Mach number maps for fast magnetosonic speed on the 
\citep[We calculate the Alfven velocity as $v_a/c = \sqrt{\sigma/(\sigma+1)}$ and the fast magnetosonic speed can be estimated as $v_f/c = \sqrt{(\sigma+c_s^2)/(\sigma+1)}$][]{2003MNRAS.339..765L}.  In the explosion-like expansion we clearly see supersonically expanding parts of the flow. 

\subsubsection{Equilibrium spheromak in external Hubble-like  flow (the wind)}

\begin{figure}[th!]
	\includegraphics[width=.49\linewidth]{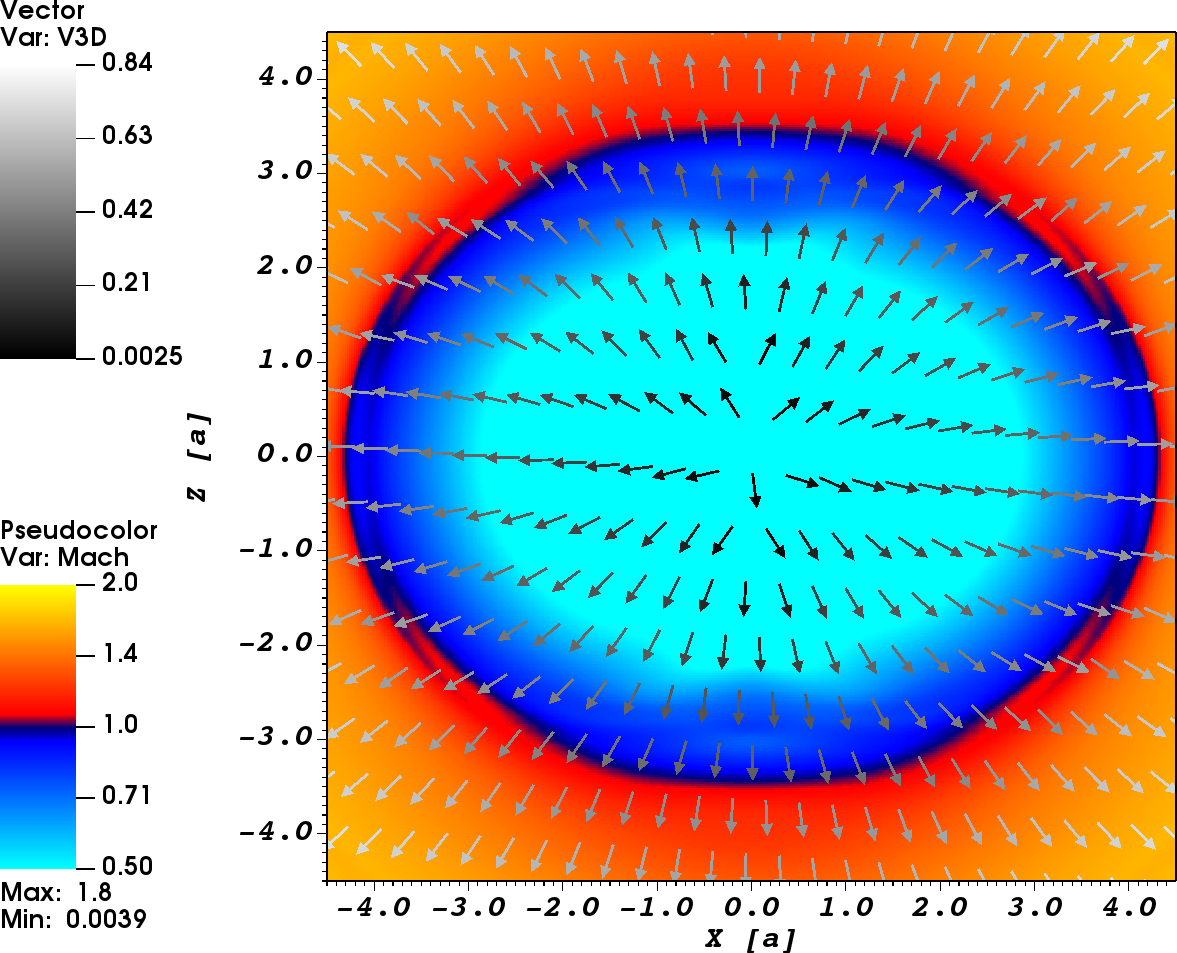}
	\caption{Expansion of  spheromak initially in equilibrium  with Hubble-expanding environment with $\eta_v= 0.3 $ for time moment  $t=6 [a/c]$.  The color shows Mach number and arrows show velocity field.  If the spheromak initially was in equilibrium even for  fast expansion speed ($\eta_v = 0.3$)  the flow remains subsonic, no explosion is generated, in contrast to over-pressured spheromaks (see Fig.~\ref{fig:shmach}). 
	}
	\label{fig:eqsh}
\end{figure}

Finally,  we investigated evolution of the spheromak in the Hubble expanding wind which initially was in pressure equilibrium. We increase the expansion speed up to $\eta_v= 0.3$ to increase the effect of expansion. The expansion takes place in the subsonic regime during all the simulation evolution.  The Mach number distribution and velocity field near the end of the simulation are presented in Fig.~\ref{fig:eqsh}. The flow outside of the spheromak is supersonic, but spheromak itself is subsonic. We did not see the change of the magnetic field topology and its significant redistribution in radial direction. 

\subsection{Higher order spheromak}
  \label{Higherorder}
  
\begin{figure}[!ht]
	\includegraphics[width=.49\linewidth]{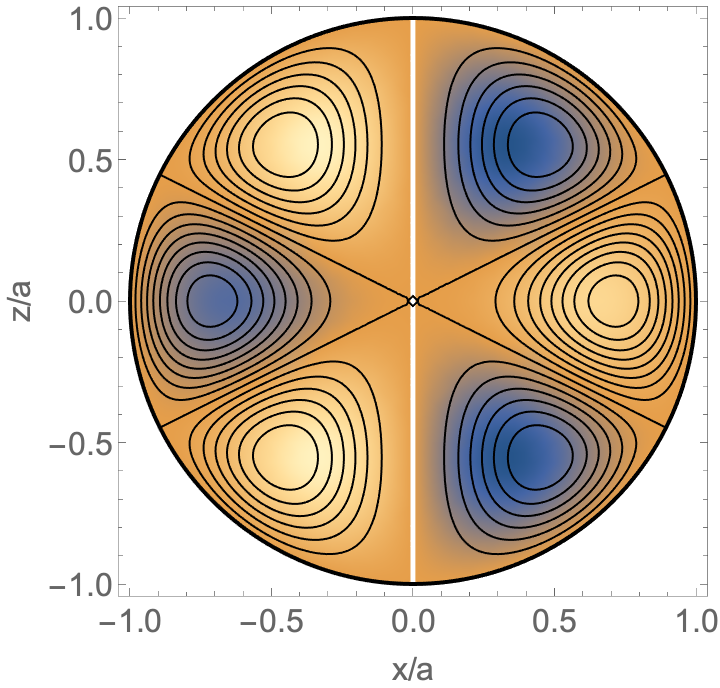}
	\caption{The 3d order spheromak ($x-z$ cut). Solid lines show  magnetic flux surfaces, color indicates the toroidal (out of the board) component of the \Bf. }
	\label{fig:3Spher}
\end{figure}

Next, we used the same setup as in the previous Section~\ref{Spheromakinthewind}, but we change topology of the magnetic field to the 3rd order spheromak with the flux function
\ba 
\Psi_3 & \propto& \left(1+5 \cos ^4(\theta )-6 \cos ^2(\theta )\right)  \nonumber \\
&\times& \left(\frac{15 a^3 \sin \left(\frac{\alpha 
   r}{a}\right)}{\alpha ^3 r^3}-\frac{15 a^2 \cos \left(\frac{\alpha  r}{a}\right)}{\alpha ^2
   r^2}-\frac{6 a \sin \left(\frac{\alpha  r}{a}\right)}{\alpha  r}+\cos \left(\frac{\alpha 
   r}{a}\right)\right)
\label{eq:ms}
\ea
where $\alpha = {6.98793}$, Fig.~\ref{fig:3Spher}. This magnetic field has three changes of magnetic polarity in the $\theta$ direction.  This configuration mimics  expected  highly tangled \Bf\ of the ejection.

\begin{figure}[th!]
	\includegraphics[width=.49\linewidth]{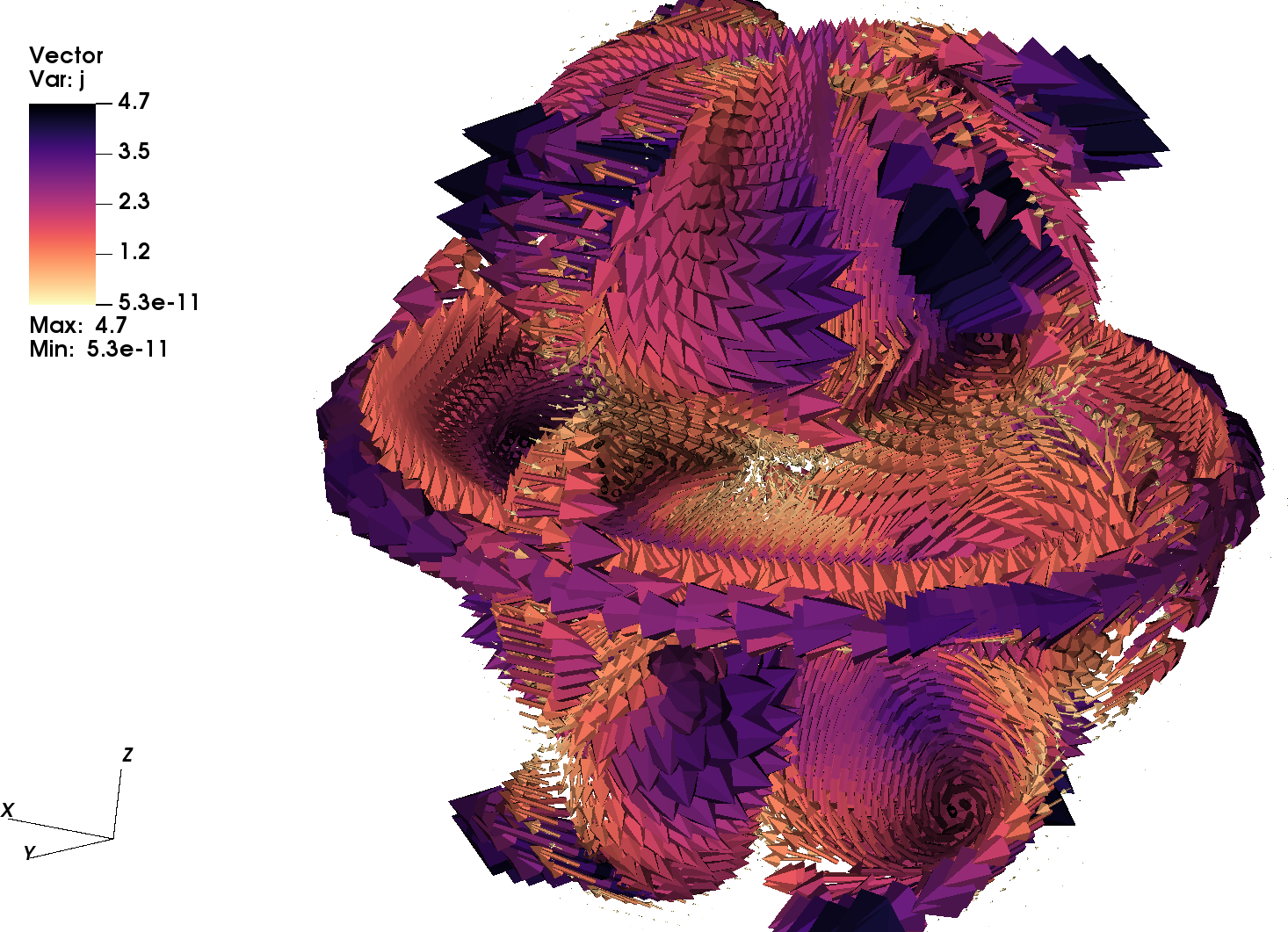}
	\includegraphics[width=.49\linewidth]{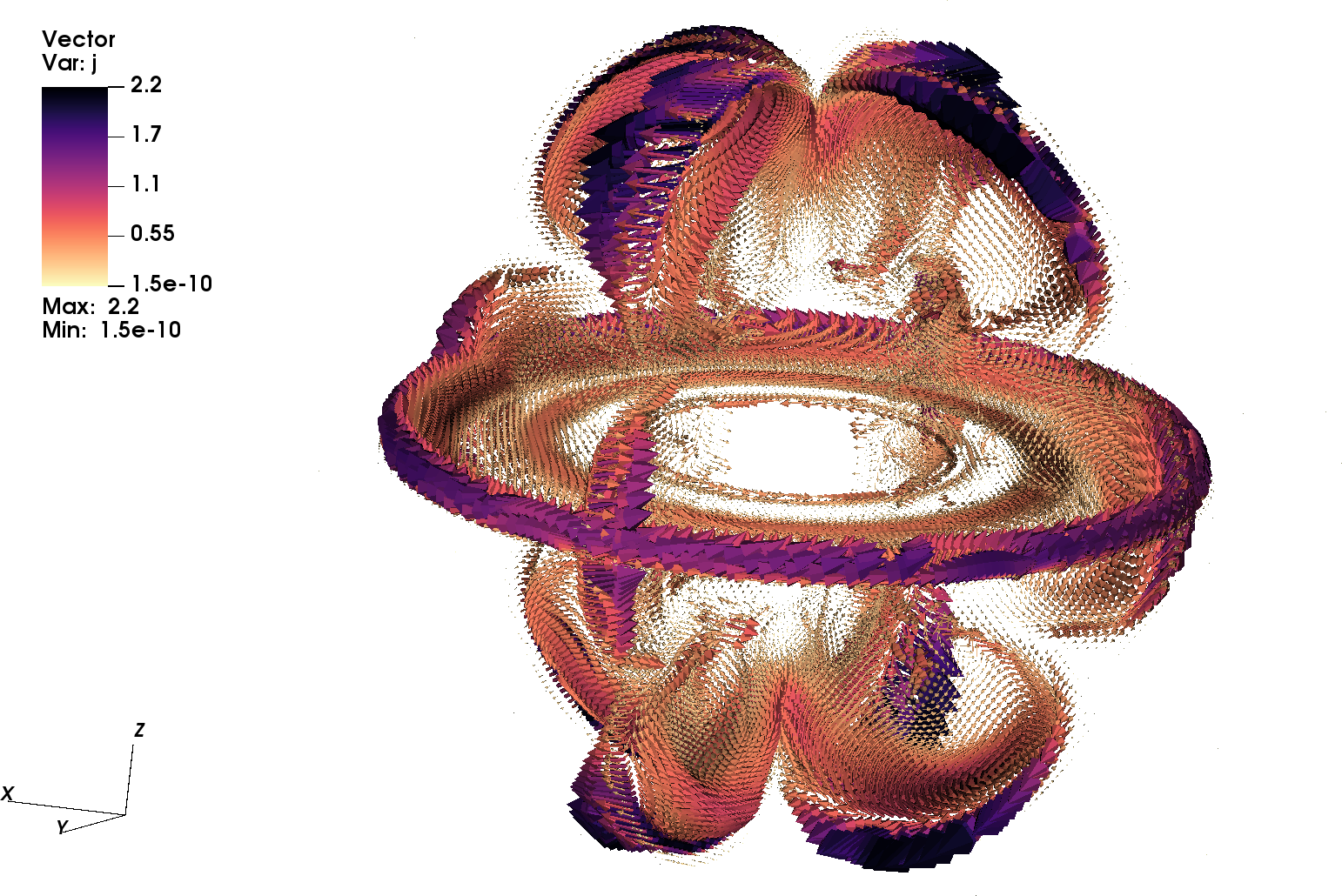}
	\caption{ Current density for higher order  spheromak  in external  for Hubble expanding environment with $\eta_v = 0.1$  (exploding)  at  $t=1.5 a/c$ (left) and $t=4.5$ (right).}
	\label{fig:msbjvec}
\end{figure}

\begin{figure}[th!]
	\includegraphics[width=.49\linewidth]{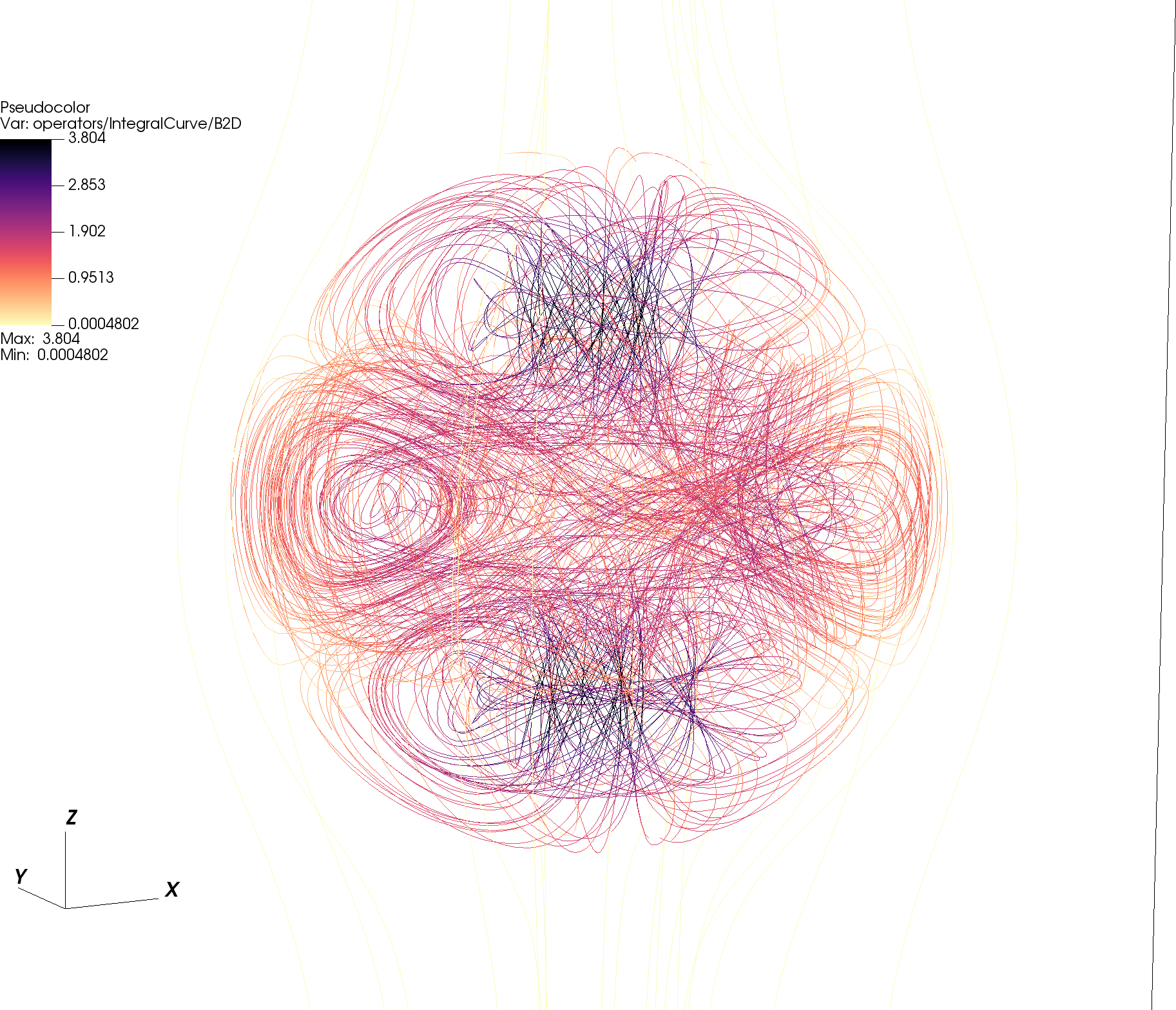}
	\includegraphics[width=.49\linewidth]{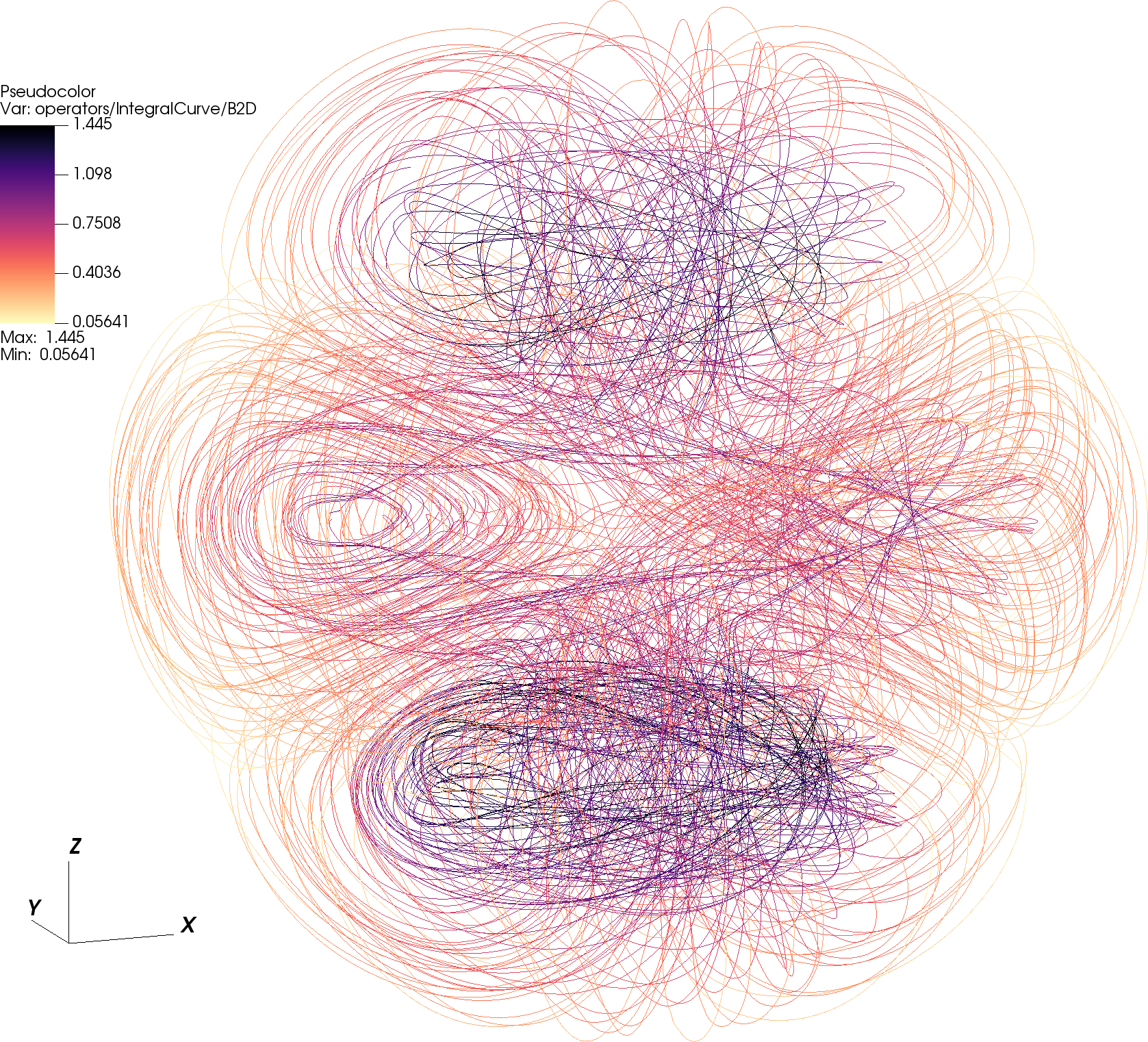}
	\includegraphics[width=.49\linewidth]{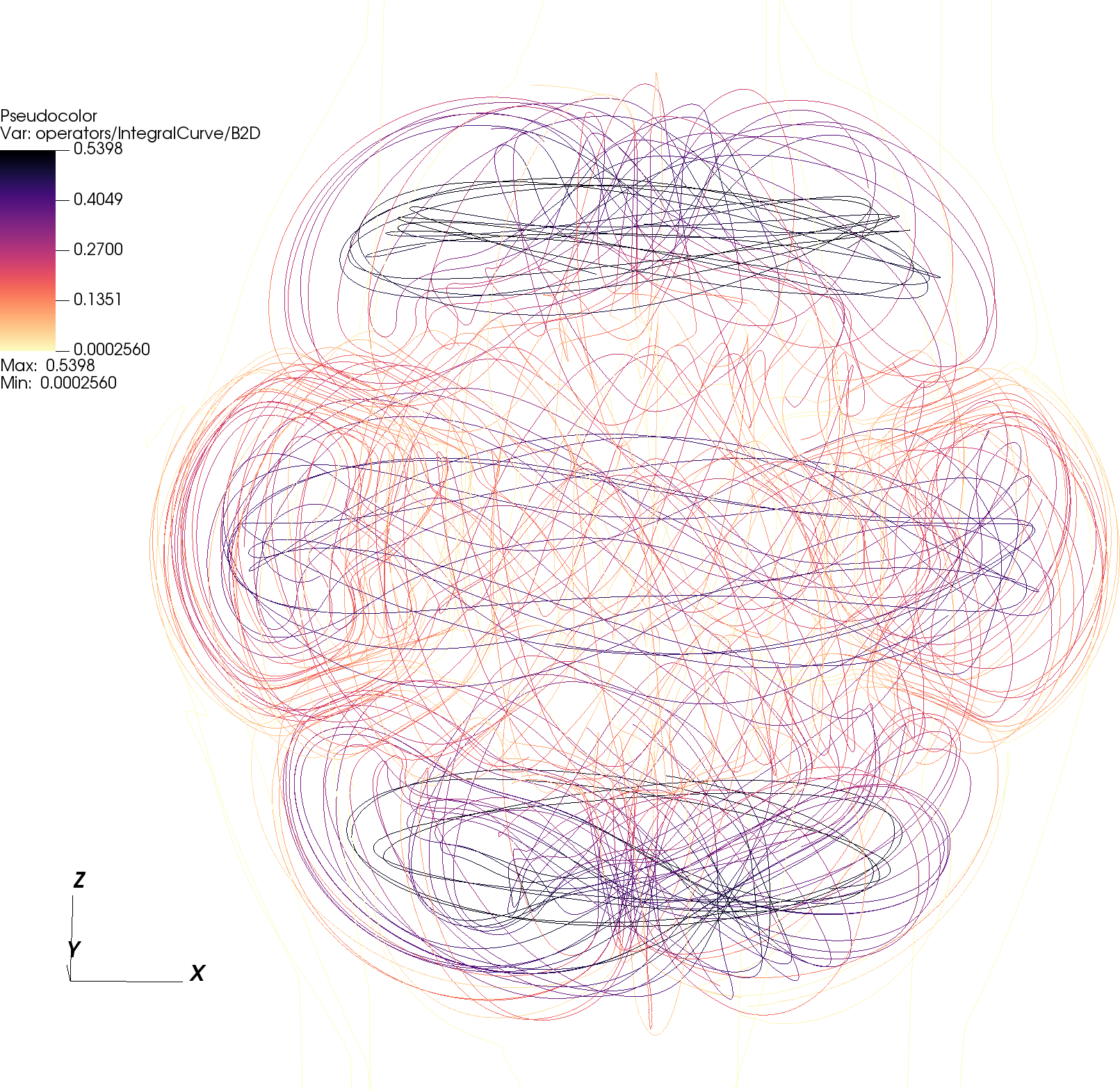}
	\includegraphics[width=.49\linewidth]{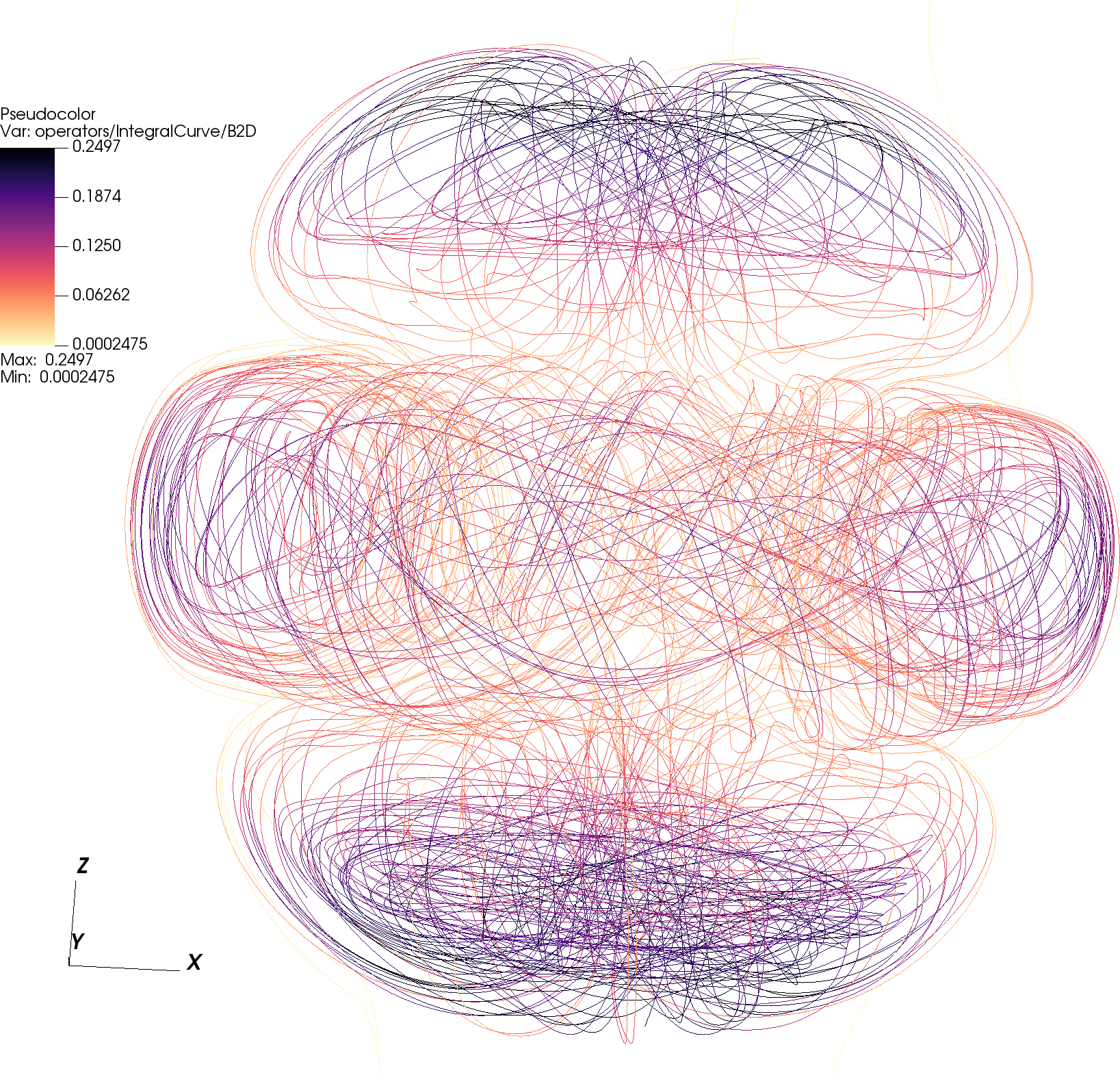}
	\caption{The magnetic field structure evolution for the ``high order'' spheromak expansion with Hubble expanding environment with $\eta_v = 0.1$ for different time moments  $t= 0 [a/c]$  (top right)  $t=1 [a/c]$ (top left),  $t= 2.5 [a/c]$  (bottom right)  $t=4.5 [a/c]$ (bottom left). The  \Bf\ strength is shown by color: notice that field is concentrated near the surface, as predicted by the analytics, see \S  \protect\ref{Prend1}.}
	\label{fig:mshubblem1fl}
\end{figure}

\begin{figure}[th!]
\includegraphics[width=.49\linewidth]{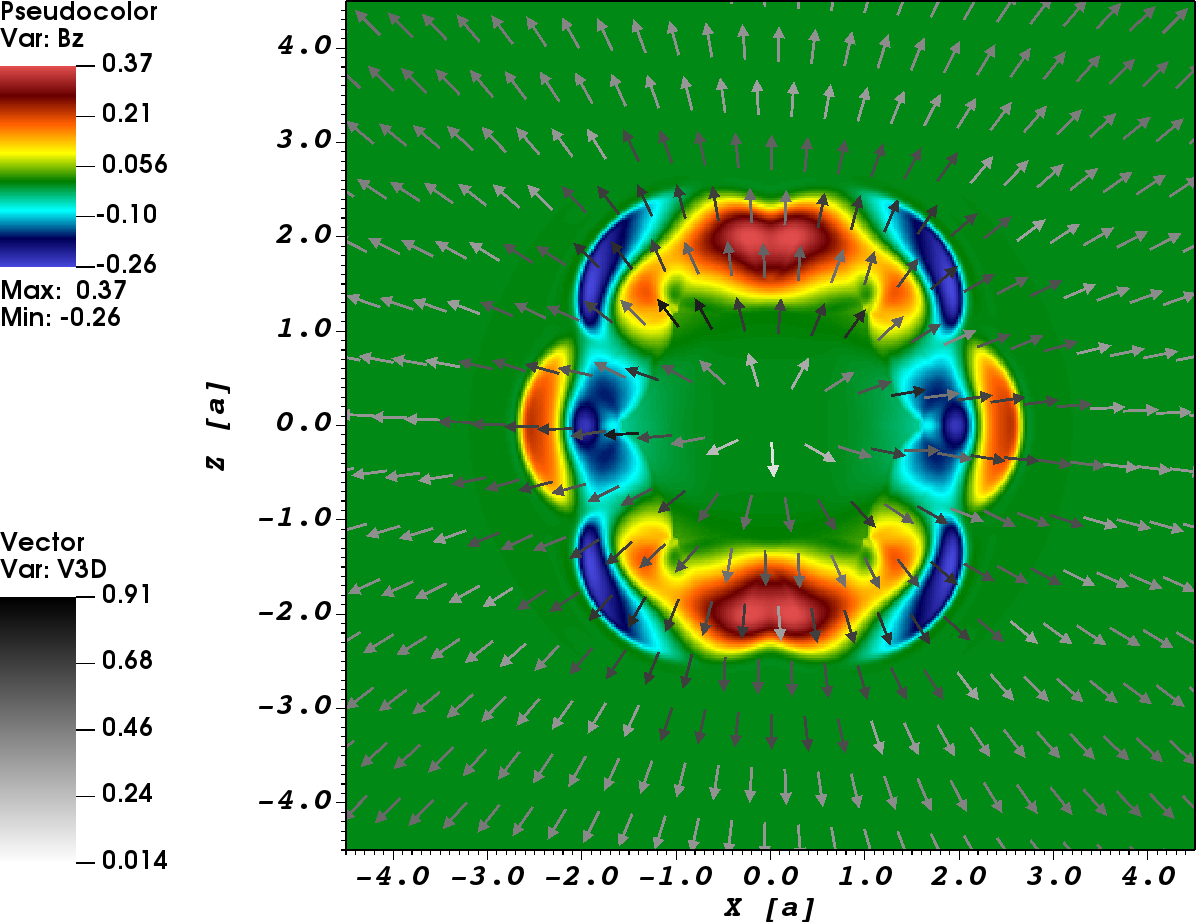}
\includegraphics[width=.49\linewidth]{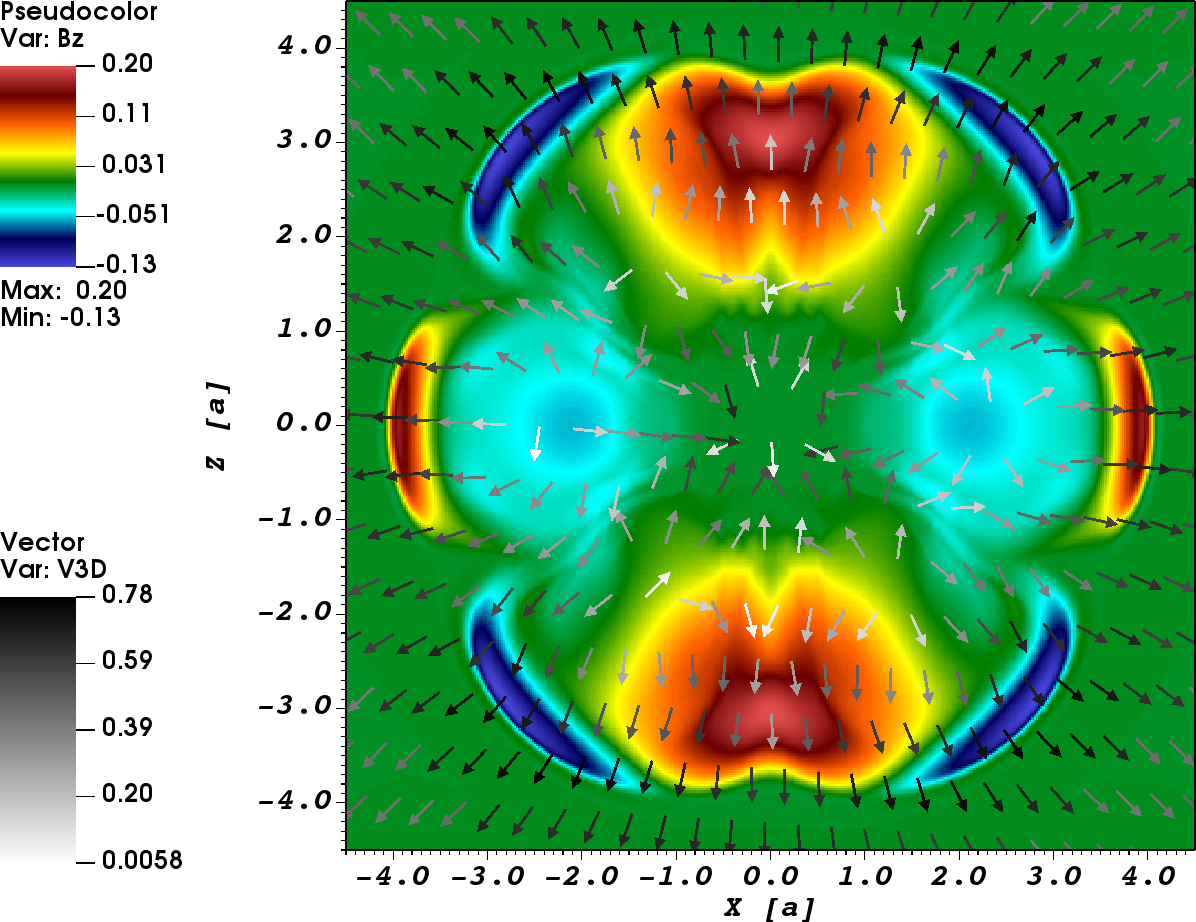}\\
\includegraphics[width=.49\linewidth]{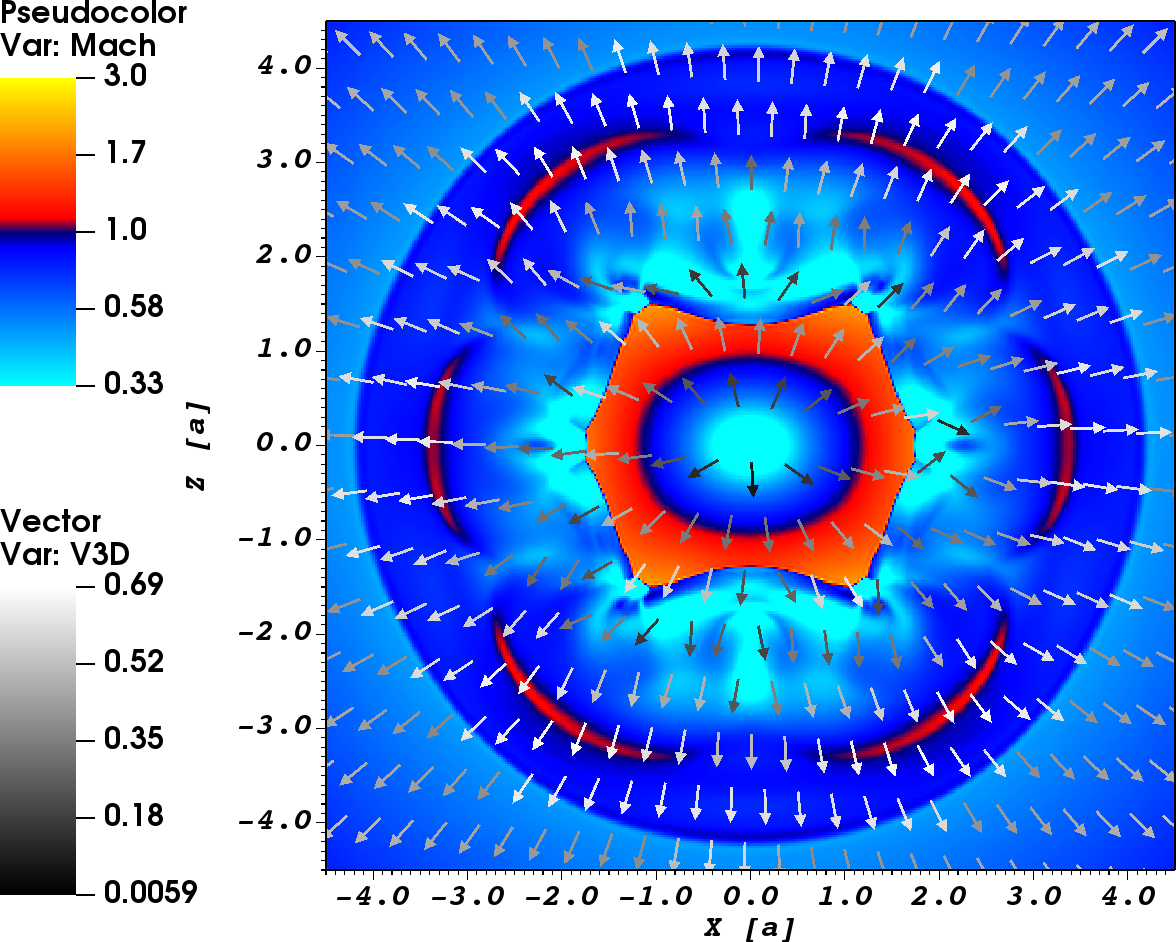}
\includegraphics[width=.49\linewidth]{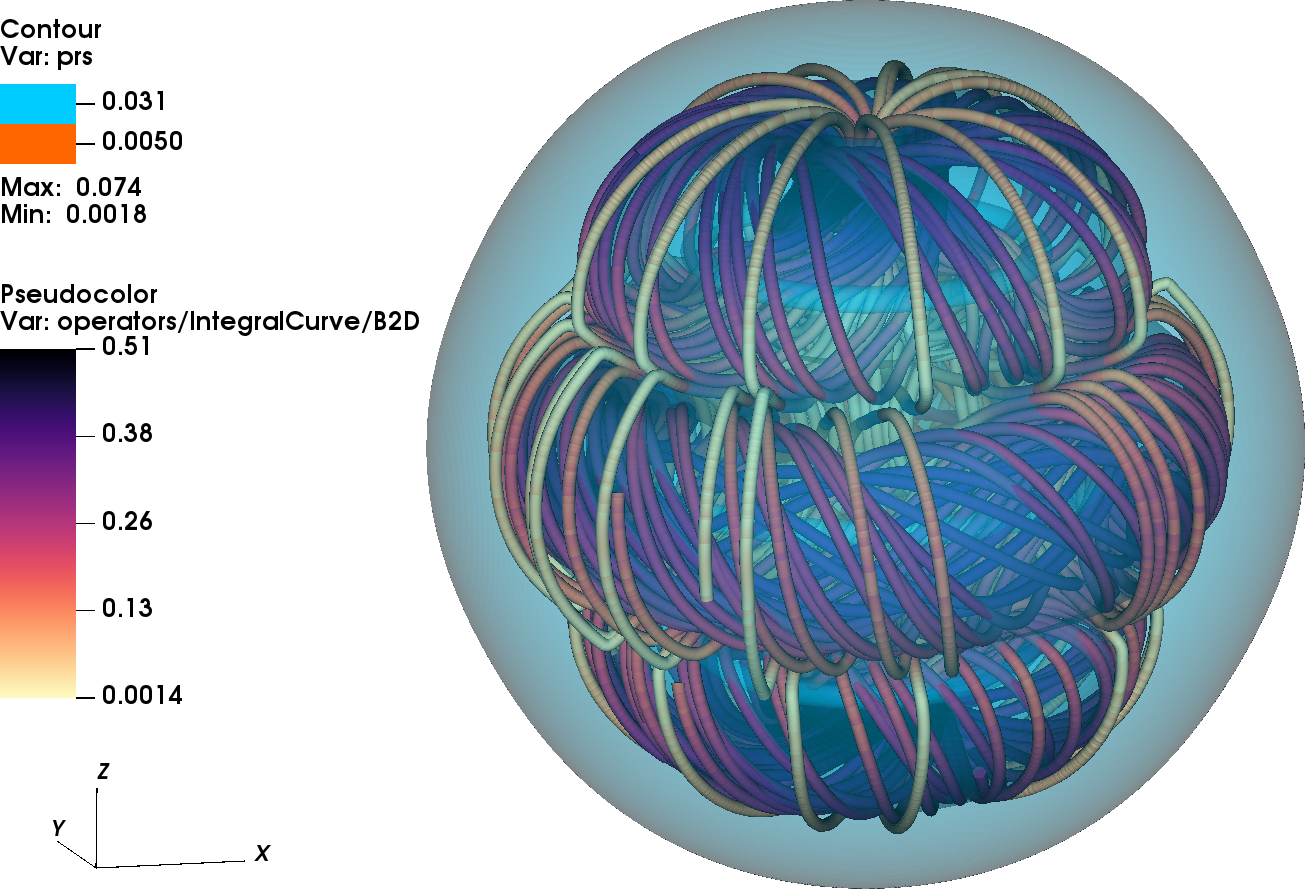}
  \caption{The ``high order'' spheromak expansion with Hubble expanding environment with $\eta_v = 0.1$ for different time moments  $t=2.5 [a/c]$  (upper left panel)  $t=4.5 [a/c]$ (upper right panel).  The color shows $B_z$ component of magnetic field and arrows show velocity field. Bottom left panel:  the Mach number for  ``high order'' spheromak with  $\eta_v = 0.1$. Bottom right panel:  pressure distribution by contours and magnetic field lines for  spheromak expansion with Hubble expanding environment with $\eta_v = 0.1$. }
\label{fig:mshubblem1}
\end{figure}

 The current density of the high order spheromak is presented in Fig.~\ref{fig:msbjvec} and the 
3D rendering of magnetic field  lines in Fig.~\ref{fig:mshubblem1fl} and slices of the \Bf\  in   Fig.~\ref{fig:mshubblem1}. We observe a behavior similar to the basic spheromak: for supersonically expanding case the \Bf\ is concentrated near the surface, while subsonically expanding case approximately  keeps the initial field structure. The expansion is more isotropic in this case.


\section{Force-free  simulations of a magnetic explosion}
\label{PHAEDRA}

In order to further verify the results of MHD simulations, we performed 2D force-free simulation of a spheromak using PHAEDRA code \citep{2012MNRAS.423.1416P}. PHAEDRA is a pseudo-spectral code developed specifically to study highly magnetized plasma regime, force-free electrodynamics, the vanishing-inertia limit of magnetohydrodynamics. The grid is defined in axisymmetric spherical coordinates (r, $\theta$), with $N_r$ points in r and $N_{\theta}$ points in $\theta$ direction. For this work, we chose $N_r=280$ and $N_{\theta}= 160$.

First, we repeat the  experiment with magnetically confined over-pressured spheromak, \S \ref{static}.
Similar to   \S \ref{MHD}, the external field is reduced with respect to the internal magnetic field. We set the external field to be a fraction  $\lambda$ of the equatorial spheromak's field on the surface. 

As a new ingredient, we  investigated different  profiles  of the external \Bf, scaling with  radius as  $\propto r^{-\gamma}$, 
\begin{equation}
{B_{0}}^{(out)} =  \lambda   \left(  \frac{a}{r} \right)^\gamma {B_{in}}(r= a )
\label{phiscaling}
\end{equation}
(This is the scaling of the flux function  $\Phi_{out}$, to conserve the div-B condition.) 
Qualitatively, this mimics the decreasing \Bf\ in the wind.

\begin{figure}[h!]
\includegraphics[width=.3\linewidth,height=0.3\textheight]{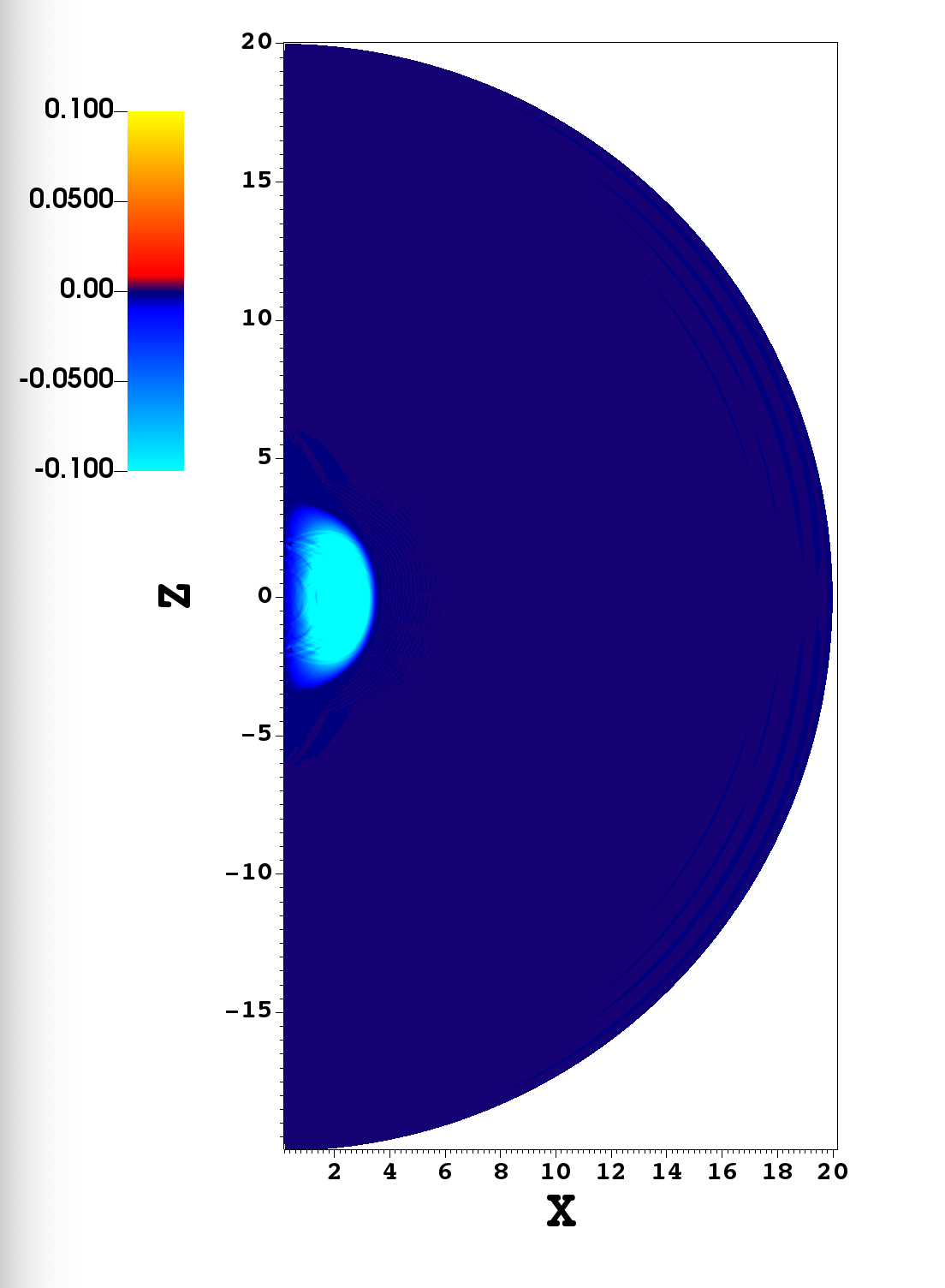}
\includegraphics[width=.3\linewidth,height=0.3\textheight]{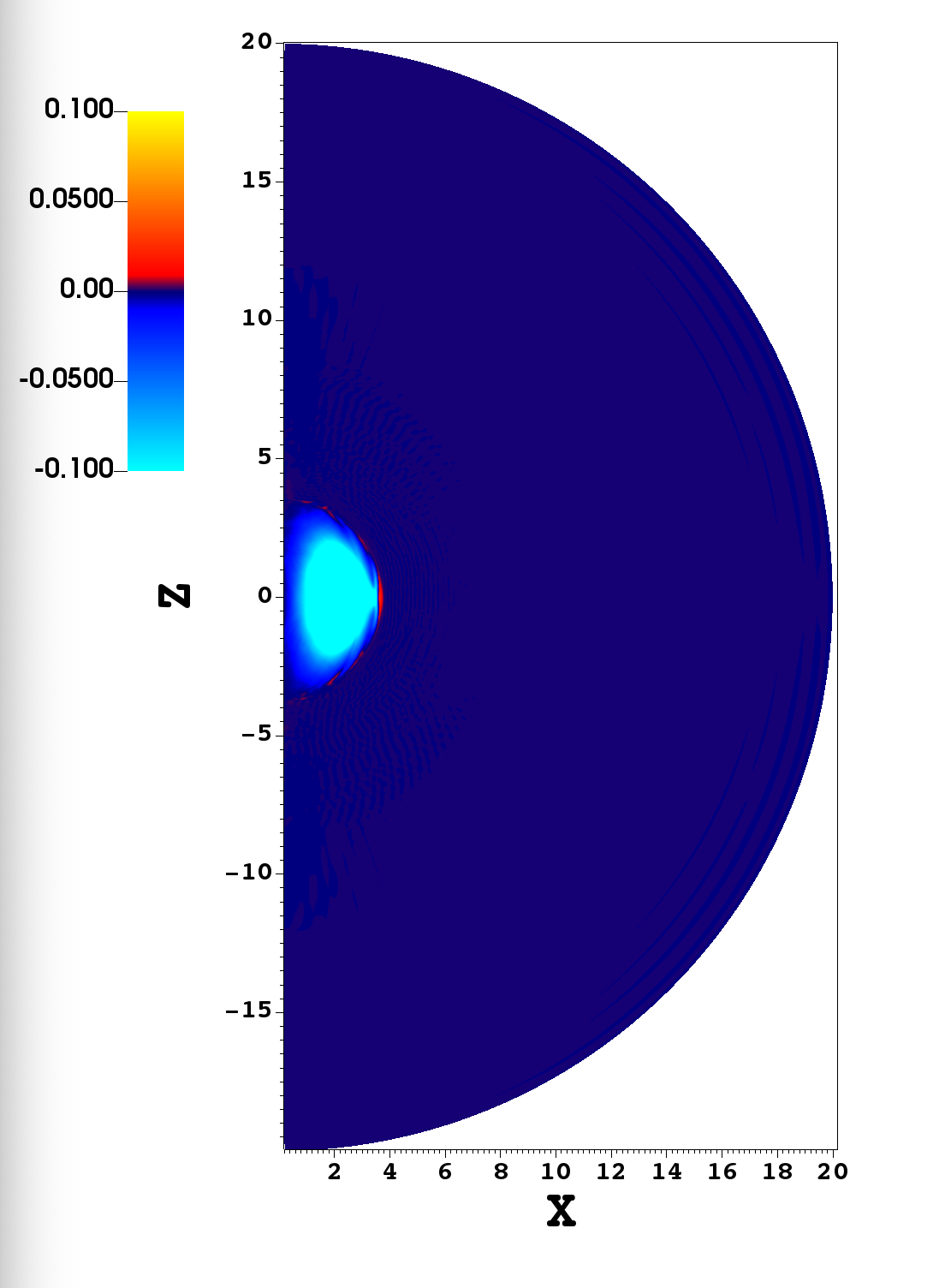}
\includegraphics[width=.3\linewidth,height=0.3\textheight]{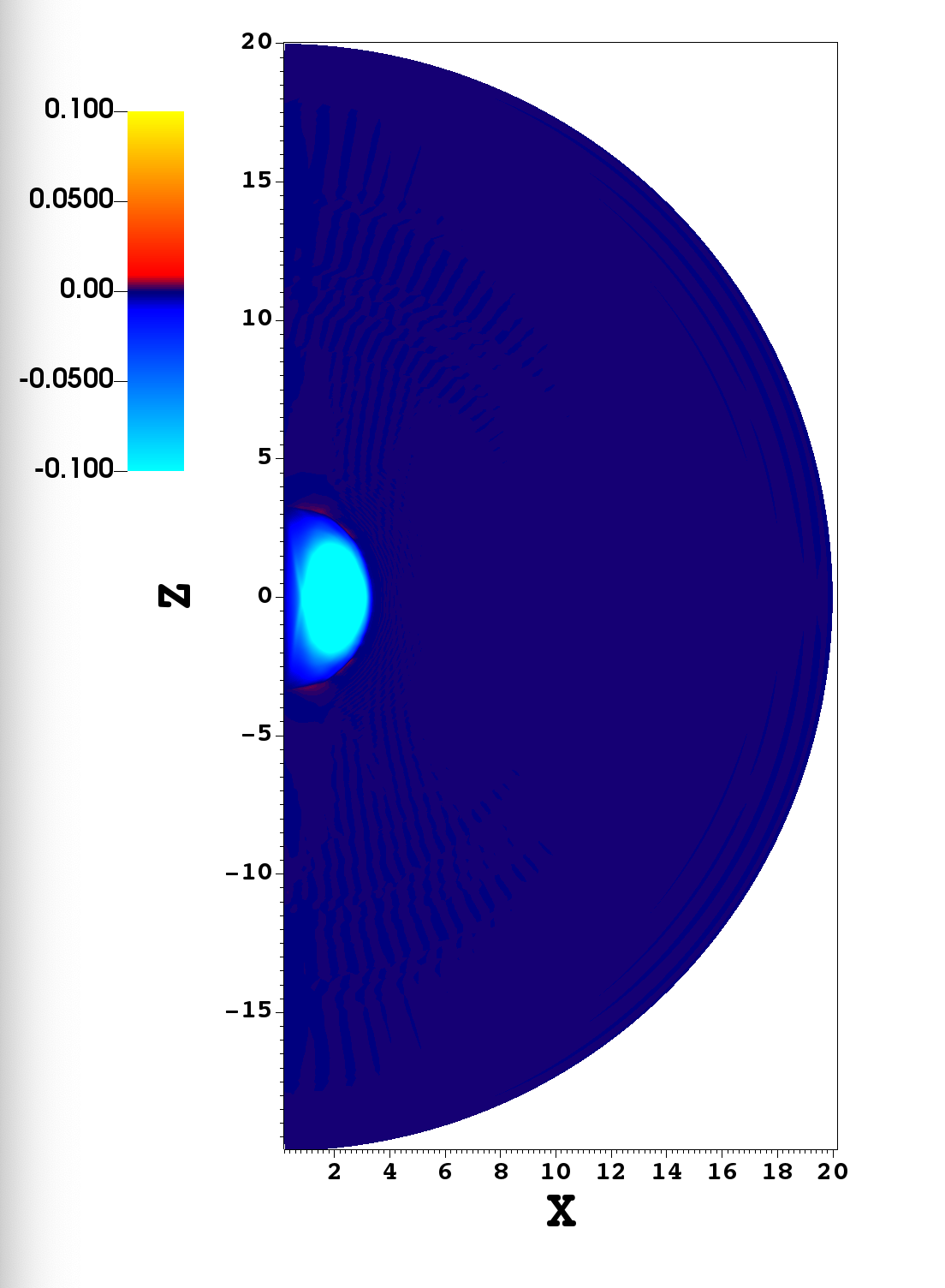}\\
\includegraphics[width=.3\linewidth,height=0.3\textheight]{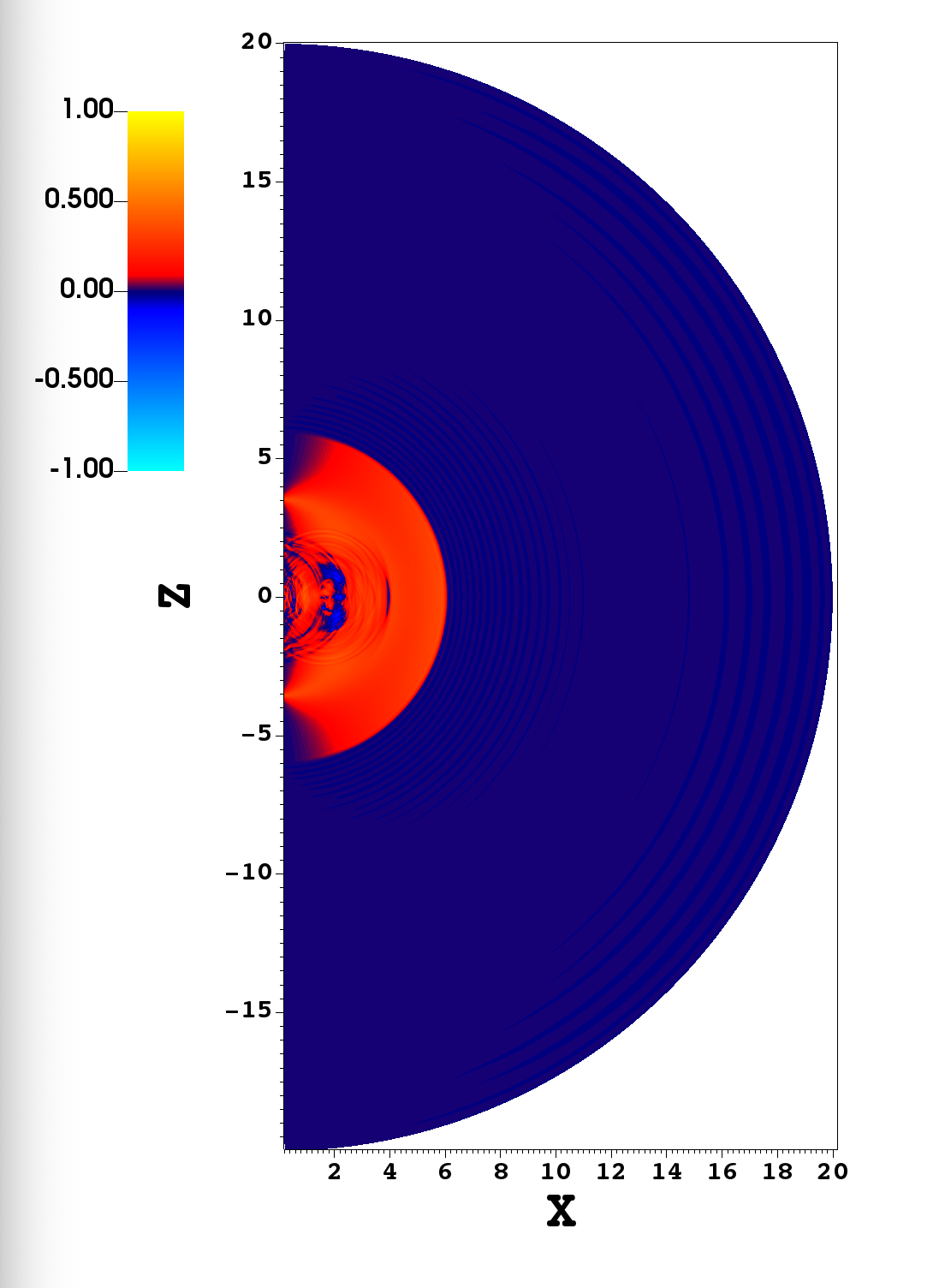}
\includegraphics[width=.3\linewidth,height=0.3\textheight]{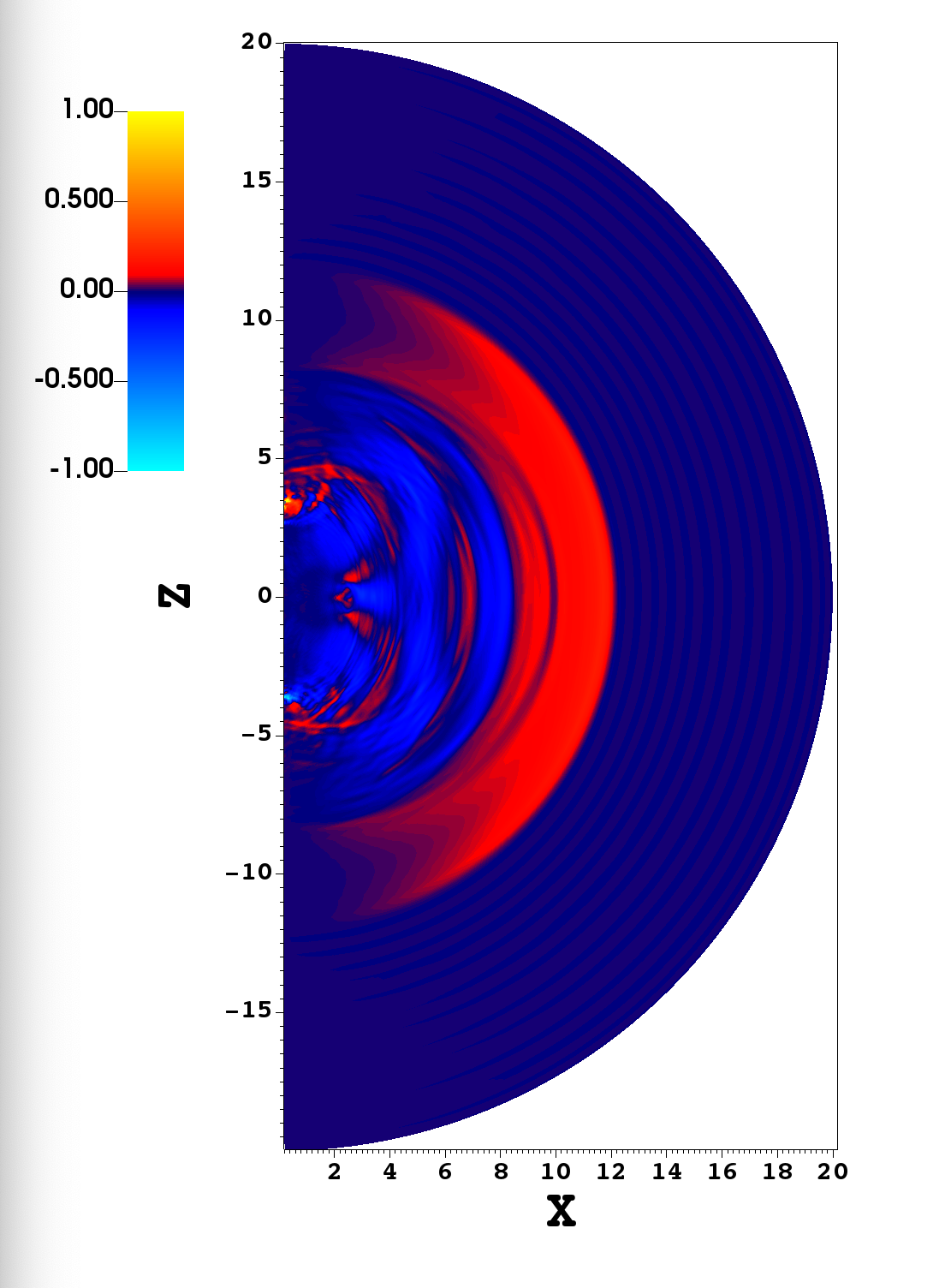}
\includegraphics[width=.3\linewidth,height=0.3\textheight]{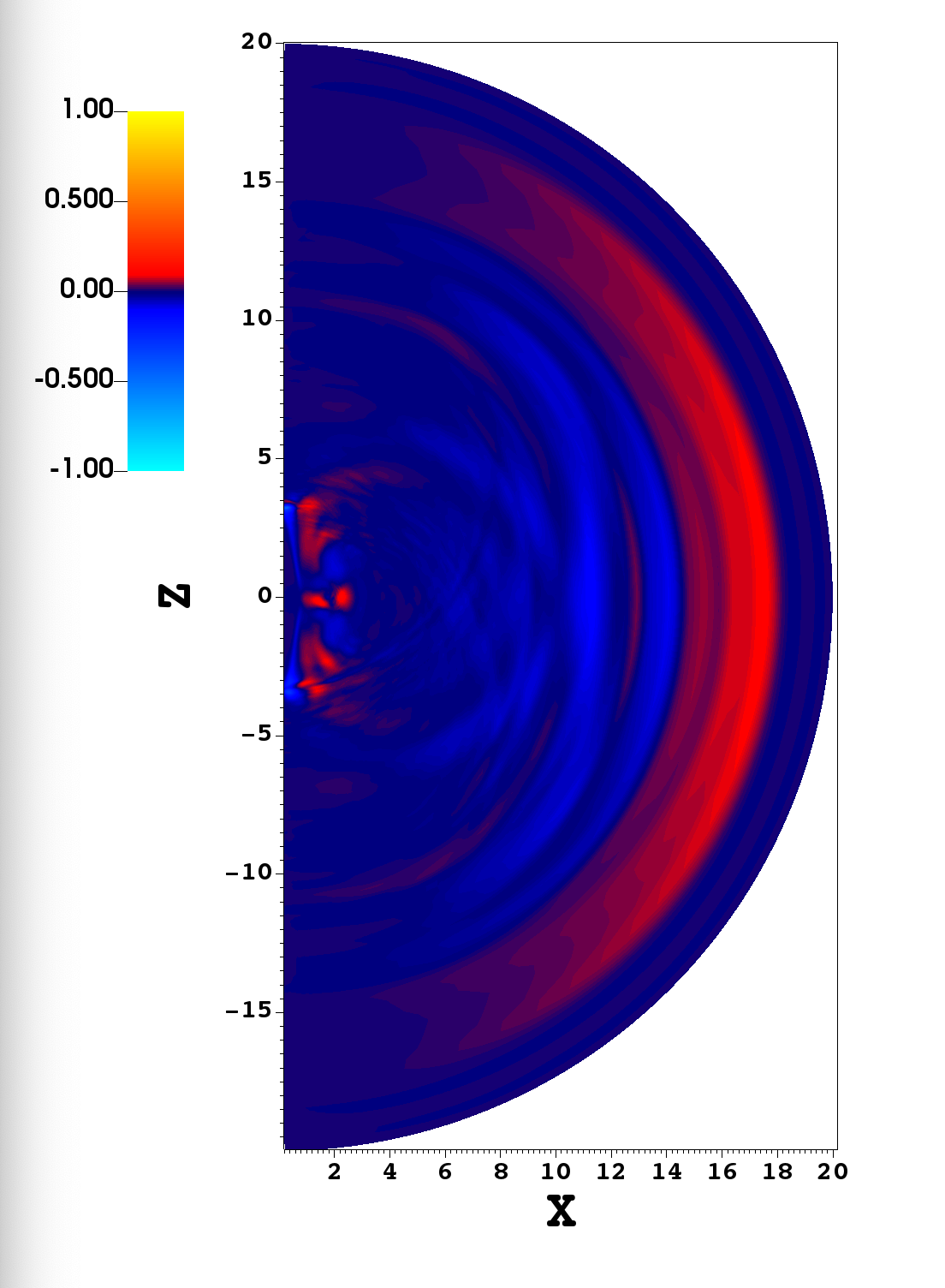}
\caption{Force-free simulations of magnetic explosion. Initially the spheromak is over-pressurized,   $\lambda=1/3$; the external \Bf\ asymptotes to a constant. Plotted are  the time slices of $r\sin\theta B_{\phi}$ (top panel) and  $p_{r}= \Gamma v_{r}$,where $v_{r}$ is the radial component of $\mathbf{E} \times \mathbf{B}$ drift velocity (bottom panel). Initial radius is $ a =2$.  The magnetic cloud ``puff-up'' and comes to a new equilibrium}
\label{pulseff1}
\end {figure}

\begin{figure}[h!]
\includegraphics[width=.3\linewidth,height=0.3\textheight]{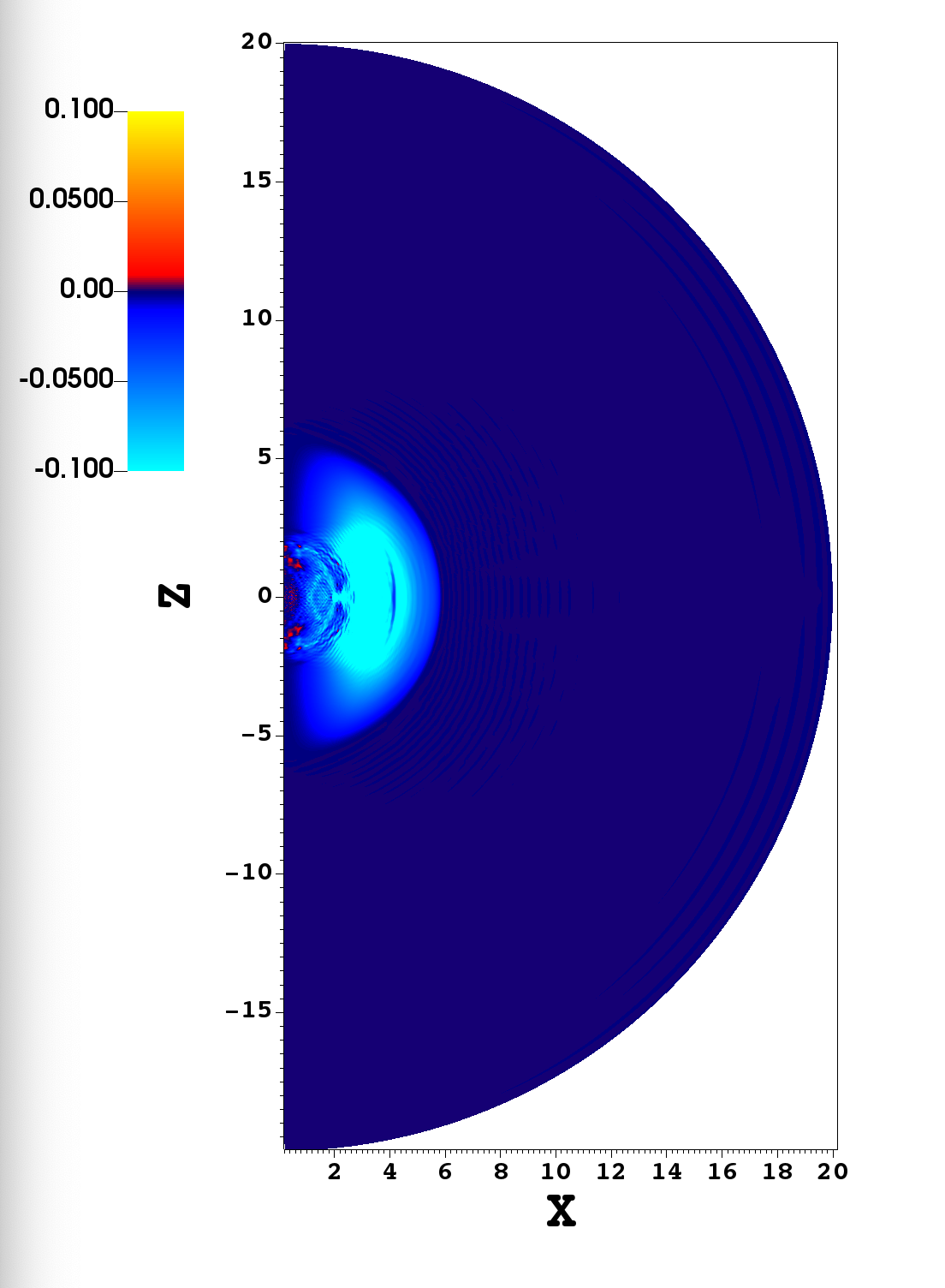}
\includegraphics[width=.3\linewidth,height=0.3\textheight]{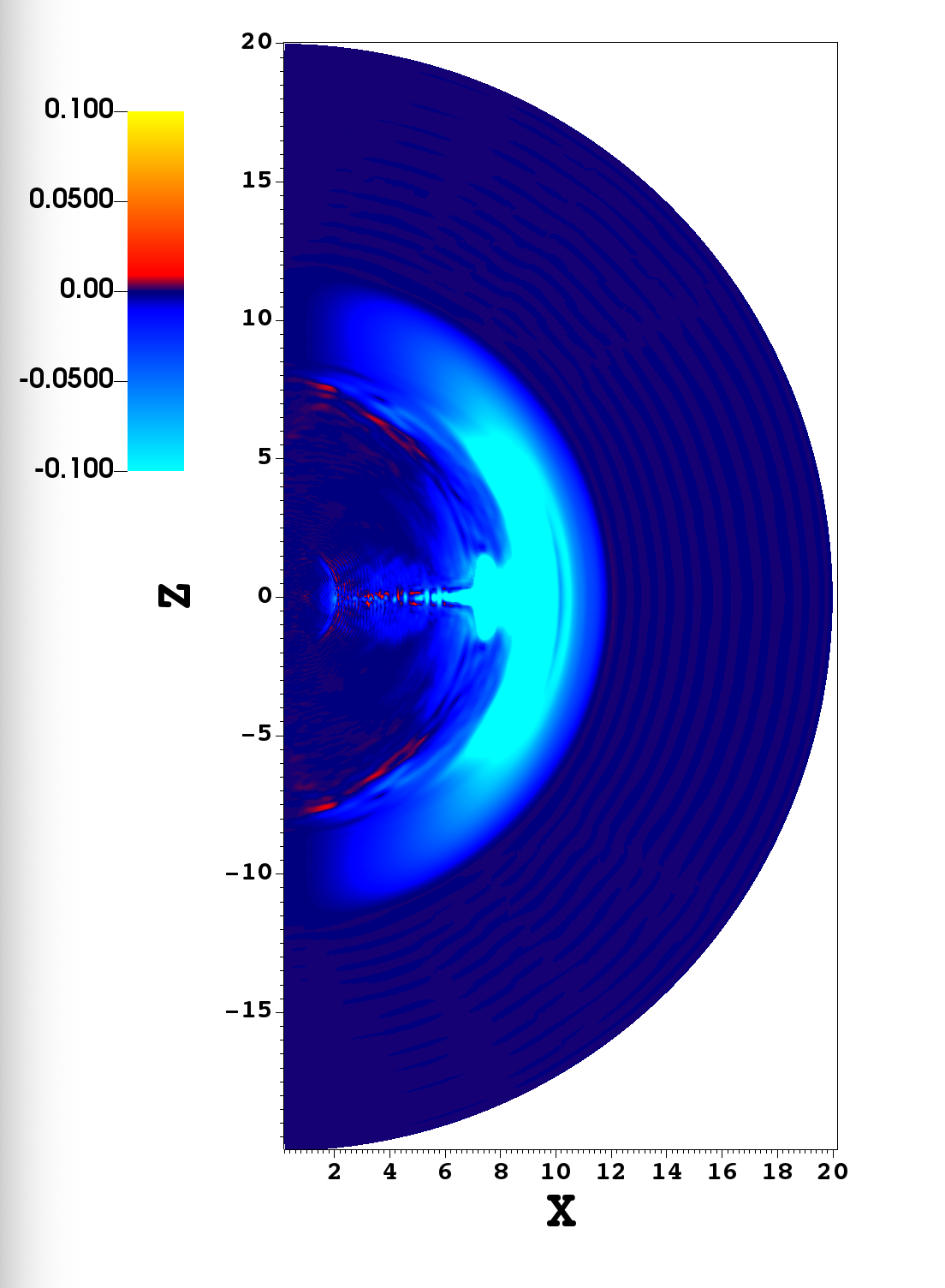}
\includegraphics[width=.3\linewidth,height=0.3\textheight]{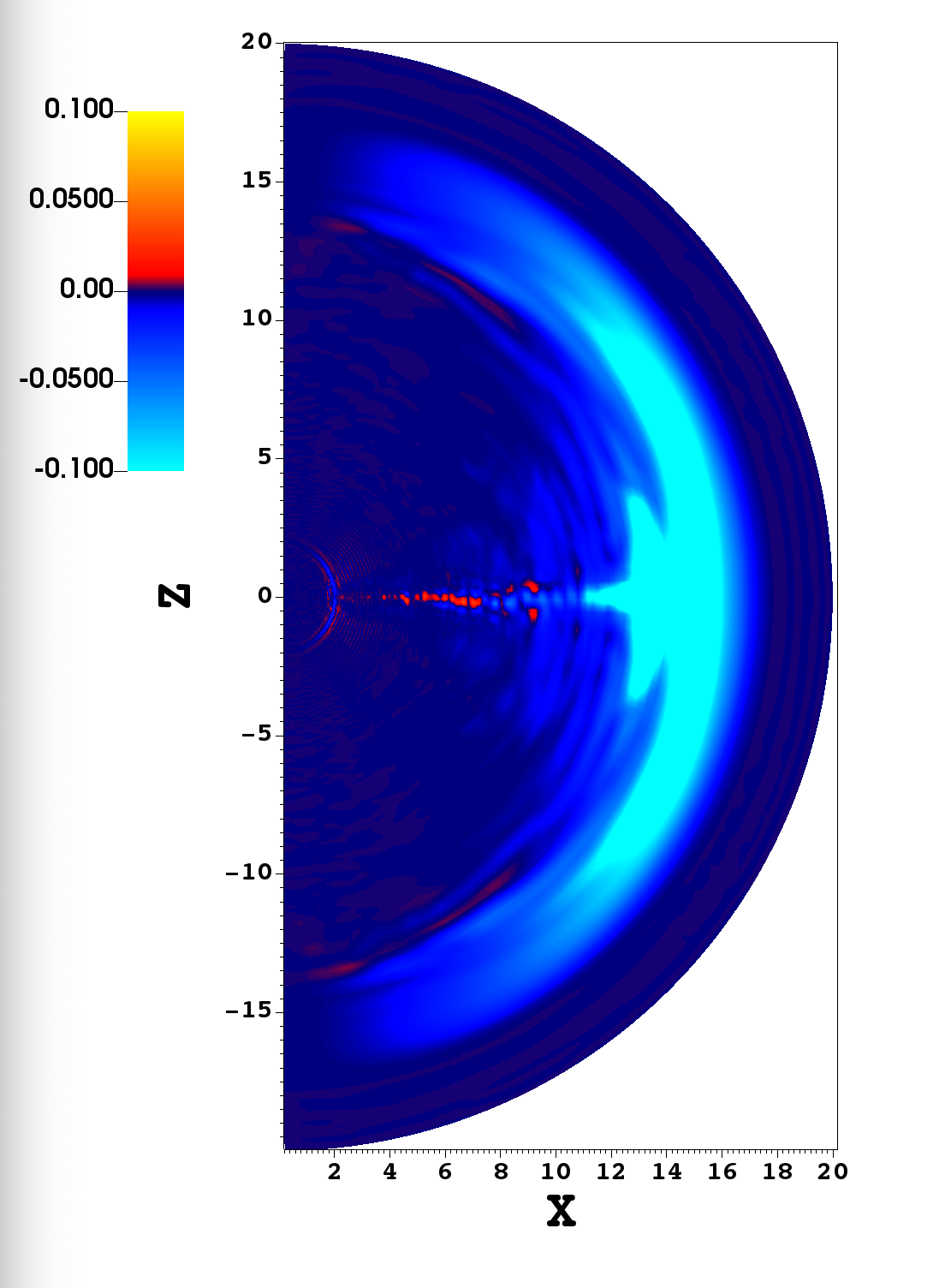}\\
\includegraphics[width=.3\linewidth,height=0.3\textheight]{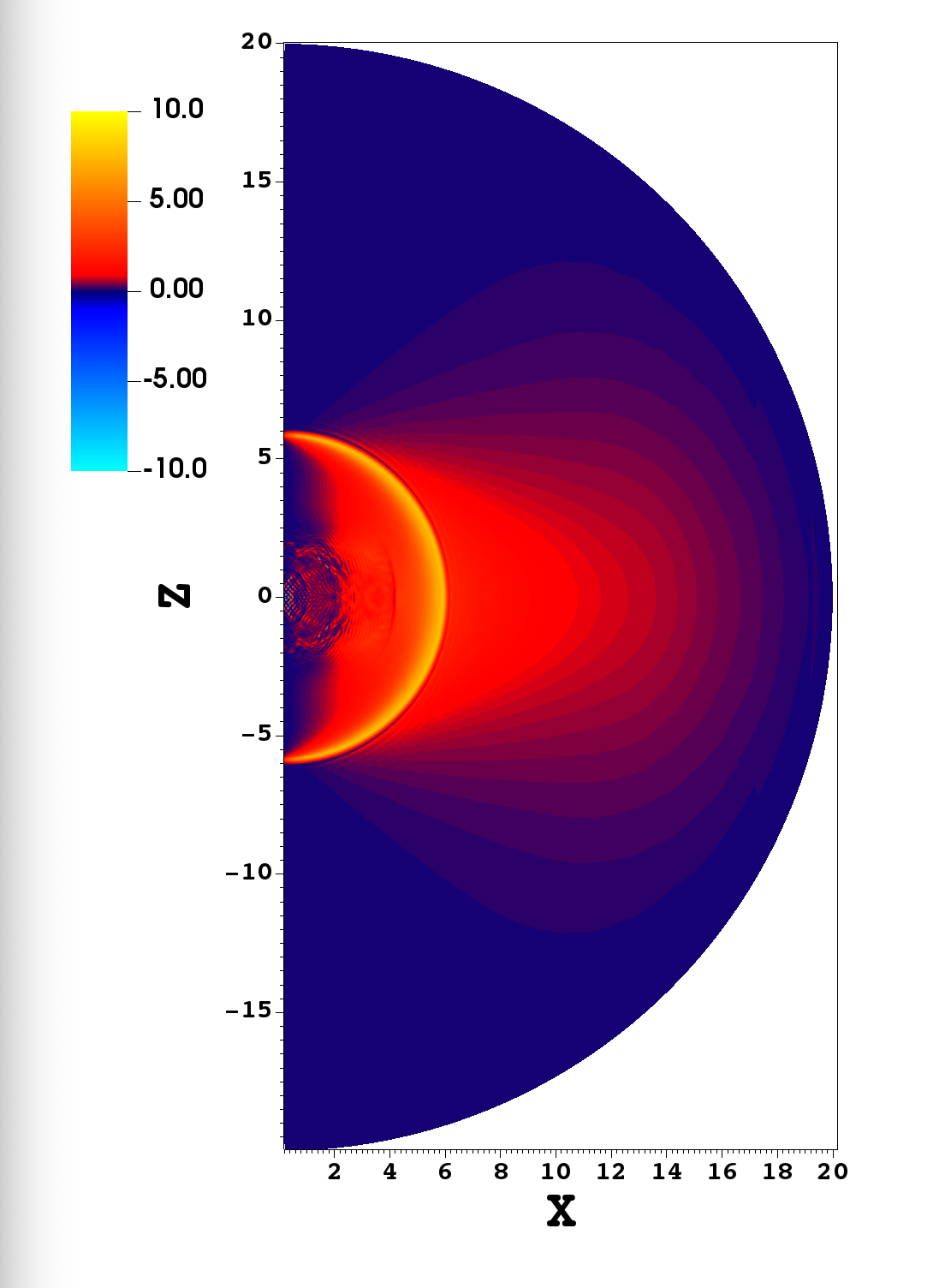}
\includegraphics[width=.3\linewidth,height=0.3\textheight]{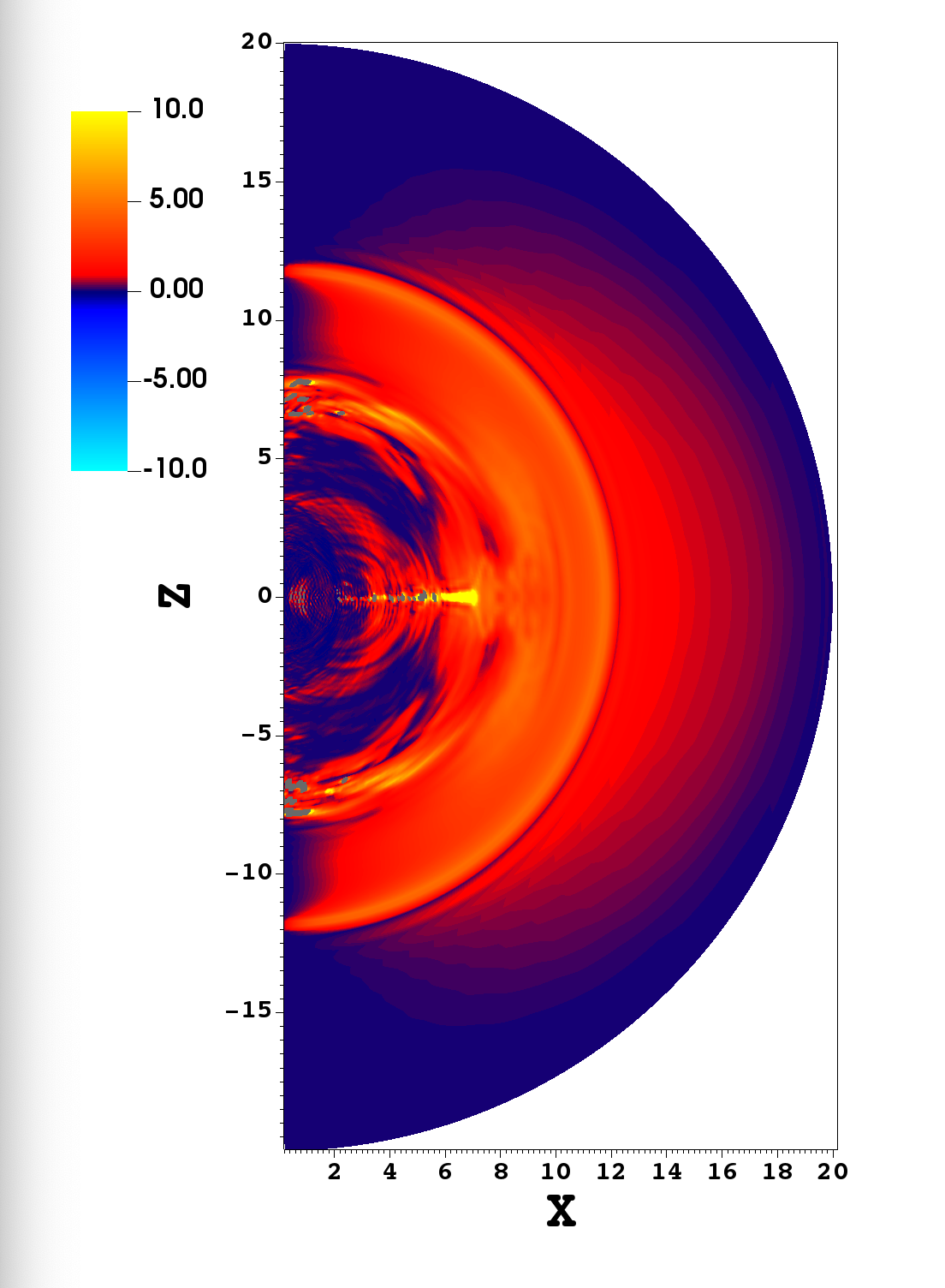}
\includegraphics[width=.3\linewidth,height=0.3\textheight]{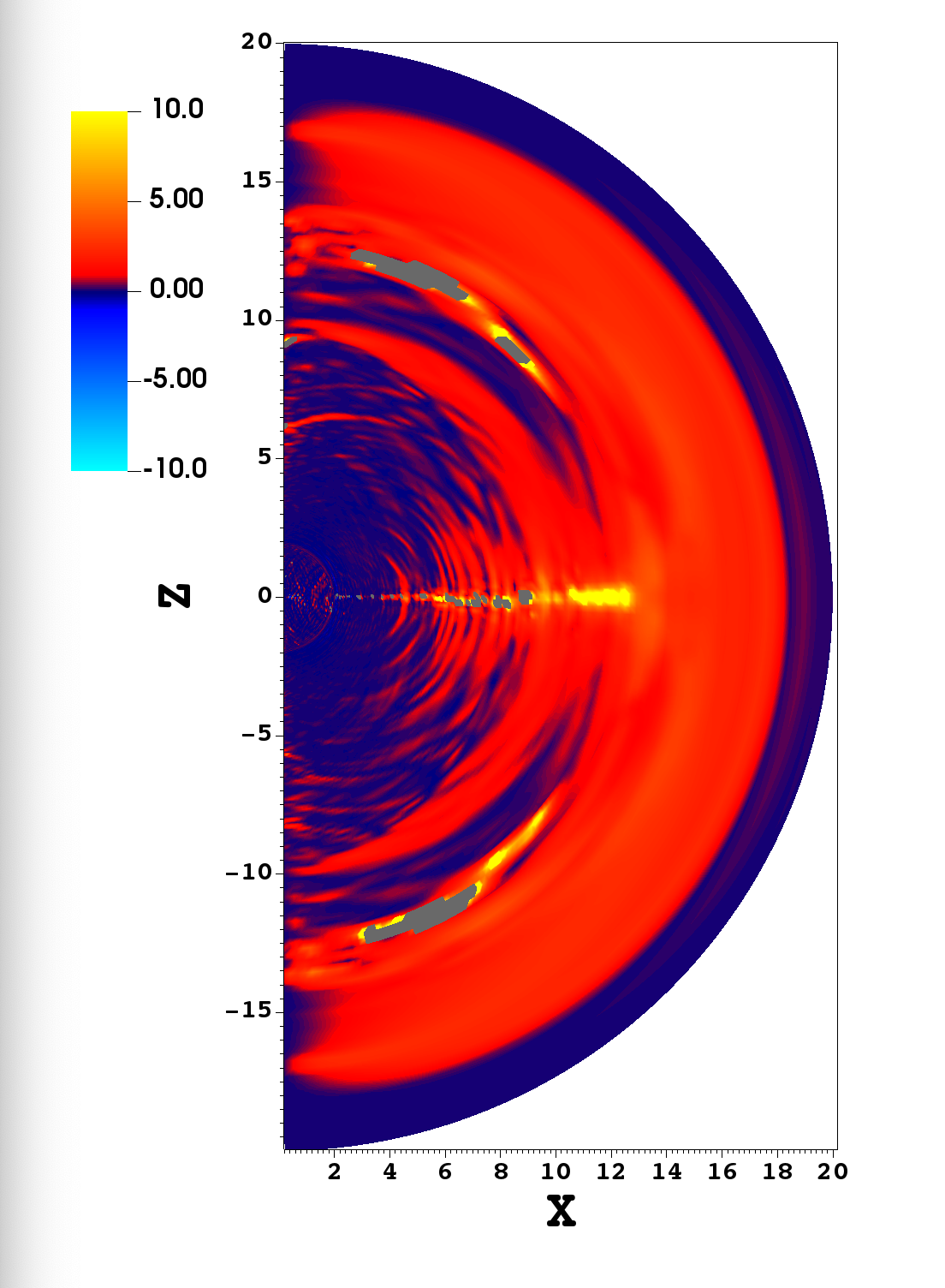}\caption{Same as Fig. \ref{pulseff1} but  the external field has an extra radial dependence of $1/r$, see Eq. (\ref{phiscaling}). This mimics \Bf\ in the wind. An expanding flux tube is generated: the spheromak first expands, and then  detonates generating causally disconnected outflow.  The spheromak is completely destroyed. (Motion in the wind ahead of the spheromak-generated \Alfven pulse, bottom left panel,  appears due to the fact that the initial configuration is not in force balance.}
\label{pulseff2}
\end {figure}

The outer boundary of the setup was fixed at $r=20$, absorbing layer beginning at $r=15$ whereas the radius of the spheromak was chosen to be $a=2$ \citep[somewhat larger initial size is chosen to avoid a quickly developing tilting instability][]{2020JPlPh..86d9007M}. The distances are in code unit. We tried various  grid sizes and found no substantial difference in the results for very high resolutions.   We ran our simulations for number of combinations of  $\lambda=1, \, 1/3$ and $\gamma=0, \, 1$. Results are plotted in Figs. \ref{pulseff1}-\ref{pulseff2}.

Results  of force-free simulations are qualitatively similar to the  MHD case: the evolution of spheromak depends on the profile  of the external field. We first  initialize the  over-pressured spheromak by bringing it out of force balance  by setting $\lambda=1/3$.  The magnetic  blob first puffs up, launches \Alfven waves in the external medium, but later  finds a new equilibrium (resembling original configuration),  Fig. \ref{pulseff1}.
 In the case of decreasing external \Bf\ $\propto 1/r$, Fig. \ref{pulseff2}, the spheromak detonates, creating an expanding flux tube. (We also did simulations for $B_{ex} \propto 1/r^{2}$, results are similar.)

\section{Analytical  solution to  the Prendergast problem}
\label{Prend1}

Next we supplement our numerical results with 
analytical solutions for  expanding  force-free spheromaks. The mathematical equation describing this problem has been derived by \cite{2005MNRAS.359..725P} who considered a system of inertial frames of reference without any electric field, but only  force-free magnetic field. Then, by transforming back to the origin frame of reference and postulating an appropriate expansion rate, he obtained the partial differential equation describing this system. Solutions that resemble expanding jets and analogues of the magnetic towers in the relativistic regime have been explored in this context \citep{GL2008,GV2010}, nevertheless, the spheromak family of these solutions has not been studied thoroughly. Here, we present a more intuitive derivation for the governing equation, where both the magnetic and electric field are accounted for an observer who is at rest with respect to the origin by considering both the magnetic and the electric field \citep{2009PhDT.......295G}.

Consider an axisymmetric time-dependent magnetic field in spherical coordinates $(r,\theta,\phi)$:
\begin{equation}
    {\bf B}= \nabla \Psi(r,t,\theta) \times \nabla \phi +T(r,t,\theta)\nabla \phi \,. 
\end{equation}
By construction the above magnetic field satisfies Gauss' law, and is expressed in terms of two scalar functions $\Psi$ and $T$ for the poloidal and toroidal field. Let the whole system expand with the following velocity profile, scaled to the speed of light $c$:
\begin{equation}
    {\bf v}= \frac{r}{ct}{\bf \hat{r}}\,
    \label{bfv}
\end{equation}
We note that this velocity is not the drift velocity defined as ${\bf E} \times {\bf B}/B^2$, as the former has components parallel to the magnetic field, whereas the latter is always normal to the magnetic field. 

Let us introduce the dimensionless parameter
\begin{equation}
    v=\frac{r}{ct}\,.
\end{equation}
We demand that the poloidal flux depends on the following combination of variables: 
\begin{equation}
    \Psi=\Psi\left(v,\theta\right)\,,
\end{equation}
and 
\begin{equation}
    T=T(r,v,\theta)\,.
\end{equation}
The induced electric field is:
\begin{equation}
    {\bf E} = -{\bf v}\times {\bf B}/c\,.
\end{equation}
We substitute into the induction equation:
\begin{equation}
    \nabla \times {\bf E} +\frac{1}{c} \frac{\partial {\bf B}}{\partial t}=0\,.
\end{equation}
The ${\bf \hat{r}}$ and $\hat{\theta}$ components are satisfied identically, but the $\hat{\phi}$ component sets the extra constraint:
\begin{equation}
    \frac{v}{r^2 \sin\theta}\left(T+r \frac{\partial T}{\partial r}\right)=0\,,
\end{equation}
which is straightforward to integrate. Thus, the expression for $T$ becomes:
\begin{equation}
    T=\frac{1}{r} W(v,\theta)\,.
        \label{eq:T}
\end{equation}
Next, we take the force equation:
\begin{equation}
    {\bf F}= \rho {\bf E} +\frac{{\bf j }\times {\bf B}}{c}\,.
\end{equation}
For the evaluation of the electric charge and current we use Gauss's law for the electric field and Amp\`ere's law with Maxwell's correction:
\begin{equation}
    \rho = \frac{1}{4\pi} \nabla \cdot {\bf E}
\end{equation}
\begin{equation}
    {\bf j} = \frac{c}{4\pi} \nabla \times {\bf B}-\frac{1}{4\pi}\frac{\partial {\bf E}}{\partial t}\,.
\end{equation}

We set all three components of the force to zero. From the $\phi$ component we take the equation:
\begin{equation}
    v(1-v^2)\frac{\partial \Psi}{\partial v}\frac{\partial W}{\partial \theta} -v(1-v^2)\frac{\partial \Psi}{\partial \theta}\frac{\partial W}{\partial v}+(1+v^2)
\frac{\partial \Psi}{\partial \theta} W=0 \,.
\end{equation}
On division by $v^2$ and we obtain the following equation:
\begin{equation}
    \frac{\partial \Psi}{\partial v}\frac{\partial }{\partial \theta}\left(\frac{1-v^2}{v}W\right) -\frac{\partial \Psi}{\partial \theta}\frac{\partial}{\partial v}\left(\frac{1-v^2}{v}W\right)=0\,.
\end{equation}
The above expression is the Jacobian of $\Psi$ and $\frac{1-v^2}{v}W(v,\theta)$, therefore we can write:
\begin{equation}
    W=\frac{v}{1-v^2}\beta(\Psi)
        \label{eq:W}
\end{equation}
with $\beta$ an arbitrary function of $\Psi$. 
We substitute into the $r$ and $\theta$ components of the force equation, we substitute $\mu=\cos\theta$ and we obtain the ``Prendergast Equation":
\begin{equation}
    v^2(v^2-1)\frac{\partial^2 \Psi}{\partial v^2}+2v^3\frac{\partial \Psi}{\partial v}-(1-\mu^2)\frac{\partial^2\Psi}{\partial \mu^2}=\left(\frac{4 \pi}{c}\right)^2 \frac{v^2}{1-v^2} \beta^{\prime}\beta\,.
    \label{Prendergast}
\end{equation}

We found fully analytic, separable, self-similar solution for a relativistically expanding force-free flow for a dipolar structure
\begin{equation}
\Psi = f(v) \sin^2 \theta\,,
\end{equation}
and linear dependence between $\Psi$ and $\beta$ in the form:
\begin{equation}
\beta=\alpha \Psi\,.
    \label{eq:beta}
\end{equation}
Under this condition, the equation for $f$ becomes:
\begin{equation}
(-2 + v^2 (2+\alpha^2)) f + 
v^2 (1-v^2) \partial_v \left ( (1-v^2) f'  \right) =0\,.
\end{equation}
The equation admits the analytical solution:
\ba &&
f= \cos \xi  - \frac{\sin\xi }{ \alpha v}, \, \xi = \alpha \, {\rm arctanh}  (v) 
\nn && 
\B = \nabla \Psi \times \nabla \phi + \frac{v}{1-v^2} \frac{\alpha \Psi} {r} \nabla \phi
\nn &&
\E = v {\bf e}_r \times \B\,.
\label{Prend}
  \ea
 Combining equation (\ref{eq:T}), (\ref{eq:W}) and (\ref{eq:beta}) we obtain the expression for $T$:
 \begin{eqnarray}
T=  \frac{v}{1-v^2} \frac{\alpha \Psi} {r}=\frac{v}{1-v^2} \frac{\alpha \Psi} {r} \,.
\label{T-final}
\end{eqnarray}
  It is then straightforward to evaluate the fields and the current:
  \ba &&
B_r = \frac{2  \cos  \theta }{r^2} f
\nn &&
B_\theta= 
\frac{\sin  \theta  \left(\left(\left(\alpha ^2+1\right) v^2-1\right) \sin \xi+\alpha  v \cos \xi \right)}{\alpha  r^2 v \left(1-v^2\right)}
   \nn &&
   B_\phi = \frac{\alpha   v \sin  \theta }{r^2
   \left(1-v^2\right)}f
   \nn &&
   E_r = 0
   \nn &&
   E_\theta = v B_\phi
   \nn &&
   E_\phi = - v B_\theta
   \nn &&
   J_r = \frac{ \alpha v}{r (1-v^2)} B_r
   \nn &&
   J_\theta = \frac{ \alpha v}{r } B_\theta
   \nn && 
    J_\phi = \frac{ \alpha v}{r } B_\phi
  \nn && 
  \div \E = \frac{2 \alpha   v^2 \cos  \theta }{r^3 \left(1-v^2\right)}f
  \ea

Even though these equations may seem over-complicated: they closely match the well know spheromak solution (\eg\ function $f$), yet the independent variable is given in terms of the variable ${\rm arctanh}v$. These are fully analytic self-similar  solutions for time-dependent axially-symmetric force-free expanding magnetic bomb. 
  \begin{figure}[th!]
\centering
\includegraphics[width=.49\textwidth]{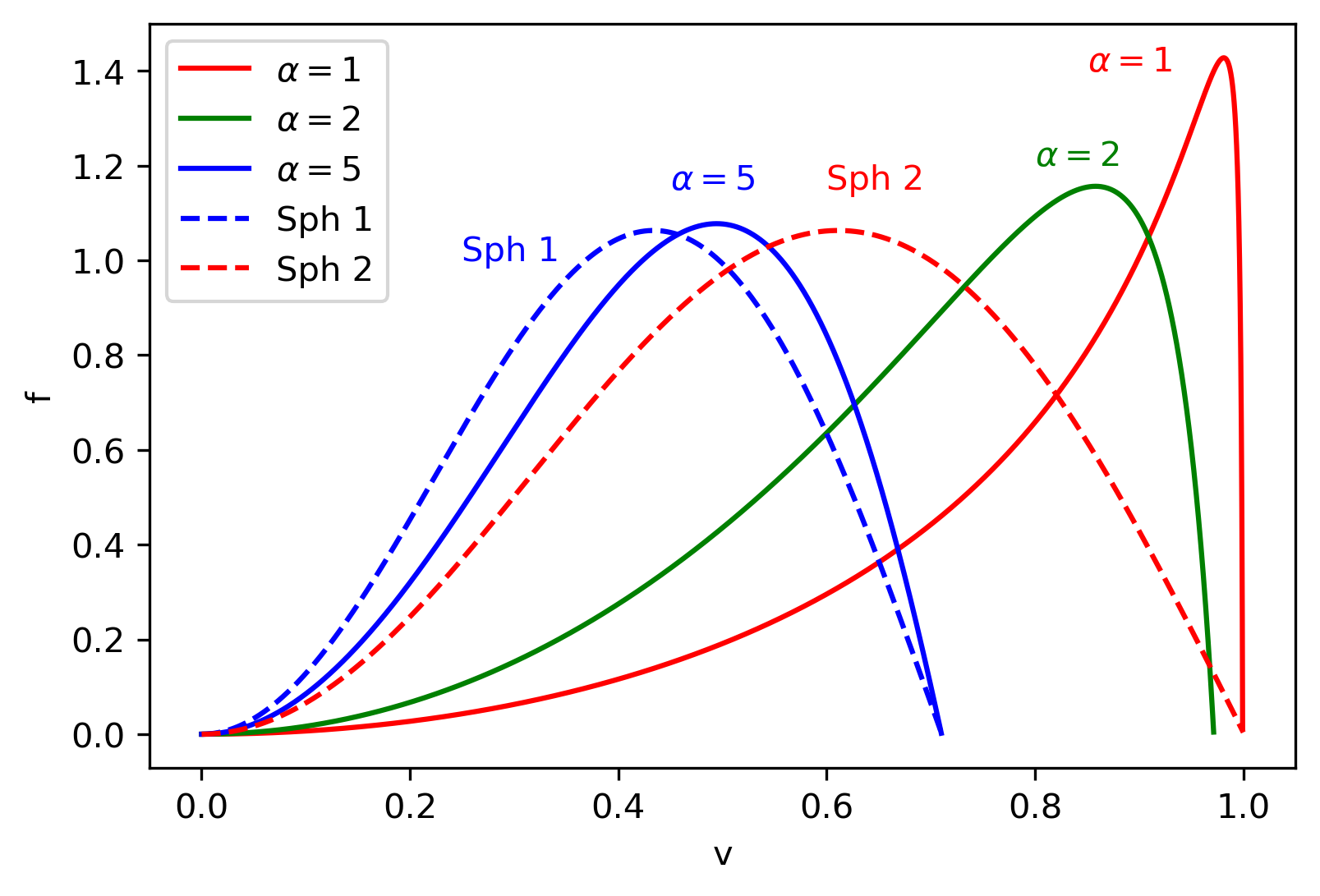} 
\caption{ The function $f$ of the Prendergast solution for three choices of $\alpha=1\,,~2\,,~5$ shown in red, green and blue lines respectively. The solution for $\alpha=1$ has its first root at $v_0=0.9992$, corresponding to maximum Lorentz factor $\Gamma_0=25.4$, the one for $\alpha=2$ has $v_0=0.9715$ and $\Gamma_0=4.2$ and the one for $\alpha=5$ has $v_0=0.71$ and $\Gamma_0=1.4$. We also plot the solutions for two stationary spheromaks appropriately scaled so that their first roots lie at $0.71$ (Sph1) and $1$ (Sph2), for comparison.}
\label{Prendergast_f} 
\end{figure}

\begin{figure}[h!]
\centering
\includegraphics[width=.49\textwidth]{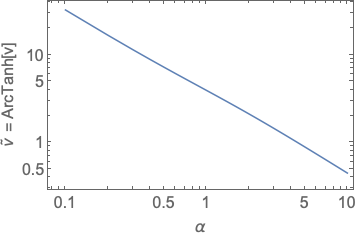} 
\includegraphics[width=.49\textwidth]{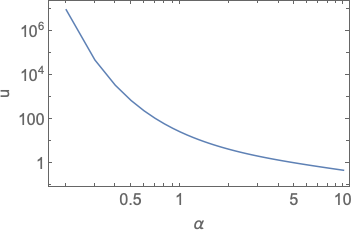}
\caption{Location of the first (innermost) surface of the expanding flow in   $\tilde{v}$ and $u = v/\sqrt{1-v^2}$ coordinates for different $\alpha$.
}
\label{vprimeofalpha} 
\end{figure}

The location of the first positive root, $v_0>0$, of the function $f(v)$ depends on the value of $\alpha$. For smaller values of $a$ it is pushed towards $v\to 1$, whereas for a larger $\alpha$ the radius of the spheromak becomes smaller. As $\alpha$ is a factor relating the poloidal current to the poloidal field, a stronger current leads to a field of smaller radius as shown in Fig.  \ref{Prendergast_f}, where we present the solution of equation (\ref{Prend}), for $\alpha=1\,,~2\,,~5$. Indeed, we can evaluate the relation between the outer boundary of the bubble with $\alpha$ for the expanding spheromak, Fig. \ref{vprimeofalpha}. Introducing
\be
v = \tanh \tilde{v}
\ee
the surface is located at 
\be
\tanh \tilde{v} = \frac{\tan( \alpha \tilde{v})}{\alpha}
\ee
For small $\alpha$ (or sufficiently large $\tilde{v}$ so that 
 $\tanh v \sim 1$) 
\ba &&
\tilde{v}= \frac{{\rm arctan} (\alpha) + \pi n}{\alpha} 
\nn &&
  v   =   \tanh \left( \frac{{\rm arctan} (\alpha) + \pi n}{\alpha}\right)  
\ea

For example:
\ba &&
\alpha =1
\nn && 
v_0=0.999223
\nn &&
\tilde{v}_0 = 3.9266
\nn &&
\Gamma_0 = \frac{1} {\sqrt{1-v_0^2}} = 25.38
\nn &&
u _0= \frac{v_0} {\sqrt{1-v_0^2}} = 25.36
\ea

\begin{figure}[th!]
\centering
a\includegraphics[width=.4\textwidth]{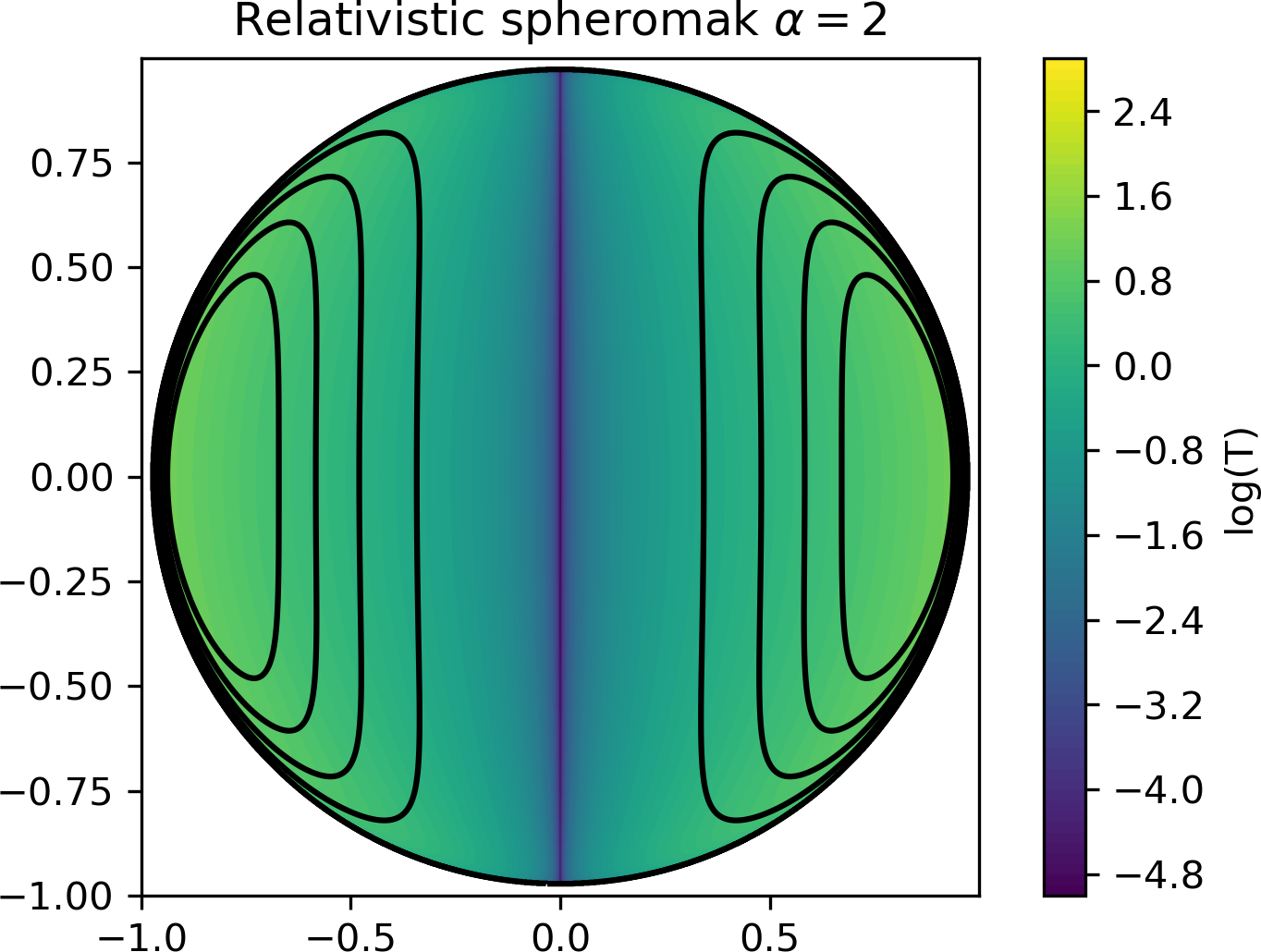}
b\includegraphics[width=.4\textwidth]{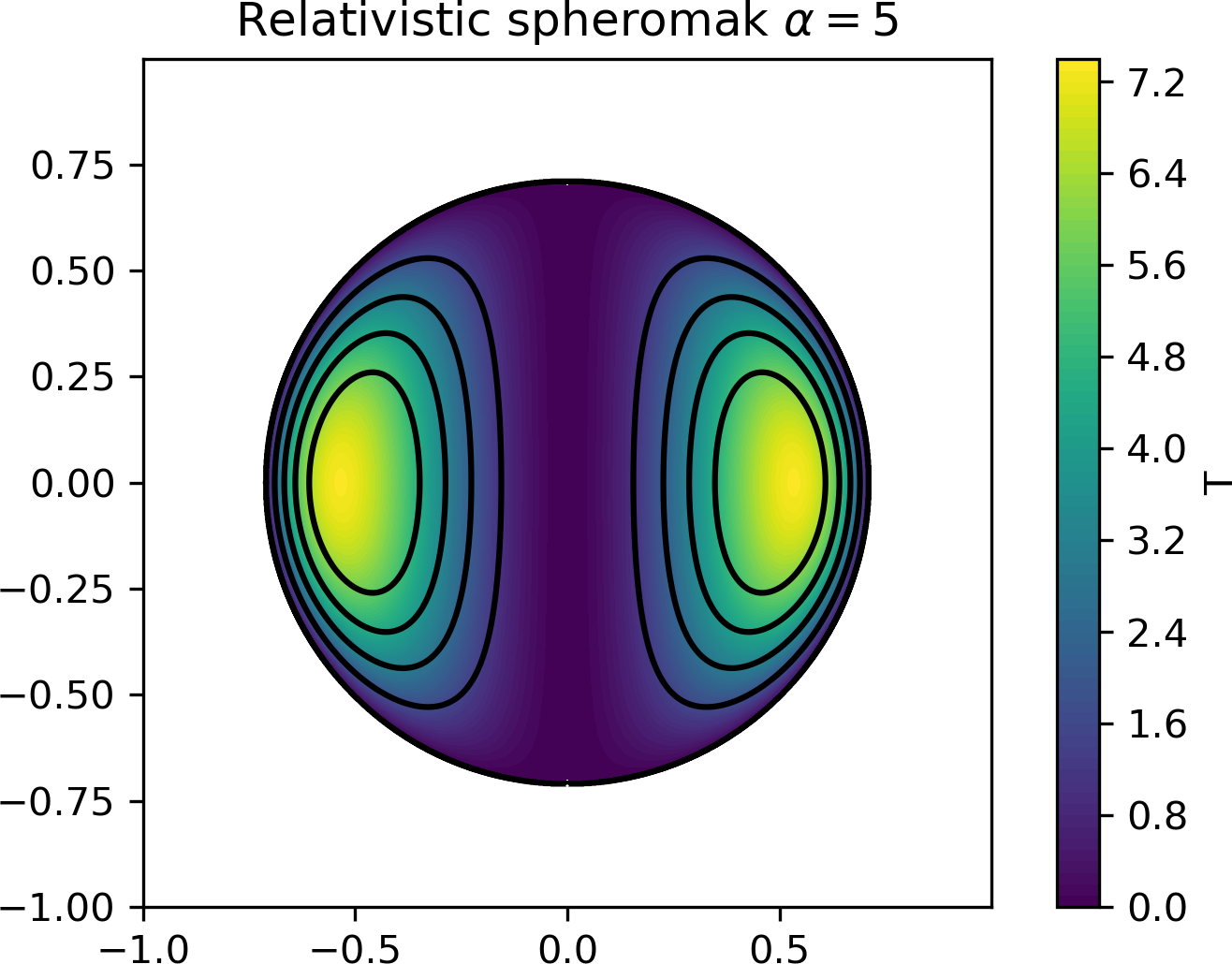}  
\caption{ The structure of the spheromak for $\alpha=2$ (a) and $\alpha=5$ (b). The poloidal field lines are shown in black, and in color we plot $\log T$ and $T$ respectively. As $T$ depends on both $v$ and $r$, equation (\ref{T-final}) the snapshot is taken at time $t=1/c$. }
\label{Prendergast_spheromak} 
\end{figure}

In the stationary spheromak solution, the change of $\alpha$ is a simple scaling factor as is evident from equation (\ref{eq:Bsp}). In the relativistic case however, a change in $\alpha$ is not a mere rescaling, but a drastic alteration in the structure of the spheromak. This is evident from Fig. \ref{Prendergast_spheromak}, where we show the structure of two relativistic spheromaks for $\alpha=2$ and $\alpha=5$, a higher value of $\alpha$ leads to a solution that is reminiscent of the stationary one, whereas a smaller $\alpha$ leads to a field where most of the field lines are concentrated near the edge of the expanding bubble and so does the energy and the pressure. We further comment, that the contours of the poloidal field lines given by $\Psi$ do not coincide to the contours of $T$, which is the case for the stationary spheromak. This difference is related to the fact that in the stationary case it is only the ${\bf j}\times {\bf B}$ term that contributes to the force, thus the electric current must flow along the field lines. On the relativistically expanding case, the contribution of the electric field and the charge in the dynamics leads to this difference.

\begin{figure}[th!]
\centering
\includegraphics[width=.3\textwidth]{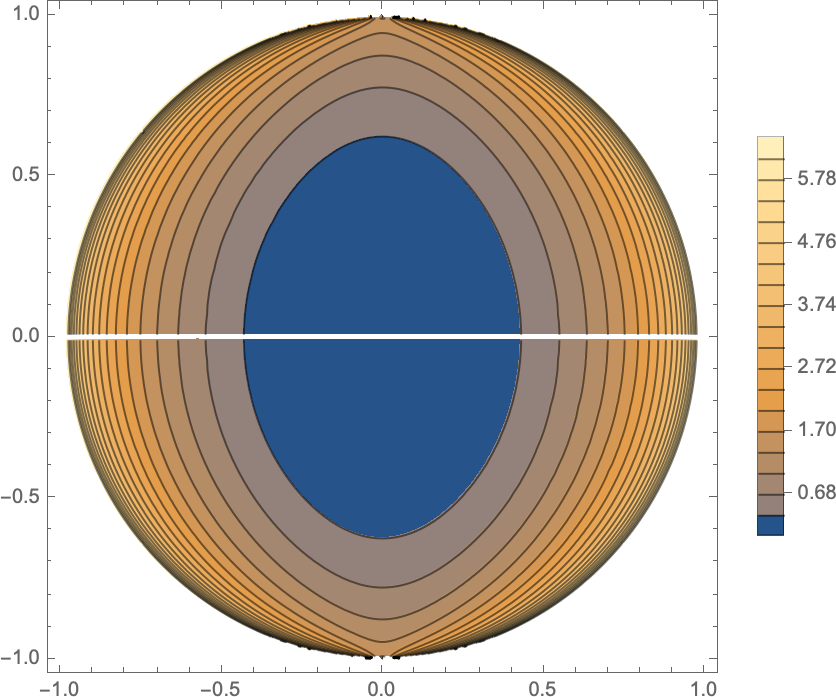} 
\includegraphics[width=.3\textwidth]{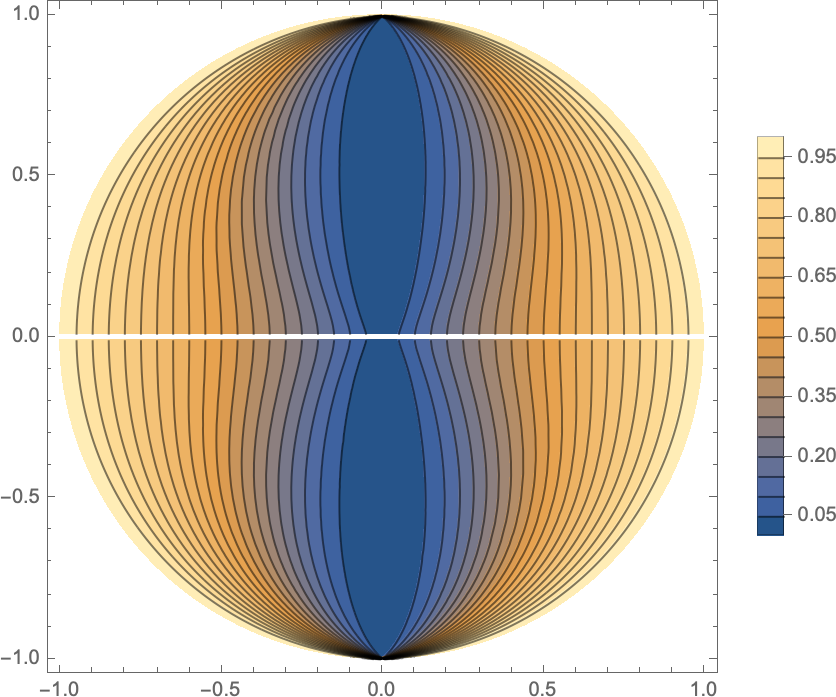}
\includegraphics[width=.3\textwidth]{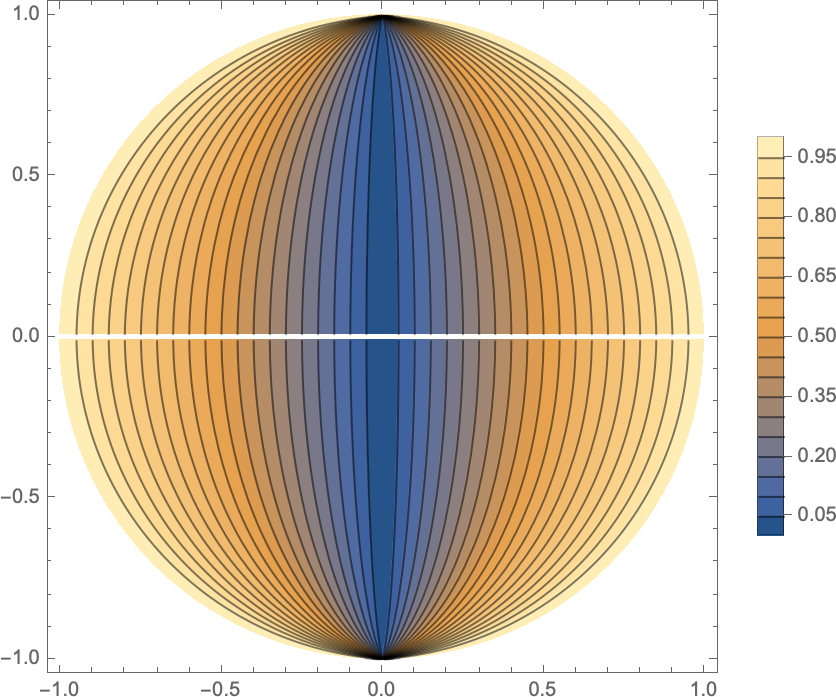}
\caption{ Values of $\ln (B^2/2)$, radial component of the velocity   $\beta_r$  and value of the total \EM\ velocity  $\beta_{EM}$ for relativistically expanding spheromak  (\protect\ref{Prend}). The surface is located at $r=0.9992$ and moves with \Lf\ $\Gamma=25.4$ (for $\alpha =1$).
}
\label{vtot} 
\end{figure}

Moreover, while, $\alpha$, the constant of proportionality between $\Psi$ and $\beta$ is smaller this does not imply that the toroidal field becomes weaker. This is because the maximum of $f$ is pushed towards higher velocities. Given the factor $v/(1-v^2)$ that appears in the form of $T$, this term dominates, leading to a strong toroidal field. The part of the solution that is beyond the first positive root, shows an oscillatory behavior. This part corresponds to field lines that are not connected to the rest of the spheromak, thus we focus our attention to the inner sphere. Overall, the field is rather strong near the boundary of the bubble, Fig. \ref{vtot}. We note that in the limit of low expansion velocities ($v\ll 1$), where only the first order term is kept, it reduces to the standard spheromak solution as ${\rm arctanh} v\approx v$. 

Even though the field within the spheromak itself is force-free, as the bubble expands the energy within the bubble decreases as $ \propto 1/t$. Indeed the total energy is 
\be
E_{tot} = \frac{2\pi}{4\pi} \int _0^\pi \sin \theta d \theta
\int_0^{v_0 t} r^2  dr (E^2 +B^2) \propto \frac {1} {t}
   \ee
The energy decrease can be clearly seen from the requirement of the conservation of the magnetic flux: $B \propto 1/R^2$, so that $B^2 R^3  \propto 1/t $. This decrease in energy is the a work done by the expanding blob on the external medium. In the origin frame the surface \Bf\ is
\be
B_\theta^{(s)} = \frac{\left(\alpha ^2+1\right) \sin  \theta  \cos \left(\alpha  \tanh
   ^{-1}\left(v_0\right)\right)}{t^2 \left(1-v_0^2\right)}
   \label{B-theta}
   \ee
The corresponding work made by the pressure due to the expansion is 
\be 
pdV \propto \frac{B_\theta^{(s),2} }{\Gamma^2} R^2
\propto \frac{1}{t^2}
\ee
Thus, the energy of the explosion evolves according to
\ba  &&
\frac{ d E_{tot}}{dt}  \propto - \frac{1}{t^2}
\nn &&
E_{tot} \propto \frac{1}{t}
\ea
We note that the magnetic pressure at the surface of the bubble is highly anisotropic as is evident by the form of the meridional component of the field there, Equation (\ref{B-theta}), which is much stronger at the equator than higher latitudes. Thus, if such an expanding spheromak is embedded within a uniform medium it will expand primarily along the equator. Such a disturbance will affect its spherical boundary and the force-equilibrium at its interior. A possible extension could be the inclusion of a thermal pressure pressure profile, which removed the electric current discontinuity which is responsible for this anisotropy \citep{2010MNRAS.409.1660G,2012MNRAS.420..505G}.

\section{Discussion}

In this work we investigate the dynamics of relativistic magnetically driven  explosions.  
Our  analytical results,  MHD and force-free simulations consistently  show a  complicated, parameter-dependent picture, with various  important effects. Most importantly, the dynamics differs qualitatively from the fluid case. 
The most exemplary case, demonstrating the differences between highly magnetized  and fluid cases,  is the magnetic explosion in static external medium, \S \ref{static}. For example, Fig. \ref{fig:sphtv} demonstrates that low-$\sigma$ over-pressurized configurations create relativistically expanding flows which accelerated with time,  while high-$\sigma$ configurations shows tendency of de-acceleration and saturation to the constant  velocity. 

Similarly, just over-pressurized highly  magnetized cloud in constant density/constant confining \Bf\ would just puff-up - no explosion is generated. 
Slightly over-pressurized spheromak in a Hubble flow just quickly expand trying to rich pressure balance with preceding wind.  Only in the case of highly over-pressurized initial configuration, and the external wind environment the magnetic cloud ``explodes" - generates supersonically expanding (causally disconnected) outflow.

In case of over-pressured spheromak expansion into external wind, if the surrounding expansion speed is small  ($\eta_v = 0.01$,  or  $\eta_v = 0.032$) the spheromak evolution is similar to the steady case. The initial growth is not so fast and we see bounce from the external matter. 
 The fast expansion  of surrounding matter ($\eta_v = 0.1$) allows fast expansion of spheromak  and  generation of explosion-like, causally-disconnected expansion.
 In such the case  the spheromak transforms to shell like structure with ``hollow'' central part.

Thus, the internal structure of the explosion depends crucially on the type of expansion. Significant redistribution of the magnetic field take place in the case of supersonic expansion - \Bf\ is concentrated near the surface of the cloud: this is seen in MHD simulations and is analytically predicted,   see \S  \protect\ref{Prend1}.  On the other hand, if expansion is subsonic, the magnetic field keeps approximately  the initial configuration.

Puffing-up solutions did not show significant explosion like events. So it is difficult to expect significant observational manifestations. In other hand, explosion like solution can form shocks and particle acceleration. We shown what for the Hubble like expansion and over-pressured spheromaks if expansion speed exceed $>0.1 cr/a$. We can interpret it as formation of the spheromak at light cylinder with radius about $r_{sh}\approx 0.1 r_l$, so the variability time can be up to $t_{flare}\approx 2\pi/0.1\sim100$ times smaller compare to a magnetar rotation period.

Qualitatively  the strong explosion condition is given by Eq.  (\ref{condition}-\ref{condition2})   \citep[see also][]{2022MNRAS.509.2689L}. For a magnetar with period $P$ this gives for the initial  size and energy
\ba && 
R_0  \geq 150\, P^{-1/2}\,  {\rm meters} 
\nn && 
E_0 \geq \frac{B_{NS}^2 R_{NS}^{9/2} \Omega^{3/2}}{c^{3/2} }  = 5 \times 10^{39}\, P^{-3/2} \, {\rm erg} 
\ea
(period $P$ in seconds.)




 \section{Acknowledgements}
We are appreciated Prof.~S.~Nagataki and Dr.~J.~Mahlmann for useful discussion.  The 3D RMHD calculations were carried out in the CFCA cluster XC50 of National Astronomical Observatory of Japan.
 This work had been supported by 
NASA grants 80NSSC17K0757 and 80NSSC20K0910,   NSF grants 1903332 and  1908590. KNG acknowledges funding from grant FK 81641, University of Patras ELKE.

\section{Data availability}
The data underlying this article will be shared on reasonable request to the corresponding author.

 

\appendix

\section{Explicit expressions for magnetic fields}
\label{explicit}
For references we  give explicit forms for the fields.
For basic spheromak: 
\ba && 
\mathbf{B}_{in} =B_0^{(in)} \left\{\frac{2 \cos \theta \left(\alpha r\cos  (\alpha r)-\sin  (\alpha r)\right)}{\alpha^{3}r^{3}},\right.
\nn &&
 \sin \theta  \frac{\left[\left( \alpha^2r^2-1 \right)\sin (\alpha r) +   \alpha r \cos  (\alpha r) \right]}{\alpha^3r^3},
\nn && 
\left. \sin \theta  \frac{   \alpha r \cos  (\alpha r) - \sin (\alpha r)  }{\alpha^2r^2} \right\}
\nn &&
\mathbf{B}^{(out)} =B_0^{(out)} \left\{\left(1-\frac{a^3}{r^3}\right) \cos  \theta , - \left(1+\frac{a^3}{2 r^3}\right) \sin
    \theta ,0\right\}
   \nn &&
   \alpha = {4.49341}/{ a }
   \nn &&
  B_0^{(in)} = B_0^{(out)} = - \frac{2}{3}  \frac{\sin \alpha}{ \alpha}   B_0 =   6.90501  B_0
\label{eq:BspB}
\ea
Basic spheromak in decreasing external field
\be
\mathbf{B}^{(out)} =B_0^{(out)} \left\{a \left( \frac{r^3-a^3}{  r^4} \right) \cos  \theta , -2 a  \left( \frac{r^3+ 2 a^3}{4 r^4} \right) \sin    \theta ,0\right\}
    \ee

3d order spheromak:
\ba && 
\label{eq:Bhosp3}
\mathbf{B}_{in,ho} = \frac{B_0}{4\alpha^5r^5} \left\{4  \left(3\cos(\theta)+5\cos(3\theta) \right)   \left[ \alpha r \left(\alpha^2r^2 - 15\right)\cos(\alpha r) + 3(2\alpha^2r^2-5)\sin(\alpha r) \right] ,\right.
\nn &&
 \left(\sin(\theta)+5\sin(3\theta) \right)   \left[ 3\alpha r \left(2\alpha^2r^2 - 15\right)\cos(\alpha r) + (21\alpha^2r^2-\alpha^4r^4-45)\sin(\alpha r) \right] ,
\nn && 
\left. 2\alpha r \sin(\theta) \left(3+5\cos(2\theta) \right)   \left[ \alpha r \left(\alpha^2r^2 - 15\right)\cos(\alpha r) + 3(2\alpha^2r^2-5)\sin(\alpha r) \right]   \right\}
\ea
where $\alpha = {6.98793}/{a}$.

\section{The magnetic field structure for Hubble expansion case. }
\label{sec:mhdBz}

\begin{figure}[!ht]
	\includegraphics[width=.45\linewidth]{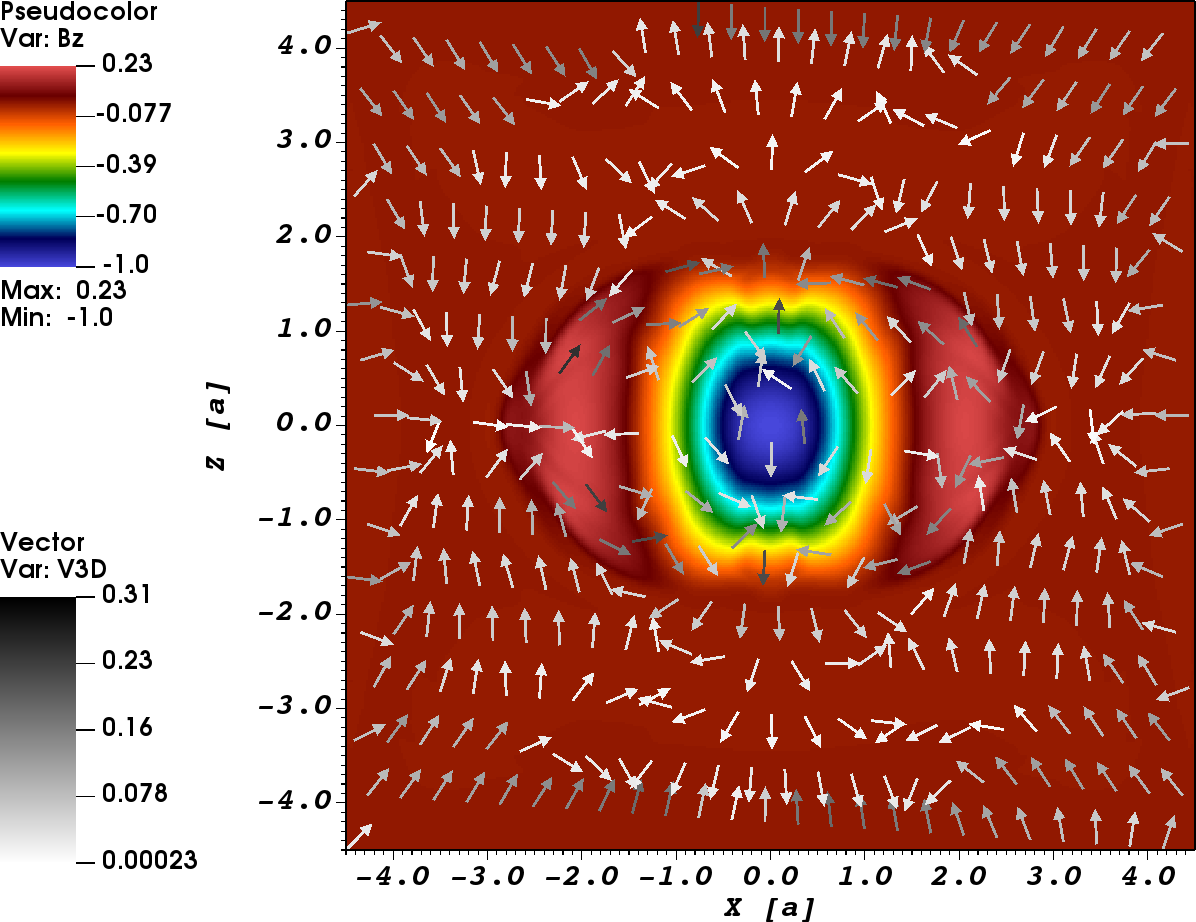}
	\includegraphics[width=.45\linewidth]{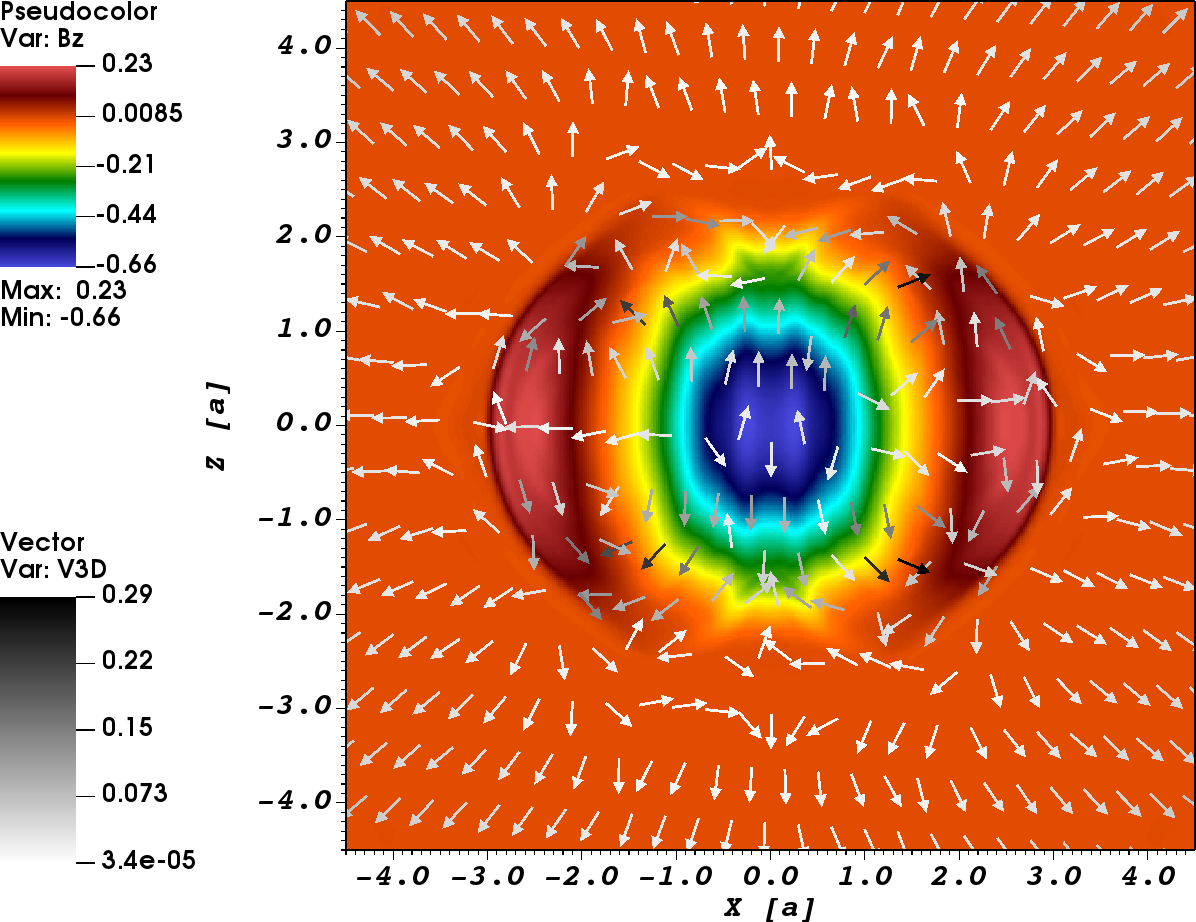}
	\includegraphics[width=.45\linewidth]{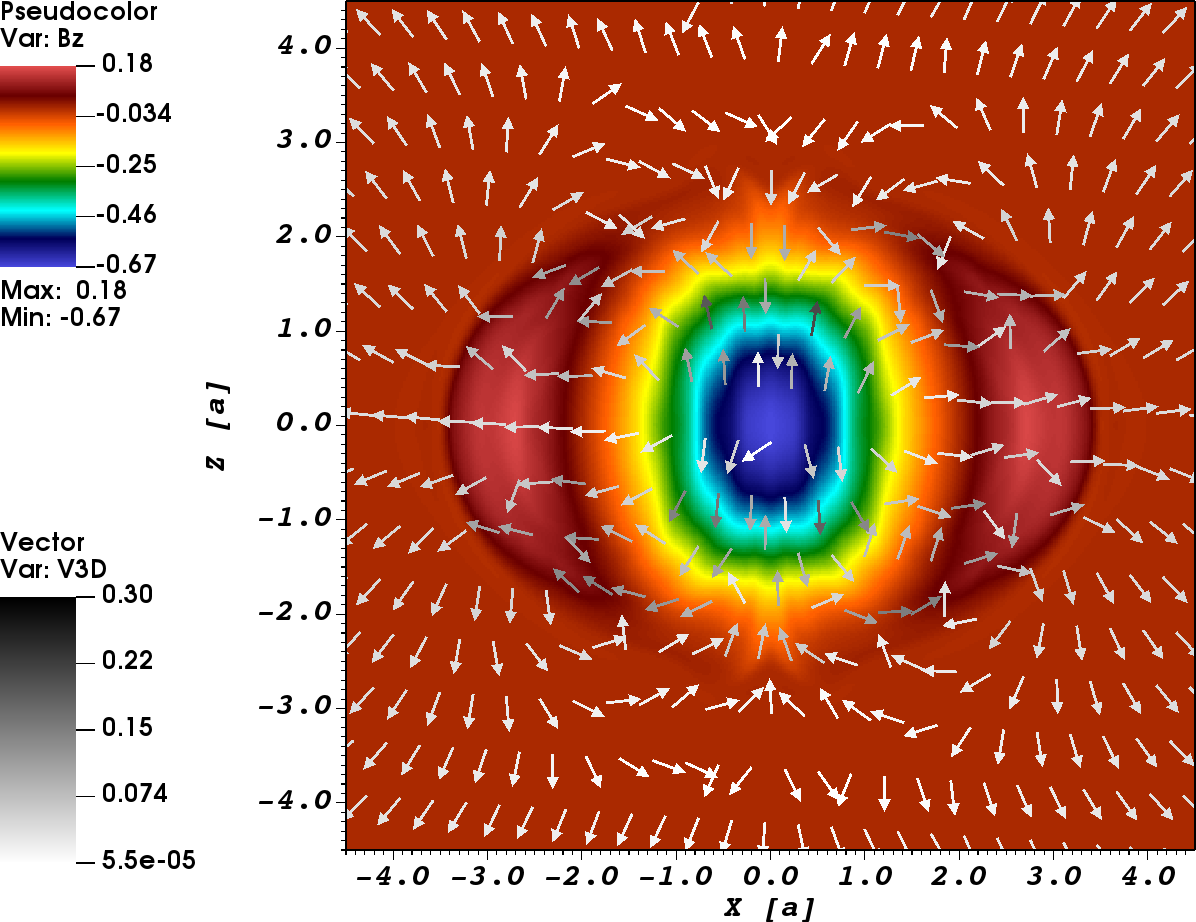}
	\includegraphics[width=.45\linewidth]{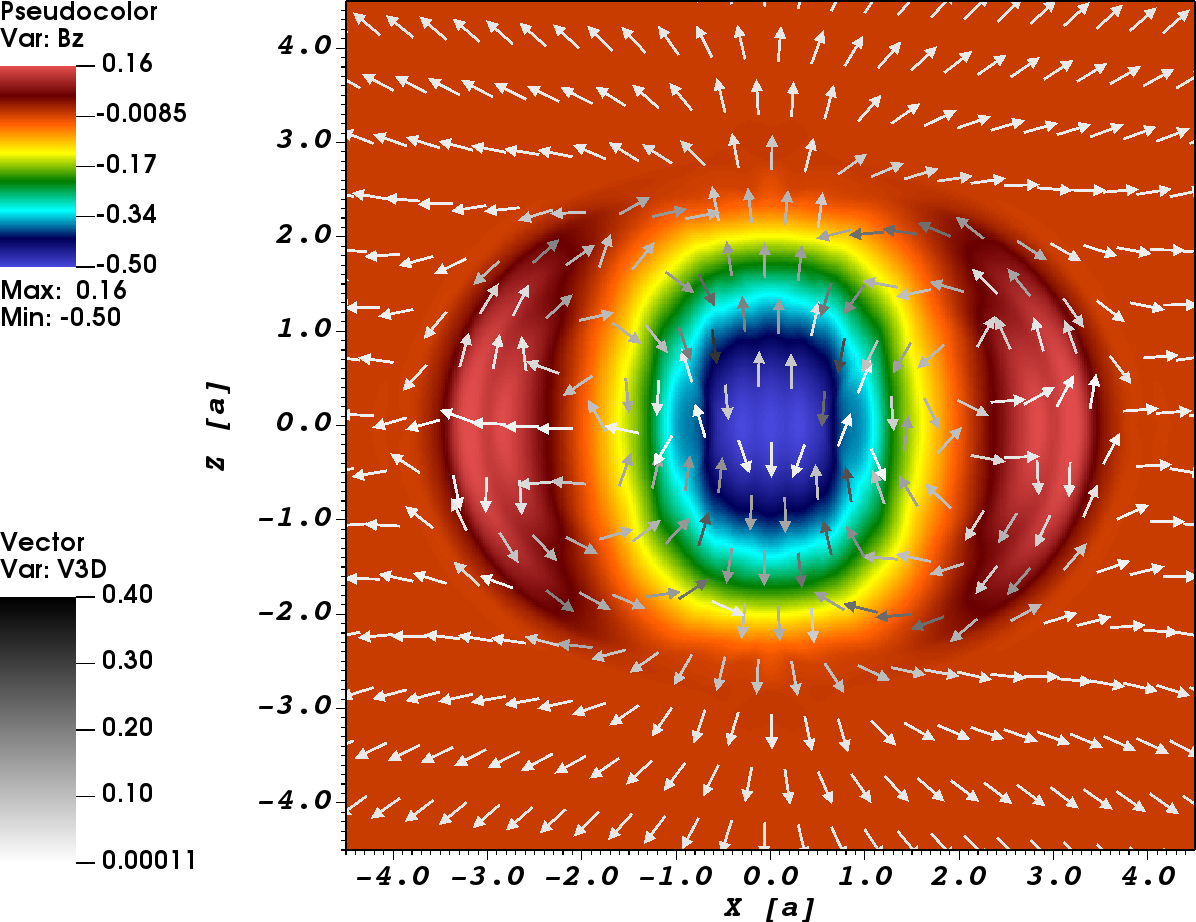}
	\includegraphics[width=.45\linewidth]{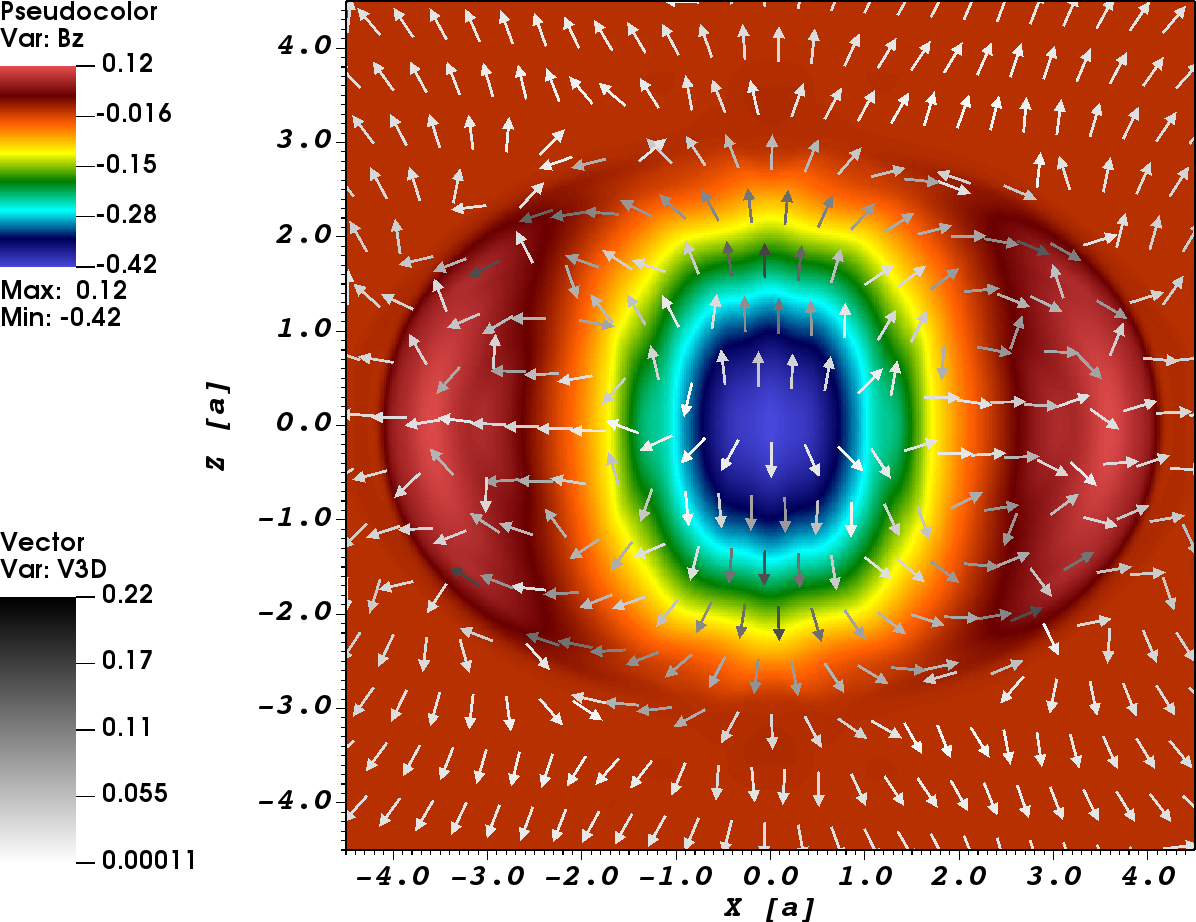}
	\includegraphics[width=.45\linewidth]{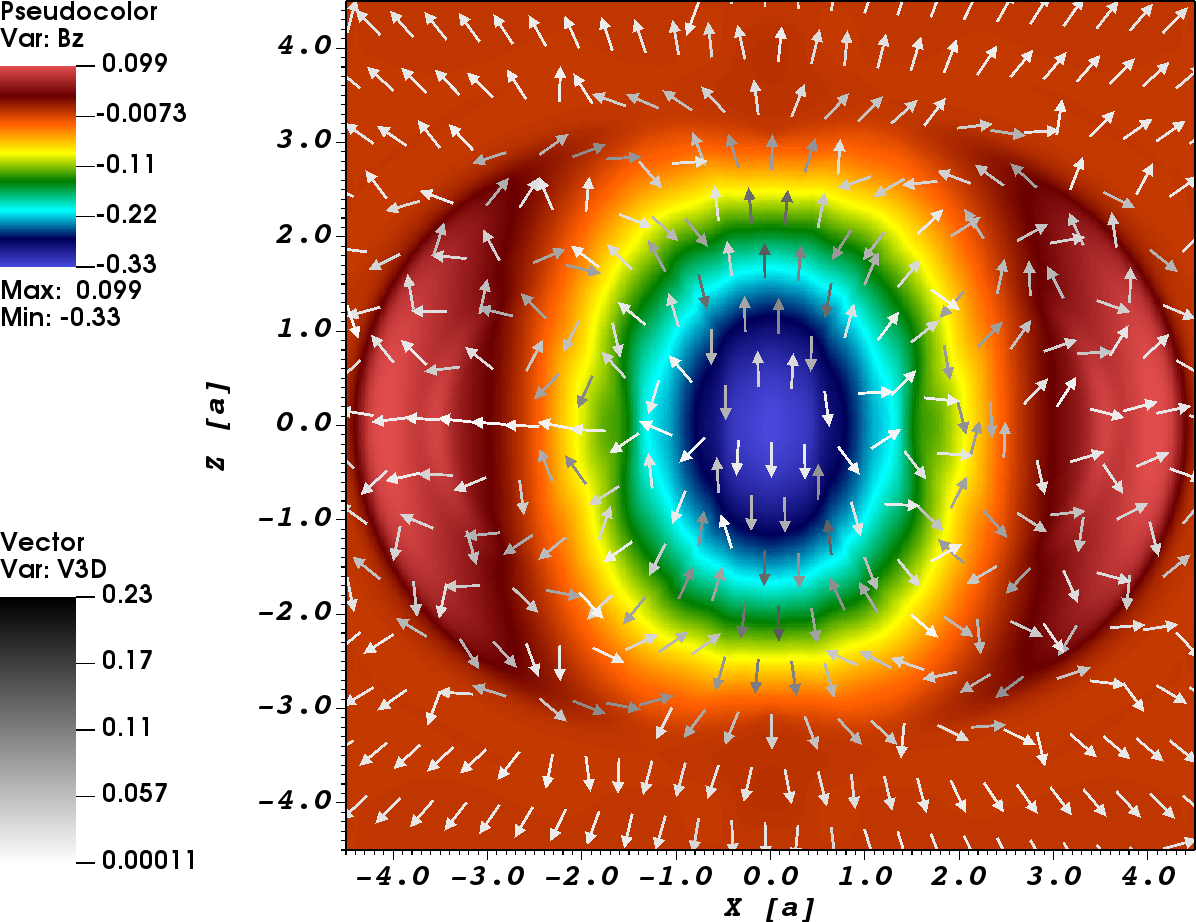}
	\caption{The spheromak expansion with Hubble expanding environment with $\eta_v=  0.01 $ for different time moments $t=10 [a/c]$   (top left), $t=20 [a/c]$   (top right), $t=30 [a/c]$   (middle left), $t=40 [a/c]$   (middle right), $t=60 [a/c]$   (bottom left), $t=80 [a/c]$   (bottom right). The color shows $B_z$ component of magnetic field and arrows show velocity field.}
	\label{fig:shubblem2}
\end{figure}  

\begin{figure}[!ht]
	\includegraphics[width=.49\linewidth]{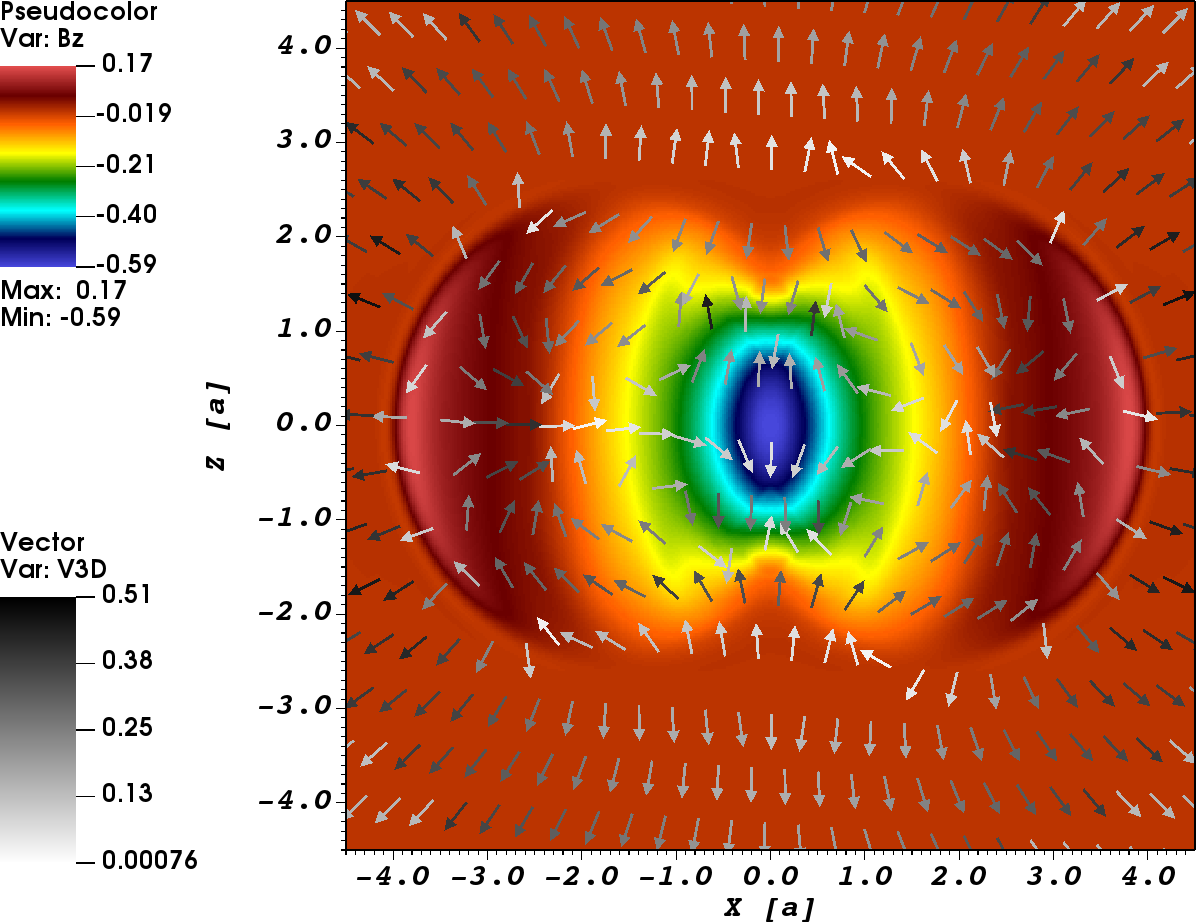}
	\includegraphics[width=.49\linewidth]{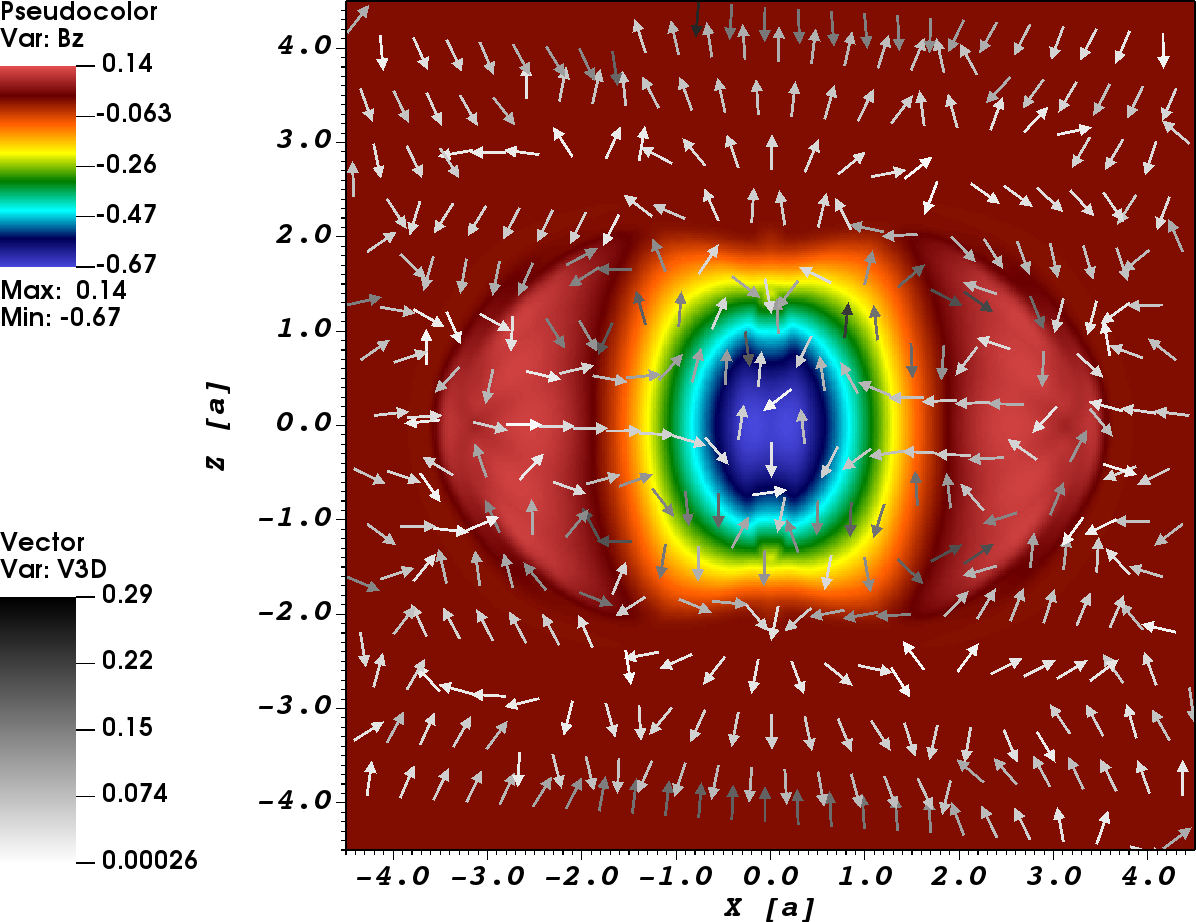}
	\includegraphics[width=.49\linewidth]{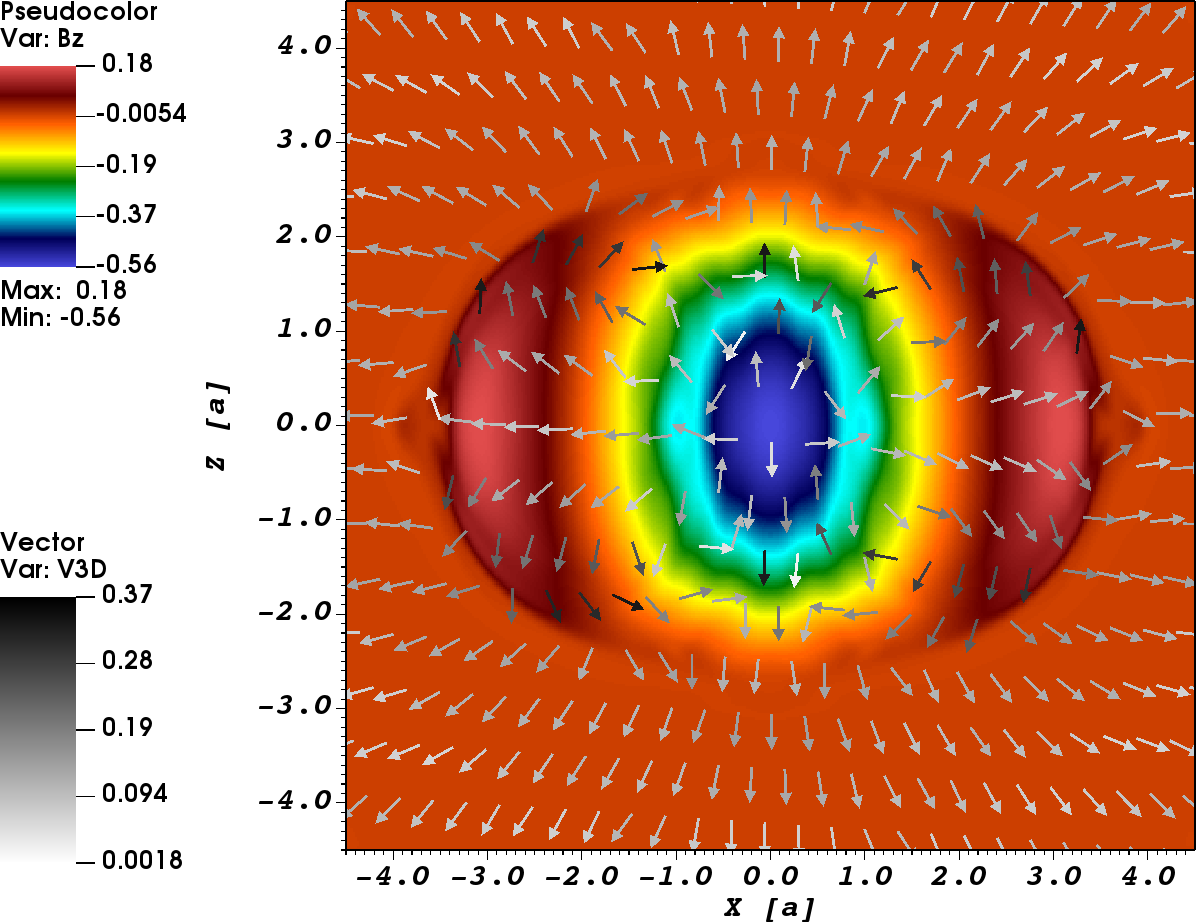}
	\includegraphics[width=.49\linewidth]{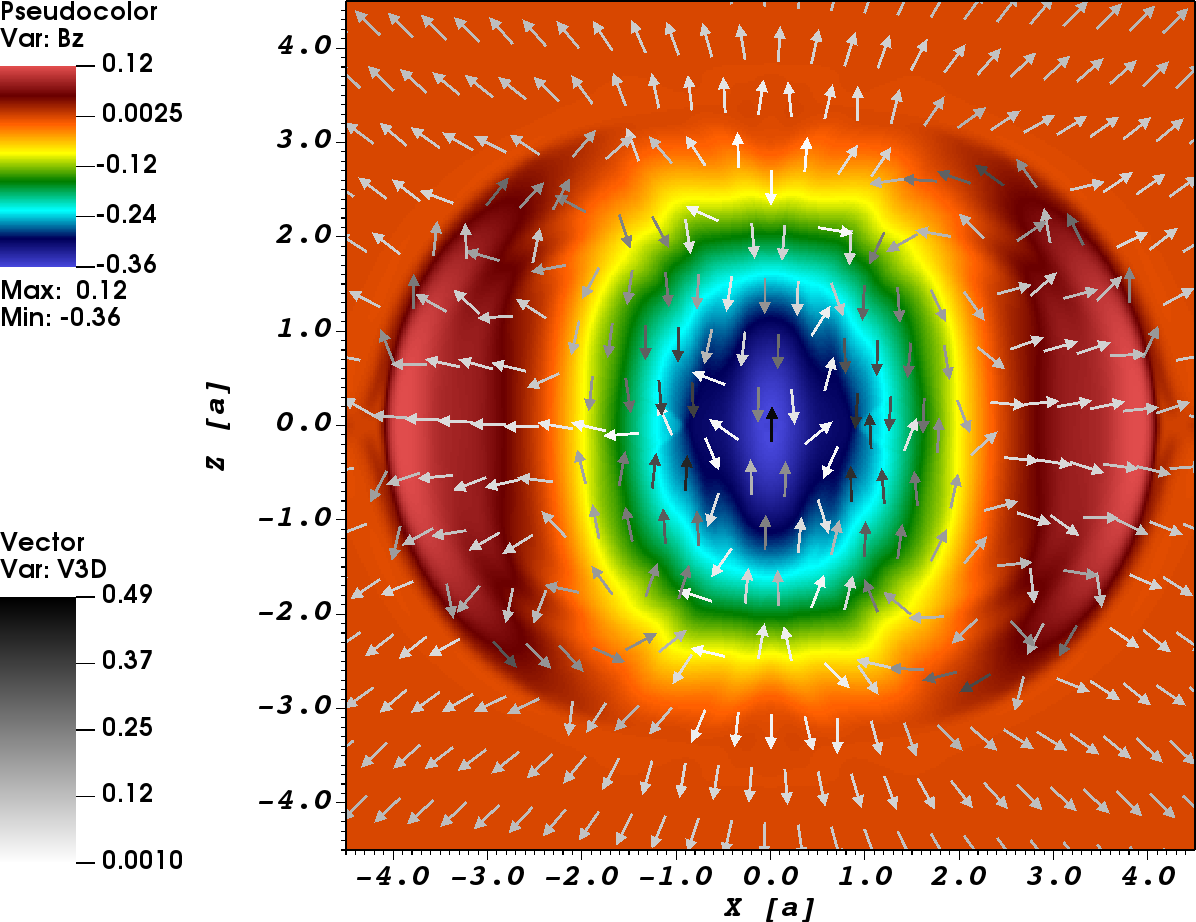}
	\caption{The spheromak expansion with Hubble expanding environment with $\eta_v = 0.032$ for different time moments  $t=5 [a/c]$ (top left),  $t=10 [a/c]$  (top right),  $t=15 [a/c]$ (bottom left),  $t=20 [a/c]$ (bottom right).  The color shows $B_z$ component of magnetic field and arrows show velocity field.}
	\label{fig:shubblem15}
\end{figure}

\begin{figure}[!th]
	\includegraphics[width=.49\linewidth]{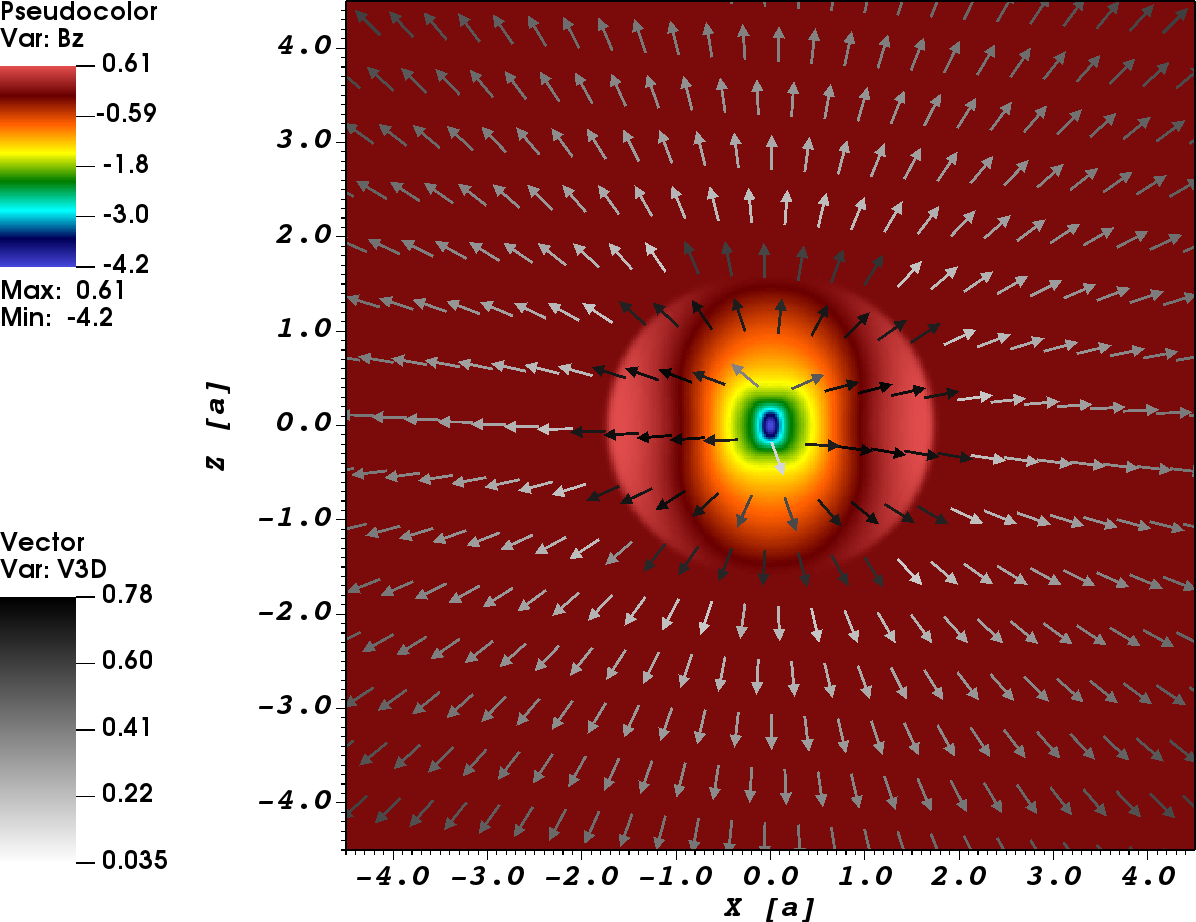}
	\includegraphics[width=.49\linewidth]{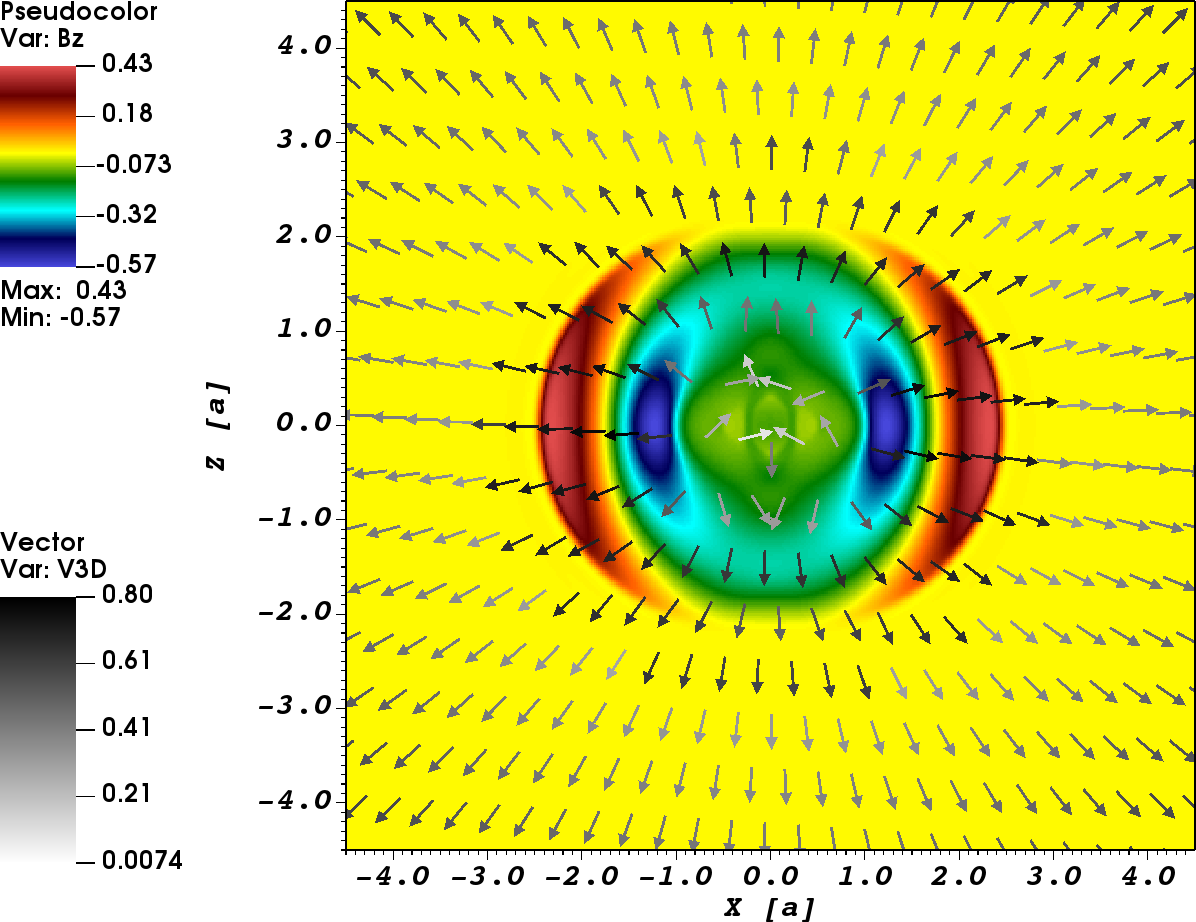}
	\includegraphics[width=.49\linewidth]{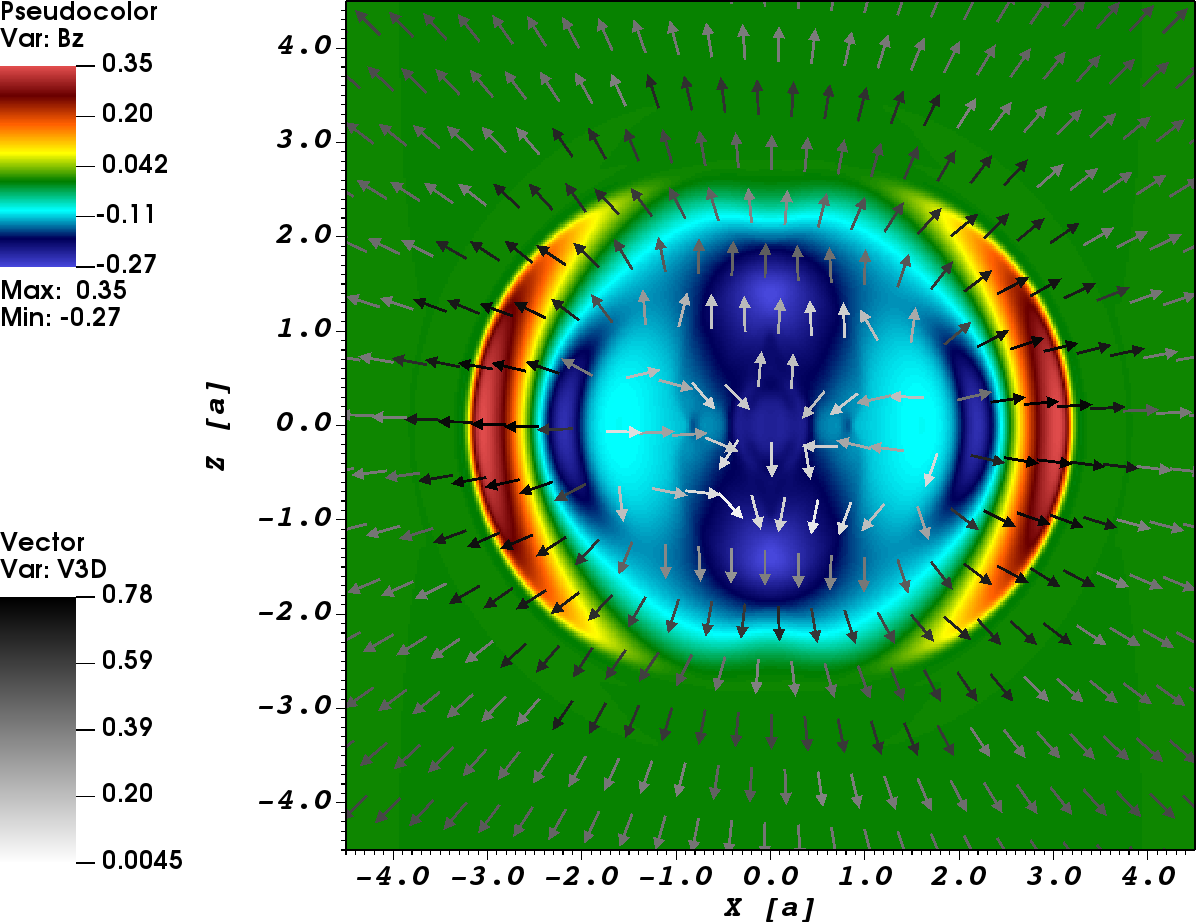}
	\includegraphics[width=.49\linewidth]{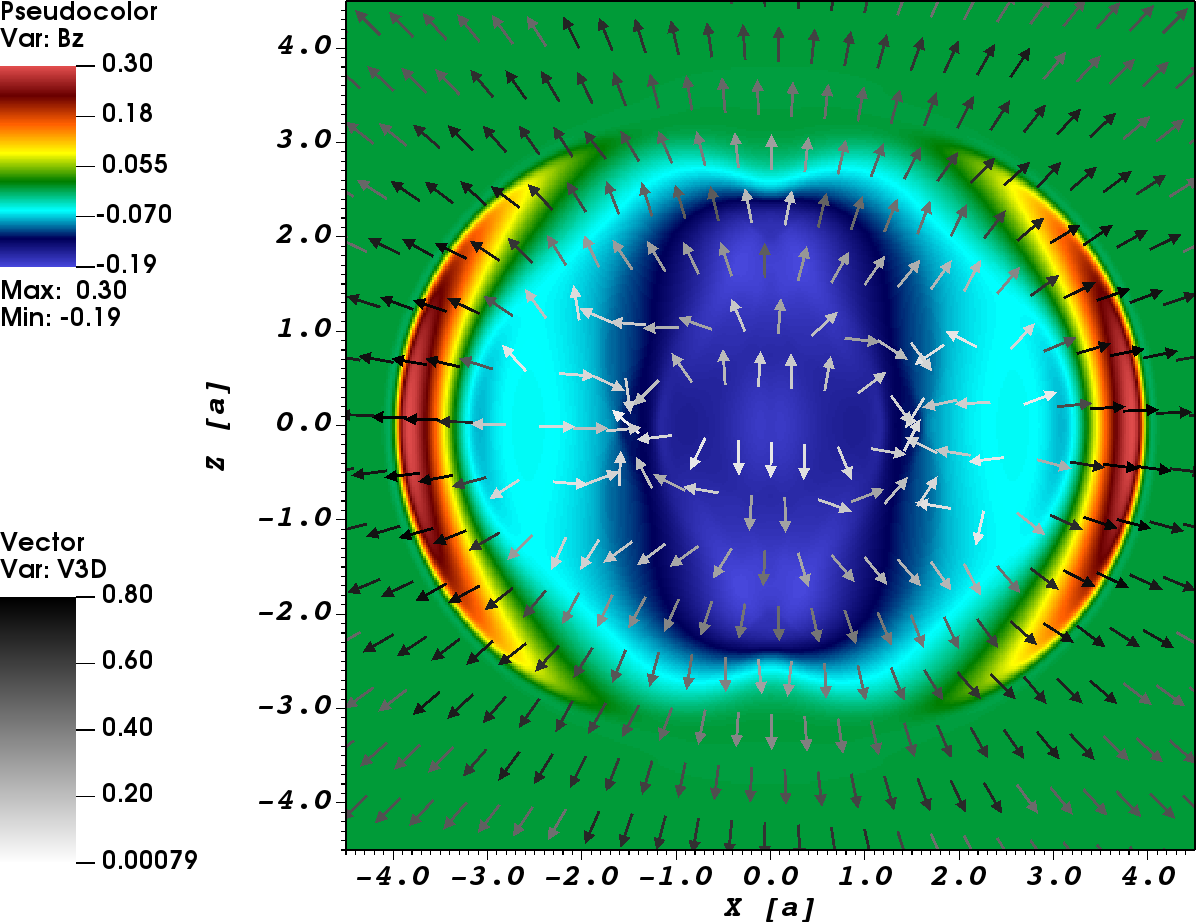}
	\caption{The spheromak expansion with Hubble expanding environment with $\eta_v = 0.1$ for different time moments $t=1 [a/c]$ (top left), $t=2  [a/c]$  (top right), $t=3 [a/c]$ (bottom left), $t=4 [a/c]$ (bottom right). The color shows $B_z$ component of magnetic field and arrows show velocity field.}
	\label{fig:shubblem1}
\end{figure}

The over-pressured spheromak evolve differently depends on Hubble expansion speed. To illustrate this point we show vertical component of the magnetic field.
If the surrounding expansion speed is small ($v_r = 0.01 \; c r/a$) Fig.~\ref{fig:shubblem2}  or ($v_r = 0.032\; c r/a$) Fig.~\ref{fig:shubblem15}  the spheromak evolution is similar to the steady case. Here the initial growth is not so fast and we see bounce from the external matter (see the velocity field evolution) and settle to quasi-static expansion. The internal structure of the magnetic field kept the same along all simulation. The relation of maximal to  minimal values  of the vertical component of the magnetic field is preserved ($\sim 0.3$).  

The fast expansion  of surrounding matter ($v_r = 0.1 \; c r/a$) leads to fast expansion of spheromak Fig.~\ref{fig:shubblem1} (see also Fig.~\ref{fig:mshubblem1}) and we expect explosion like expansion. In the fast expansion  case the internal structure of the spheromak changes significantly. The relation of  the maximal to  minimal values  of the vertical component of the magnetic field grows from $\sim 0.3$ till $\sim 1.5$.

\end{document}